\documentclass[11pt,a4paper]{article}
\usepackage[utf8]{inputenc}
\usepackage[english]{babel}
\usepackage{setspace}
\usepackage{amsmath,bm}
\usepackage{slashed}
\usepackage{amsfonts}
\usepackage{amssymb}
\usepackage{multirow}
\usepackage{cancel}
\usepackage{soul}
\usepackage[numbers,sort&compress]{natbib}
\usepackage{makeidx}
\usepackage{float}
\usepackage{booktabs}
\usepackage{mathtools}
\usepackage{tabu}
\usepackage{graphicx}
\usepackage{titlesec}
\usepackage{caption}
\usepackage{physics}
\usepackage{xcolor}
\usepackage{mathrsfs}
\usepackage{wasysym}
\usepackage{comment}
\usepackage{dsfont}
\usepackage{subcaption}
\usepackage{authblk}
\usepackage[left=2.5cm,right=2.5cm,top=2.5cm,bottom=2.5cm]{geometry}

\usepackage[colorlinks=true,
            linkcolor=blue,
            urlcolor=blue,
            citecolor=blue,
            anchorcolor=blue]{hyperref}
\usepackage{orcidlink}

\allowdisplaybreaks
\date{}

\title{{\Large\bfseries
Electroweak phase transitions in a $\bm{U(1)_D}$ extension of the standard model with dimension-six operators: Gravitational waves and LHC signatures
}}

\author[a]{\bf Arka Bhattacharyya\orcidlink{0009-0002-5851-5018}\thanks{\href{mailto:arka.bhattacharyya.phy23@gm.rkmvu.ac.in}{arka.bhattacharyya.phy23@gm.rkmvu.ac.in}}}

\author[a]{\bf Sanjoy Biswas\orcidlink{0000-0003-1305-8896}\thanks{\href{mailto:sanjoy.phy@gm.rkmvu.ac.in}{sanjoy.phy@gm.rkmvu.ac.in}}}

\author[b]{\bf Saurabh Niyogi\orcidlink{0009-0008-2355-8847}\thanks{\href{mailto:saurabhphys@gmail.com}{saurabhphys@gmail.com}}}

\affil[a]{{\it
Department of Physics, Ramakrishna Mission Vivekananda Educational and Research Institute,  
Belur Math, Howrah 711202, India}}

\affil[b]{{\it
Department of Physics, Gokhale Memorial Girls' College,  
1/1 Harish Mukherjee Road, Kolkata 700020, India}}

\begin{document}

\maketitle
\thispagestyle{empty}

\begin{abstract}
We investigate the possibility of realizing strong first-order electroweak phase transition (SFOEWPT) in an effective field theory framework where the Standard Model is extended with a complex scalar singlet ($\phi$) charged under a local $U(1)_D$ gauge group. The tree-level scalar potential contains a dimension-six term of the form $|H|^2|\phi|^4$. We show that this higher-dimensional operator plays a crucial role in the phase transition dynamics by weakening the correlation between the Higgs--singlet portal coupling and the scalar mixing angle that typically constrains singlet-extended models. Consequently, SFOEWPT can be achieved over a significantly extended region of parameter space. The strength of the phase transition is primarily driven by the vacuum expectation value (VEV) of the singlet scalar which plays a central role in this analysis.

We analyze the phase transition in this model and identify regions of parameter space consistent with SFOEWPT. The resulting phase transition can generate stochastic gravitational-wave signals potentially observable at future interferometers. The extended scalar sector in presence of the dimension-six operator also leads to distinctive multi-scalar production signatures at the LHC, intimately correlated with the singlet scalar VEV.
\end{abstract}

\clearpage
\pagenumbering{arabic}
\pagestyle{plain}


\section{Introduction}
The transition of the Universe from the $SU(2)_L \times U(1)_Y$ gauge symmetric phase to a broken phase characterized by a non-zero vacuum expectation value (VEV) of the Higgs field is referred to as the electroweak phase transition (EWPT). It is a pivotal epoch in the early Universe when the electromagnetic force and the weak force split apart to become the two distinct forces we see today. 
There is a strong motivation to connect EWPT with baryogenesis, the mechanism of generating the observed matter-antimatter imbalance in the Universe. In particular, a strong first order electroweak phase transition (SFOEWPT) provides a suitable environment in the early Universe which provides the ``departure from equilibrium" condition~\cite{Kuzmin:1985mm, Funakubo:1996dw}, one of the three key criteria laid down by Sakharov~\cite{Sakharov:1967dj} required for successful baryogenesis. These criteria suggest that sources of baryon number violation, $C$ and $CP$ violations, are also needed.

Within the standard model (SM), to have SFOEWPT one requires the mass of the Higgs to be below 40 GeV~\cite{Quiros:1999jp}. However, the Higgs-like scalar observed at the LHC with a mass of 125 GeV disfavors the realization of a SFOEWPT within the SM framework. In fact, in the SM the EWPT is a  smooth crossover where the Universe always stays in the minima of the Higgs potential. Therefore, one needs to go beyond the SM in order to achieve a SFOEWPT which is a necessary condition for sustainable electroweak baryogenesis.

The new physics (NP) effect can facilitate SFOEWPT by modifying the scalar potential. This can be achieved via introducing new degrees of freedom, adding higher dimensional operators in the scalar potential or both. Numerous studies have explored these possibilities. Scalar extensions of the SM is one of the most popular and well-studied frameworks in this context. The real singlet scalar ($S$) extension of the SM with exact $\mathcal{Z}_{2}$ symmetry, for example, has been considered in references \cite{Espinosa:1993bs, Espinosa:2007qk, Profumo:2007wc, Espinosa:2011ax, Barger:2011vm, Curtin:2014jma, Kurup:2017dzf, Chen:2025ksr} where the scalar field interacts with the SM via portal interaction of the form $|H|^2 S^2$, the so called $\mathcal{Z}_2$ symmetric Higgs-portal models. Such models can accommodate strong first order electroweak phase transition. Similarly, models with explicit $\mathcal{Z}_2$ symmetry breaking with a real or a complex scalar singlet have also been studied in the literature \cite{Profumo:2007wc, Espinosa:2011ax, Barger:2011vm, Choi:1993cv, Ham:2004cf, Noble:2007kk, Ashoorioon:2009nf, Das:2009ue, Profumo:2014opa, Kotwal:2016tex, Cho:2021itv, Chiang:2017nmu, Zhang:2023mnu, Biekotter:2025npc, Ghosh:2025rbt, Das:2026zuo}. These scenarios typically contain a larger set of free parameters and predict viable regions of parameter space consistent with  SFOEWPT.
In singlet scalar extended theories with spontaneous $\mathcal{Z}_2$ symmetry breaking \cite{Carena:2019une}, a SFOEWPT can be achieved for sizable scalar mixing angle and relatively light additional singlet scalar masses. However, both of these parameters are strongly constrained by the Higgs signal strength measurements and limits on the invisible decay branching ratio of the Higgs boson. This setup also serves as a prototype for complex singlet scalar extension of the SM, charged under the gauge symmetry of the dark sector. For example, in \cite{Hashino:2018zsi} the author considered an abelian $U(1)_D$ gauge symmetry in the dark sector which is spontaneously broken once the complex scalar field acquires a non-zero VEV and different phase transition patterns accrued from such a scenario.

Models that extend beyond renormalizable dimension-four operators, i.e., incorporate higher dimensional operators, do also have important implications in the realization of a FOEWPT. In particular, FOEWPT within the Standard Model Effective Field Theory (SMEFT) framework has been extensively explored by various authors~\cite{Barger:2011vm, Huang:2015tdv, Chung:2012vg, Noble:2007kk, Grojean:2004xa, Delaunay:2007wb, Cai:2017tmh, Chala:2018ari, Ellis:2018mja, Postma:2020toi, Camargo-Molina:2021zgz, Wagner:2023vqw, Gazi:2024boc, Banerjee:2024qiu, Jahedi:2025yjz}. Some of these studies considered operators beyond dimension six as well. For example, the contributions of dimension-eight operators in the context of EWPT in the SMEFT framework and their role in elevating the rather low cut-off scale required for EWPT have been investigated in \cite{Postma:2020toi, Banerjee:2024qiu}. An effective field theory (EFT) framework for the real singlet extension of the Standard Model has been studied in \cite{Oikonomou:2024jms}. Notably, in this scenario, the singlet scalar field does not acquire a VEV. This analysis suggests a moderate improvement in terms of the allowed parameter space consistent with SFOEWPT in the low mass region without any significant changes in the high mass regime compared to the $\mathcal{Z}_2$ symmetric Higgs portal scenario. However, both of these scenarios demand large quartic coupling involving the Higgs and the additional scalar field.

In this work, we investigate the possibility of achieving SFOEWPT in an EFT framework where the particle content of the SM is extended by a complex scalar singlet ($\phi$) charged under a local $U(1)_D$ symmetry. The gauge boson associated with this additional symmetry, commonly referred to as the dark photon which receives mass via a Higgs-like mechanism in the dark sector. The zero temperature tree-level scalar potential of the model contains dimension six terms in addition to the renormalizable dimension four terms permitted by the symmetries of the theory. In particular, we concentrate on the dimension six operator of the form $\frac{1}{\Lambda^2}|H|^2|\phi|^4$ where $\Lambda$ is the cut-off scale of the theory. The presence of dimension six operator helps to weaken the correlation between the scalar mixing angle ($\sin\theta$) and the scalar portal coupling ($\lambda_{hs}$), thus allowing a broader region of parameter space compatible with SFOEWPT compared to the case without the dimension-six contribution~\cite{Carena:2019une}. Moreover, SFOEWPT compatible parameter space is obtained for $|\lambda_{hs}|\sim 1$ without threatning the perturbative unitarity. Importantly, the phase transition pattern in this scenario is sensitive to the ratio $\frac{w}{\Lambda}$, where $w$ is the VEV of the additional singlet scalar. Consequently, the effects of the higher-dimensional operator does not necessarily decouple for large $\Lambda$ as long as the ratio $\frac{w}{\Lambda}$ is held fixed. This is in stark contrast to the SMEFT framework, where the electroweak VEV is fixed at 246 GeV, and from extensions with higher-dimensional operators involving multiple scalar fields, in which the additional scalar does not acquire a VEV \cite{Oikonomou:2024jms} and consequently decouple in the large $\Lambda$ limit. We identify the region of parameter space spanned by the scalar mixing angle, mass and VEV of the singlet scalar that not only consistent with theoretical and existing experimental constraints but also predicts SFOEWPT.

Theories predicting a first-order phase transition (FOPT) also have observable consequences in the form of stochastic gravitational waves (GW) generated by bubble nucleation, expansion, and subsequent collisions. With the rapid advancement of gravitational wave detectors, the detection of feeble and highly relic GW backgrounds may soon be within reach of next-generation interferometer experiments such as LISA~\cite{LISA:2017pwj}, BBO~\cite{Yagi:2011wg}, DECIGO~\cite{Nakayama:2009ce}, $\mu$Ares~\cite{Sesana:2019vho},  ultimate-DECIGO (UD), DECIGO-corr (DC) and ultimate-DECIGO-corr (UDC)~\cite{Nakayama:2009ce}. The model presented in this work can also be confronted with future GW observations.

Moreover, this scenario is testable at the high luminosity LHC (HL-LHC), through measurements of multi-scalar production, in particular di-scalar ($h_1h_1$, $h_1 h_2$, $h_2 h_2$) and triple-Higgs ($h_1 h_1 h_1$) production. We show that the singlet scalar VEV plays a key role in this context and the cross-section of the above processes are intimately correlated with it. This is unlike in the $\mathcal{Z}_2$ symmetric Higgs-portal framework with or without the effective operator considered in the literature where collider signal is challenging as the additional scalar does not mix with the SM Higgs.

The manuscript is organized as follows: in Section~\ref{model}, we introduce the model under consideration and discuss the relevant theoretical and experimental constraints on its parameter space. In Section~\ref{phase_transition}, we study the electroweak phase transition in this model, identify the region of parameter space that supports SFOEWPT and examine its dependence on the model parameters. Section~\ref{nucleat} presents a discussion of the associated bubble nucleation process. Section~\ref{gravwav} is dedicated to the analysis of the GW signals generated by the SFOEWPT. In Section~\ref{multisc}, we study di-scalar and triple-Higgs production at the LHC. Finally, we summarize our findings and present our concluding remarks in Section~\ref{conc}.

\section{The model}
\label{model}
We have considered an effective field theory where the SM gauge group is extended with a local $U(1)_D$ symmetry. The local $U(1)_D$ symmetry is spontaneously broken by the non zero vacuum expectation value (VEV) of a additional complex scalar field $\phi$ singlet under SM gauge interactions. As a consequence the corresponding gauge boson, to be referred as dark photon, becomes massive.
The scalar and the electroweak gauge sector of the low energy effective theory is given by 
\begin{align}
    & {\mathcal{L}}_{\rm gauge} = -\frac{1}{4}W_{\mu\nu}^i W^{i,\mu\nu}-\frac{1}{4}B_{\mu\nu} B^{\mu\nu}+\frac{\varepsilon}{2\cos\theta_W}B_{d,\mu\nu}B^{\mu\nu}-\frac{1}{4}B_{d,\mu\nu} B_d^{\mu\nu}, \label{lgauge} \\
    & \mathcal{L}_{\text{scalar}} = |D_{\mu}H|^2+|D_{\mu}\phi|^2 - V_0(H,\phi), \label{lscalar}
\end{align}

where $W_{\mu\nu}^i$ with $i=1,2,3$ are the field strength tensors of the  $SU(2)_L$ gauge group, $B_{\mu\nu}$ is that of the $U(1)_Y$  and $B_{d,\mu\nu}$ corresponds to the field strength tensor of the additional local $U(1)_d$ group. The strength of the kinetic mixing involving the field strength tensors of two abelian gauge groups is parametrized by $\frac{\varepsilon}{\cos\theta_W}$, where $\cos\theta_W$ is the cosine of the Weinberg angle.

The tree-level scalar potential introduced above, in presence of the dimension six term which plays a crucial role in the context of the electroweak phase transition is given by
\begin{equation}
	V_0(H,\phi)=	-\mu _h ^2 \abs{H}^2 + \lambda _h \abs{H}^4 -\mu _s ^2 \abs{\phi}^2 + \lambda _s \phi ^4 + \lambda _{hs}\abs{H}^2 \phi ^2 + \frac{c_6}{\Lambda^2}\abs{H}^2 \phi ^4, \label{v01}
\end{equation}
where $\Lambda$ represents the cut-off scale of the theory. The Higgs doublet $H$ and the complex singlet $\phi$ can be explicitly written as
\begin{equation}
	H= \frac{1}{\sqrt{2}}\begin{pmatrix}
		\chi_1 + i \chi _2 \\
		\phi_1 + i \chi _3
	\end{pmatrix} \qquad \phi = \frac{1}{\sqrt{2}} \left(\phi_2 + i \chi _4 \right), \label{hs}
\end{equation}
where $\chi_i$'s ($i=1,2,3,4$) are the would-be Goldstone modes. 
Since we are working in a scenario where the low energy theory contains massive electroweak gauge bosons and a dark photon, therefore, the zero temperature ground state field configuration should be such that both the scalar fields always acquire non-zero vacuum expectation values.
Hence, in this scenario the region of parameter space of interest to us is such that $(v,w)$ with $v>0 \ \text{and} \ w>0$ corresponds to the global minima of the tree-level potential, where $v$ and $w$ represent the vacuum expectation values of the SM Higgs and additional scalar fields, respectively.
In the unitary gauge the tree-level potential takes the form 
\begin{equation}
	V_0(\phi_1, \phi_2) =-\frac{1}{2} \mu_h^2 \phi_1^2 + \frac{1}{4} \lambda_h \phi_1^4 - \frac{1}{2} \mu_s^2 \phi_2^2 + \frac{1}{4} \lambda_s \phi_2^4  + \frac{1}{4} \lambda_{hs} \phi_1^2 \phi_2^2 + \frac{1}{8\Lambda^2} \phi_1^2 \phi_2^4 \label{v0},
\end{equation}
assuming $c_6=1$.
The minimization conditions of the above potential which determine $v$ and $w$ can be written as
\begin{align}
	&\frac{1}{v} \left. \frac{\partial V_0 (\phi_1, \phi_2)}{\partial \phi_1} \right|_{\substack{\phi_1= v \\ \phi_2 = w }} = -\mu_h^2 + \lambda_h v^2 + \frac{1}{2} \lambda_{hs} w^2 + \frac{w^4}{4 \Lambda^2}=0 ,\label{vev1} \\
	&\frac{1}{w} \left. \frac{\partial V_0 (\phi_1,\phi_2)}{\partial \phi_2} \right|_{\substack{\phi_1 = v \\ \phi_2 = w }} = -\mu_s^2 + \frac{1}{2} \lambda_{hs} v^2 + \lambda_s w^2 + \frac{v^2 w^2}{2 \Lambda^2} =0. \label{vev2}
\end{align}

In particular, we require $v=v_{_{\text{EW}}}= 246 $ GeV at zero temperature. Around this minima ($v_{_{\text{EW}}},w$) one can expand the field $H$ and $\phi$ in terms of the fluctuations $h$ and $s$, respectively as (in the unitary gauge )
\begin{equation}
	H= \frac{1}{\sqrt{2}}\begin{pmatrix}
		0 \\
		v_{_{\text{EW}}} + h
	\end{pmatrix} \qquad \phi = \frac{1}{\sqrt{2}} \left(w + s \right) .\label{hs1}
\end{equation}

The tree-level scalar mass matrix evaluated at this minima ($v_{_{\text{EW}}},w$) can be written as (using  Equation \ref{vev1} and \ref{vev2} )
\begin{equation}
	M^2 = \begin{pmatrix}
		2 v_{_{\text{EW}}}^2 \lambda_h & v_{_{\text{EW}}} w \left(\frac{w^2}{\Lambda^2} + \lambda_{hs}\right) \\
		v_{_{\text{EW}}} w \left(\frac{w^2}{\Lambda^2} + \lambda_{hs}\right) & -v_{_{\text{EW}}}^2 \lambda_{hs} + 2 \mu_s^2
	\end{pmatrix}
    \label{massmatrix}
\end{equation}

The basis in which the mass matrix is diagonal with corresponding eigenvalues $M_{h_1}^2$ and $M_{h_2}^2$, representing the physical masses of Higgs and additional scalar particle, is related to the $h-s$ basis by the following orthogonal transformation.
\begin{equation} 
    \begin{pmatrix}
	 h \\
	s
	\end{pmatrix} = 
	\begin{pmatrix}
	\cos\theta & \sin\theta \\
	-\sin\theta & \cos\theta
	\end{pmatrix} \begin{pmatrix}
	h_1\\
	h_2
	\end{pmatrix} 
    \label{rotation}
\end{equation}
where $\sin\theta$ is the mixing angle between the two scalars.

The relevant physical parameters of this sectors are $\{M_{h_1}, M_{h_{2}}, \sin \theta , v_{_{\text{EW}}}, w, \Lambda \}$. The Lagrangian parameters
$\{\mu_h, \mu_s, \lambda_h, \lambda_{s}, \lambda_{hs} \}$ can be recast in terms of the physical parameters and listed in Appendix~\ref{appendix1}.

\subsection{Constraints}
\subsubsection{Theoretical constraints}
\label{thcons}
\begin{figure}[h!]
  \centering

  \begin{subfigure}[b]{0.48\textwidth}
    \centering
    \includegraphics[width=\textwidth]{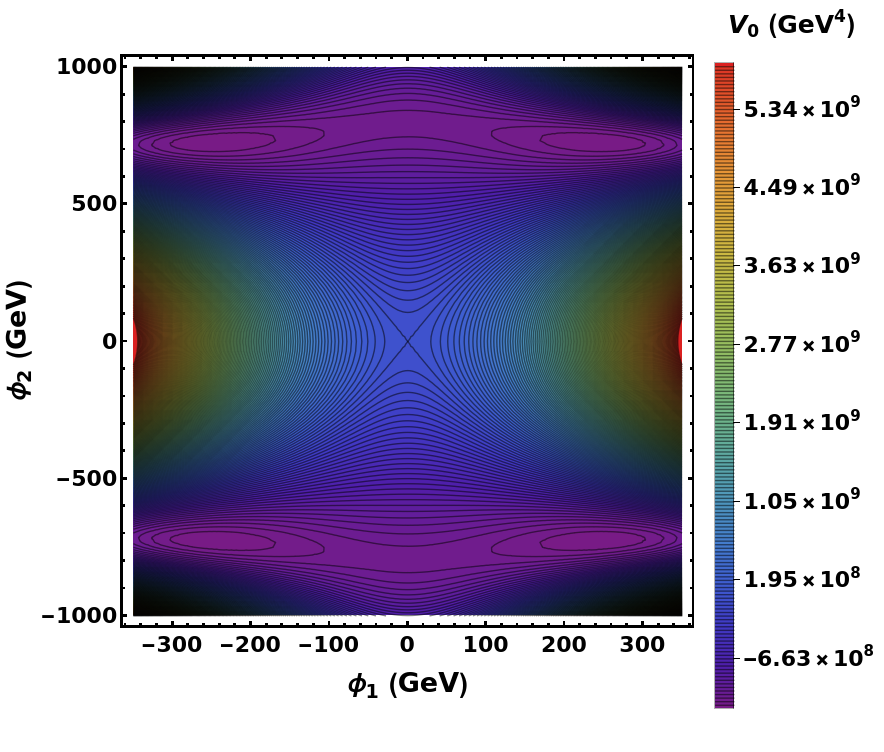}
    \caption{}
    \label{pota}
  \end{subfigure}
  \hfill
  \begin{subfigure}[b]{0.48\textwidth}
    \centering
    \includegraphics[width=\textwidth]{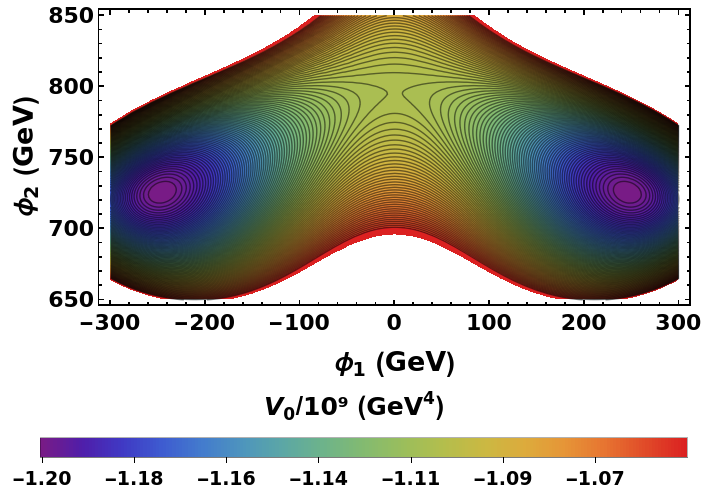}
    \caption{}
    \label{pota1}
  \end{subfigure}

  \vspace{0.8em}

  \begin{subfigure}[b]{0.48\textwidth}
    \centering
    \includegraphics[width=\textwidth]{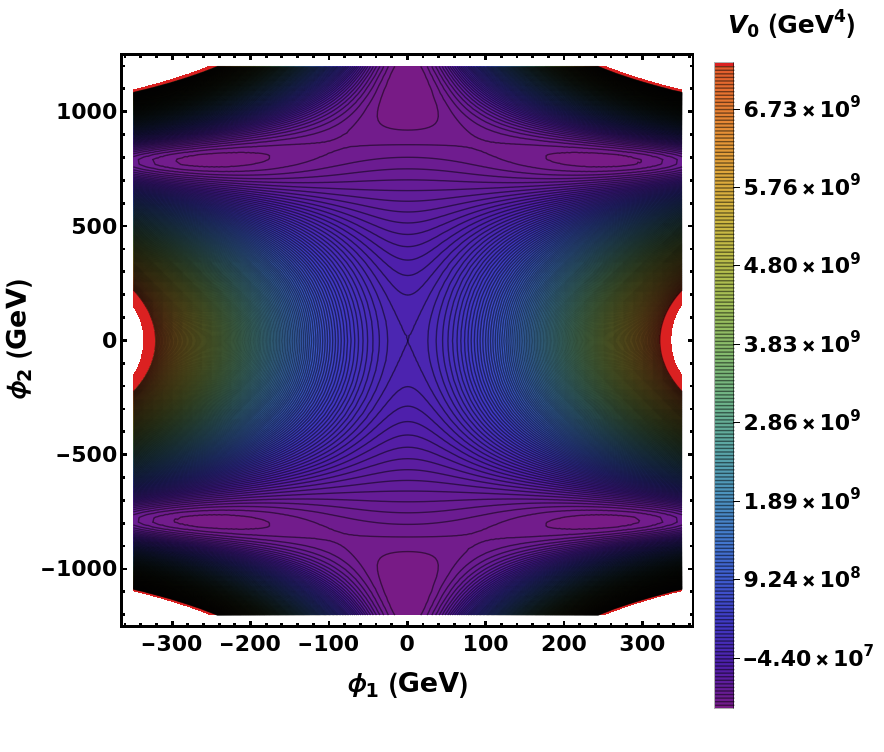}
    \caption{}
    \label{potb}
  \end{subfigure}
  \hfill
  \begin{subfigure}[b]{0.48\textwidth}
    \centering
    \includegraphics[width=\textwidth]{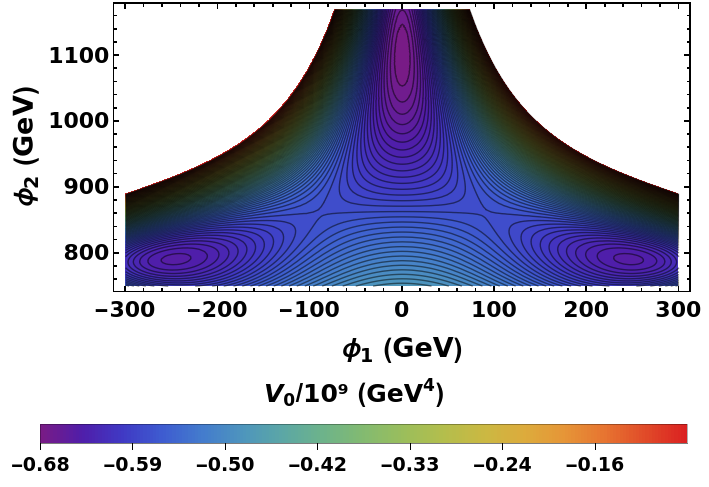}
    \caption{}
    \label{potb1}
  \end{subfigure}

  \caption{Vacuum structure of the tree-level potential given in Equation~\ref{v0} for two representative sets of Lagrangian parameters. Panels (a) and (b) correspond to ($\mu_h^2 = -89746$ GeV$^2$, $\mu_s^2 = 7002$ GeV$^2$, $\lambda_h =0.1378$, $\lambda_s= 0.011$, $\lambda_{hs} = -0.784 $) where $(246~\text{GeV},725.7~\text{GeV})$ is the global minima with a secondary extrema at $(0,795.73~\text{GeV})$. While panels (c) and (d) represent
  ($\mu_h^2 = -133321$ GeV$^2$, $\mu_s^2 = 2231$ GeV$^2$, $\lambda_h =0.1378$, $\lambda_s= 0.00184$, $\lambda_{hs} = -0.942 $) where $(246~\text{GeV},790.89~\text{GeV})$ is a local minima with the global minima at ($0, 1101.65~\text{GeV}$). The right panels show a rescaled $\phi_2$ axis for better visualization. The value of the potential $V_0$ is indicated by the color gradient. The cutoff scale is fixed at $\Lambda = 800~\mathrm{GeV}$.}
  \label{pot}
\end{figure}

\begin{itemize}
\item We will now analyze various constraints on the Lagrangian parameters. First, we demand the tree-level scalar potential should be bounded from below at large field  values or at least it should display a pattern consistent with it for field values less than the cut-off scale in an effective field theory framework. This is ensured in this model by the following requirement:
\begin{equation}
    \lambda_s , \ \lambda_h >0 \ \text{and} \ c_6 >0 \label{bbcond}
\end{equation}

\item  The validity of the EFT approach additionaly require $v, w < \Lambda$ for $c_6 \sim 1$ \cite{Falkowski:2015fla}. For the rest of our discussions we will work with $c_6 = 1$.

\item We require the tree-level potential to have a global minima at ($v_{_{\text{EW}}},w$). However, the tree-level potential in principal can have minima at $(0, u= \sqrt{\frac{\mu_s^2}{\lambda_s}})$ or at ($0,0$) as well at zero temperature. We illustrate this in Figure \ref{pot}, where the upper panel (Figure~\ref{pota} and \ref{pota1}) display a vacuum structure consists of a global minima at ($v_{_{\text{EW}}},w$). While the lower panel (Figure~\ref{potb} and \ref{potb1} ) display the same but with a global minima at ($0,u$). To ensure that ($v=v_{_{\text{EW}}}, w $) is the global minima at zero temperature, the model parameters must satisfy the following conditions in addition to the boundedness from below condition mentioned above :
\begin{align}
    &2 v_{_{\text{EW}}}^{4}\lambda_h - 2 u^{4}\lambda_s + 2 w^{4}\lambda_s
+ v_{_{\text{EW}}}^{2}\!\left(\frac{w^{4}}{\Lambda^{2}} + 2 w^{2}\lambda_{hs} - 4\mu_h^{2}\right)
+ 4 u^{2}\mu_s^{2} - 4 w^{2}\mu_s^{2} <0 , \label{bbcond1}\\
    & 2 v^{4}\lambda_h + 2 w^{4}\lambda_s
+ v^{2}\!\left(\frac{w^{4}}{\Lambda^{2}} + 2 w^{2}\lambda_{hs} - 4\mu_h^{2}\right)
- 4 w^{2}\mu_s^{2}<0. \label{bbcond2}
\end{align}

 The first one ensures $V_0(v_{_{\text{EW}}},w )< V_0(0, u)$  while the second one guarantees $V_0(v_{_{\text{EW}}},w)< V_0(0,0)$. Out of these two conditions, the first one gives more stringent restrictions on the choice of parameter space. 
In terms of physical parameters ( $M_{h_1}, M_{h_2}, v_{_{\text{EW}}}, w, \sin \theta, \Lambda$) Equation~\ref{bbcond1} can be re written as (see Appendix~\ref{appendix1})

\begin{equation}
V_0(v_{_{\text{EW}}},w )- V_0(0, u)=
\frac{
v_{_{\text{EW}}}^2
\left(
- M_{h_1}^2 M_{h_2}^2
+ \frac{v_{_{\text{EW}}}^2 w^2}{\Lambda^2}
\left(
M_{h_1}^2 \cos^2 \theta
+ M_{h_2}^2 \sin^2 \theta
\right)
\right)
}{
8 \left(
M_{h_1}^2 \sin^2 \theta
+ M_{h_2}^2 \cos^2 \theta
- \frac{v_{_{\text{EW}}}^2 w^2}{\Lambda^2}
\right)
} <0
\label{gm}
\end{equation}

For a given $M_{h_2}$ and $\sin\theta$, the upper bound set on the $w$ by the above equation supersedes that previously obtained by the restriction $w<\Lambda$.\\
In the limit either $M_{h_2}\to M_{h_1}$ or $\sin\theta\to 0 $ tree-level potential difference $V_0(v_{_{\text{EW}}},w )- V_0(0, u)$ takes the simple form
\begin{equation}
    \lim_{M_{h_2}\to M_{h_1}} V_0(v_{_{\text{EW}}},w )- V_0(0, u)=  \lim_{\sin\theta \to 0} V_0(v_{_{\text{EW}}},w )- V_0(0, u)= - \frac{M_{h_1}^2 v_{_\text{EW}}^2}{8} \ , \label{gmlimit}
\end{equation}
which is constant for a given $M_{h_1}$ and $v_{_{\text{EW}}}$, and consistent with Equation~\ref{gm}.

\end{itemize}

\begin{figure}[h!]
	\centering
	\begin{subfigure}[b]{0.49\textwidth}
		\centering
		\includegraphics[width=\textwidth]{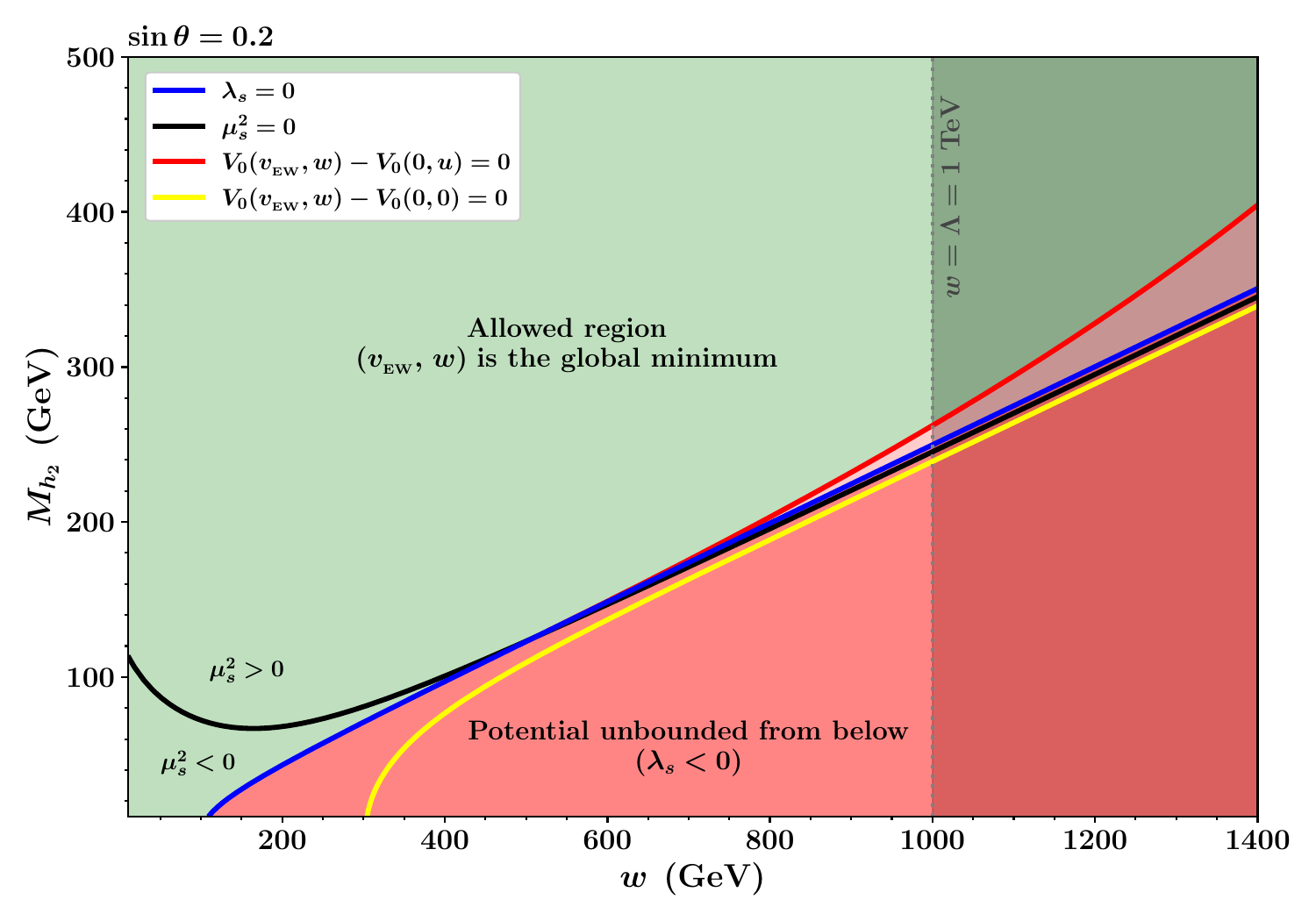}
        \caption{}
        \label{reg1}
	\end{subfigure}
	\hfill
	\begin{subfigure}[b]{0.49\textwidth}
		\centering
		\includegraphics[width=\textwidth]{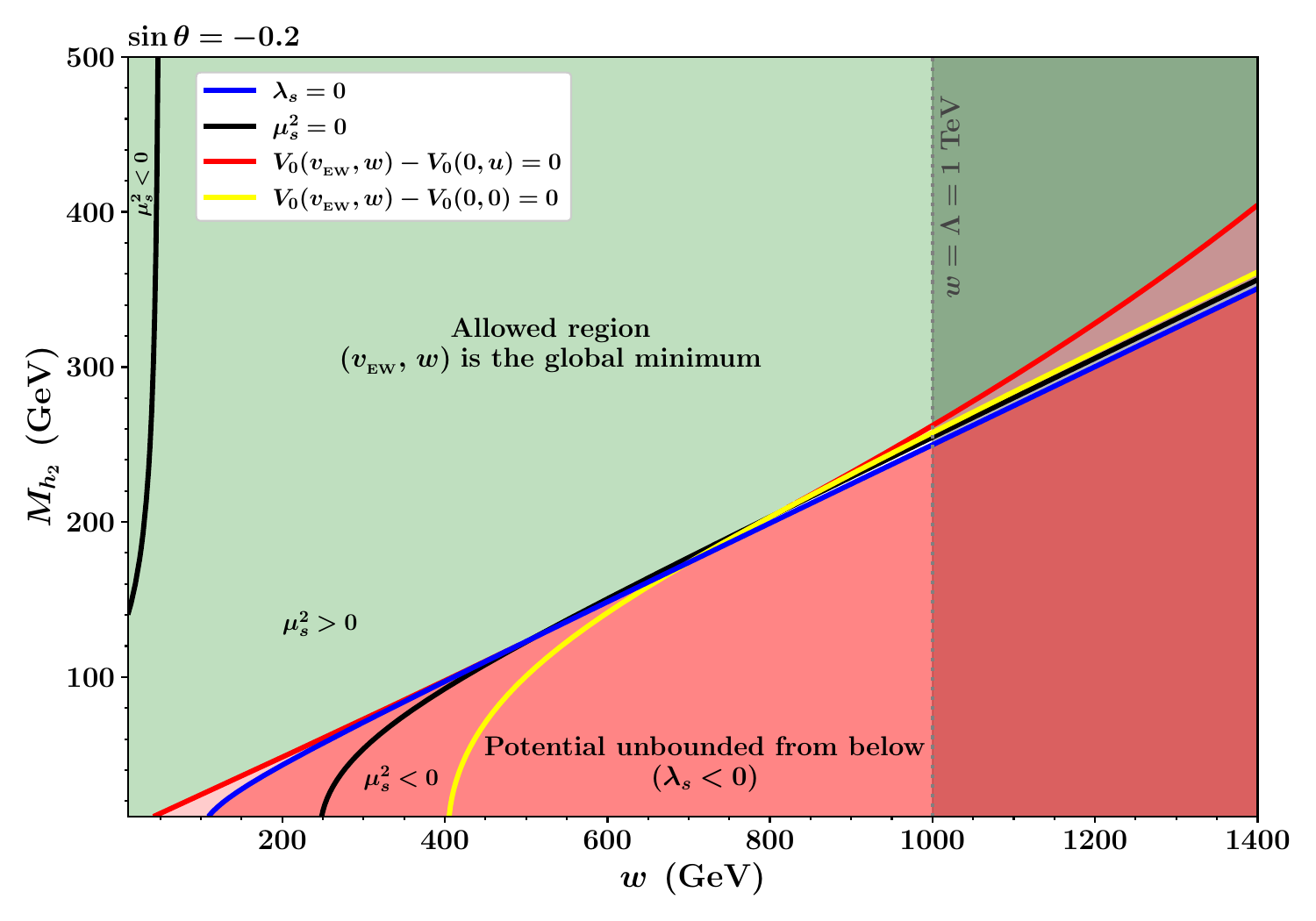}
        \caption{}
        \label{reg2}
	\end{subfigure}
    
	\caption{Allowed parameter space in the $M_{h_2}-w$ plane consistent with various theoretical constraints such as $(v_{_{\text{EW}}},w)$ is the global minima, potential is bounded from below as discussed in Section~\ref{thcons}. In the region shaded with green colour $(v_{_{\text{EW}}},w)$ is the global minima. The gray patch on the right is excluded by requirement $w<\Lambda$.}
    \label{region}
\end{figure}

 In Figure \ref{region}, we show the different region of parameter space where one or more of these requirements mentioned above are met. The shaded region marked with red color is ruled out by the condition that the potential is bounded from below, in particular, in this region the quartic coupling $\lambda_s$ becomes negative. The remaining region can be divided into two parts. The region shaded with green color corresponds to ($v_{_{\text{EW}}},w$) is the global minima of the tree-level potential and a small strip between the red and green shaded region  corresponding to $(0,u)$ or $(0,0)$ as the global minima. In this parameter space we further display the boundary corresponding to $\mu_s^2 =0$ with black solid line. The red and yellow solid lines are obtained when  ($v_{_{\text{EW}}},w$) is degenerate with $(0,u)$ and $(0,0)$, respectively at tree level.

\subsubsection{Experimental constraints}
\begin{figure}[h!]
    \centering
    \begin{subfigure}{0.48\textwidth}
        \centering
        \includegraphics[width=\linewidth]{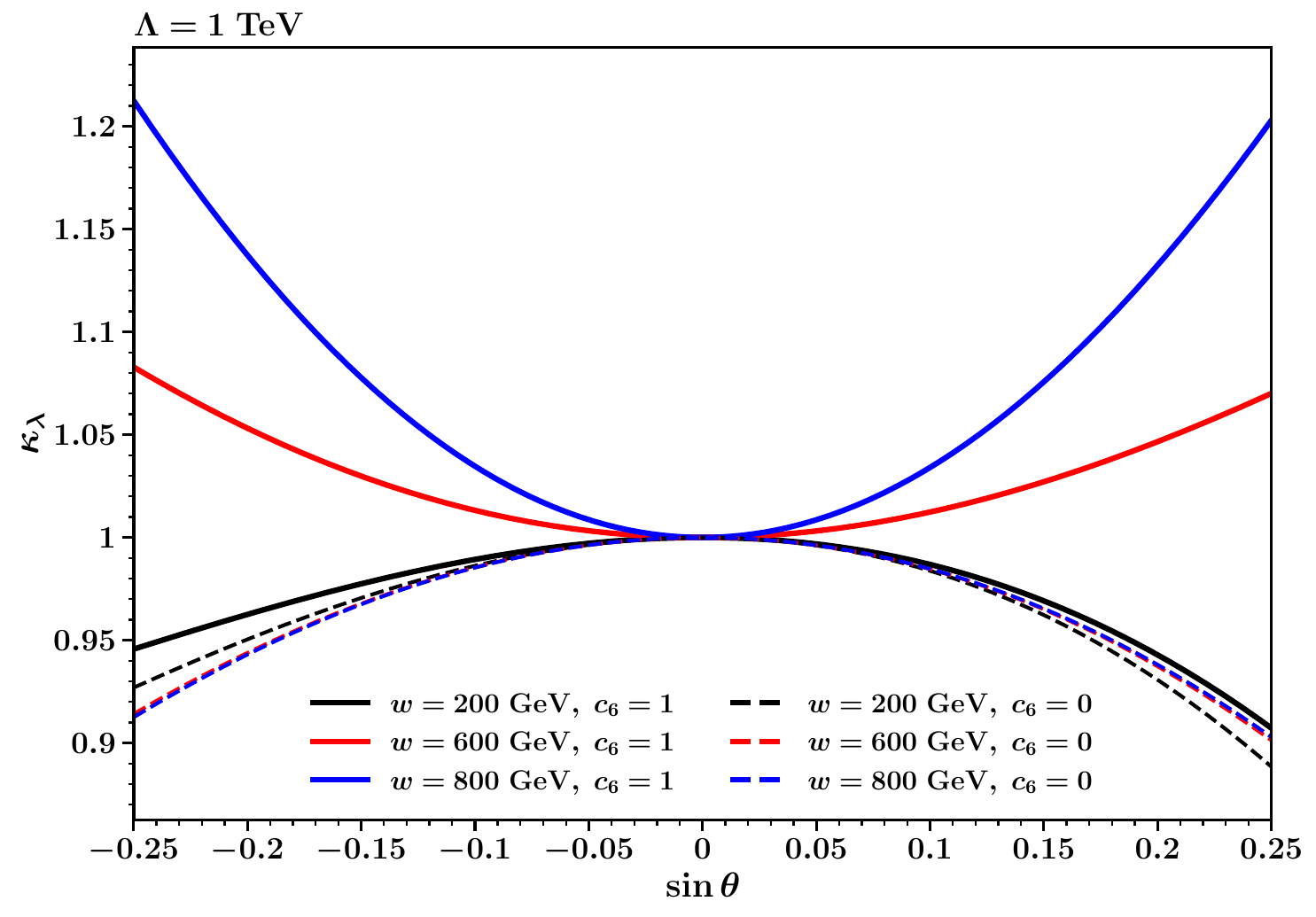}
        \caption{}
        \label{klambda1}
    \end{subfigure}\hfill
    \begin{subfigure}{0.48\textwidth}
        \centering
        \includegraphics[width=\linewidth]{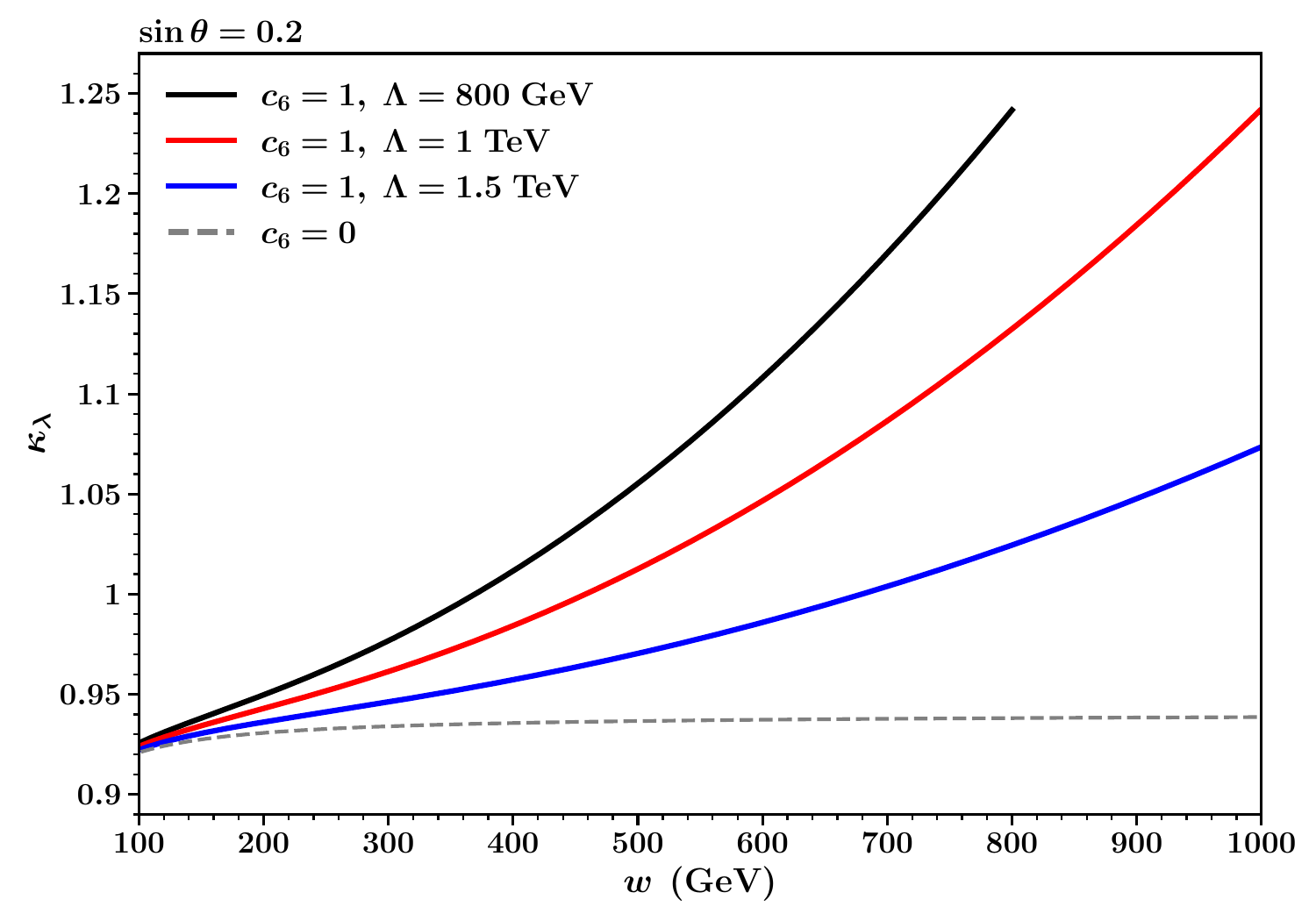}
        \caption{}
        \label{klambda2}
    \end{subfigure}
    
    \caption{ Variation of $\kappa_{\lambda}$ as a function of (a) $\sin\theta$ for various choices of $w$ and fixed cut-off scale $\Lambda= 1$ TeV, and  (b) $w$ for different choices of the cut-off scale $\Lambda$ and fixed $\sin\theta = 0.2$. The solid and dashed lines correspond to the $c_6=1$ and $c_6 =0$, respectively.}
    \label{klambda}
\end{figure}

The parameter space of the local $U(1)_D$ extension of the standard model augmented with a scalar sector is constrained by various observations, such as Higgs signal-strength measurements, di-Higgs searches at the LHC. These constraints are more relevant and readily applicable in scenarios where the singlet scalar acquires a non-zero VEV and thus introduces Higgs and singlet scaler mixing. Below we discuss these constraints in the context of the model under consideration.
\paragraph{Higgs signal-strength measurements: }
The scalar mixing angle $\sin\theta$ is constrained by Higgs signal strength measurement at the LHC and also the invisible decay of the Higgs boson. For $M_{h_2} < \frac{M_{h_1}}{2}$ the measurement of the invisible decay branching ratio of the Higgs at the LHC sets stringent limit on $\sin\theta$ and the current allowed value of $\sin\theta$ in this range is $0.01$ for $w\sim 100$ GeV \cite{ATLAS:2023tkt, CMS:2023sdw} . For $M_{h_2}$ in the range $100-200$ GeV Higgs signal strength measurement at the LHC rules out $|\sin\theta|>0.1$ for $w \sim 0.1 v_{_{\text{EW}}}$ \cite{Robens:2022cun}. For $M_{h_2}> 200$ GeV scalar mixing angle beyond $0.2$ is disfavoured by the LHC data \cite{Adhikari:2022yaa, Papaefstathiou:2022oyi}.
\paragraph{Di-Higgs searches: }\label{dihiggs}
 The new physics contribution to the di-Higgs production can enter via either the modification of the trilinear Higgs coupling or due to the presence of new resonance or both. The LHC collaborations have independently explored the first two possibilities. The CMS data \cite{CMS:2024ymd} corresponding to an integrated luminosity of $138 \ {\rm fb}^{-1}$ and $\sqrt{s} = 13 $ TeV sets $95\%$ confidence limit (C.L.) on $\kappa_\lambda$, where $\kappa_\lambda$ is defined as
\begin{equation}
    \kappa_\lambda= \frac{\lambda_{h_1 h_1 h_1}}{\lambda_{hhh}^{^{\text{SM}}}} = \left[
 \cos^3\theta
-\frac{v_{_{\text{EW}}}}{w}\sin^3\theta
+2\frac{v_{_{\text{EW}}}^2}{M_{h_1}^2}\frac{w^2}{\Lambda^2}\cos\theta \sin^2\theta
\right],
\end{equation}
 which measures the deviation of trilinear  higgs coupling (see Appendix~\ref{trip}) from its standard model value due to the presence of new physics contribution. The allowed values of $\kappa_\lambda$ at $95 \%$ C.L. is $-1.39 < \kappa_\lambda < 7.02$. In Figure~\ref{klambda1},  we plot $\kappa_\lambda$ as a function of $\sin \theta$ for various choices of $w= 200, 600, 800$ GeV and fixed cut-off scale ($\Lambda=1$ TeV). Figure~\ref{klambda2} illustrates the variation of $\kappa_{\lambda}$ as a function of $w$ for various choices of the cut-off scale $\Lambda= 800$ GeV, $1$ TeV, $1.5$ TeV and
$\sin\theta =0.2$. Both of these plots suggest that the region of parameter space relevant for the present study is consistent with the above limit on $\kappa_{\lambda}$. Note that the effect of the dimension-six operator assumed for this analysis enters as a combination of $\frac{w}{\Lambda}$, both of which are free parameters of the theory and additionally suppressed by the square of the scalar mixing angle $\sin\theta$.

\begin{figure}[h!]
    \centering
    \begin{subfigure}{0.48\textwidth}
        \centering
        \includegraphics[width=\linewidth]{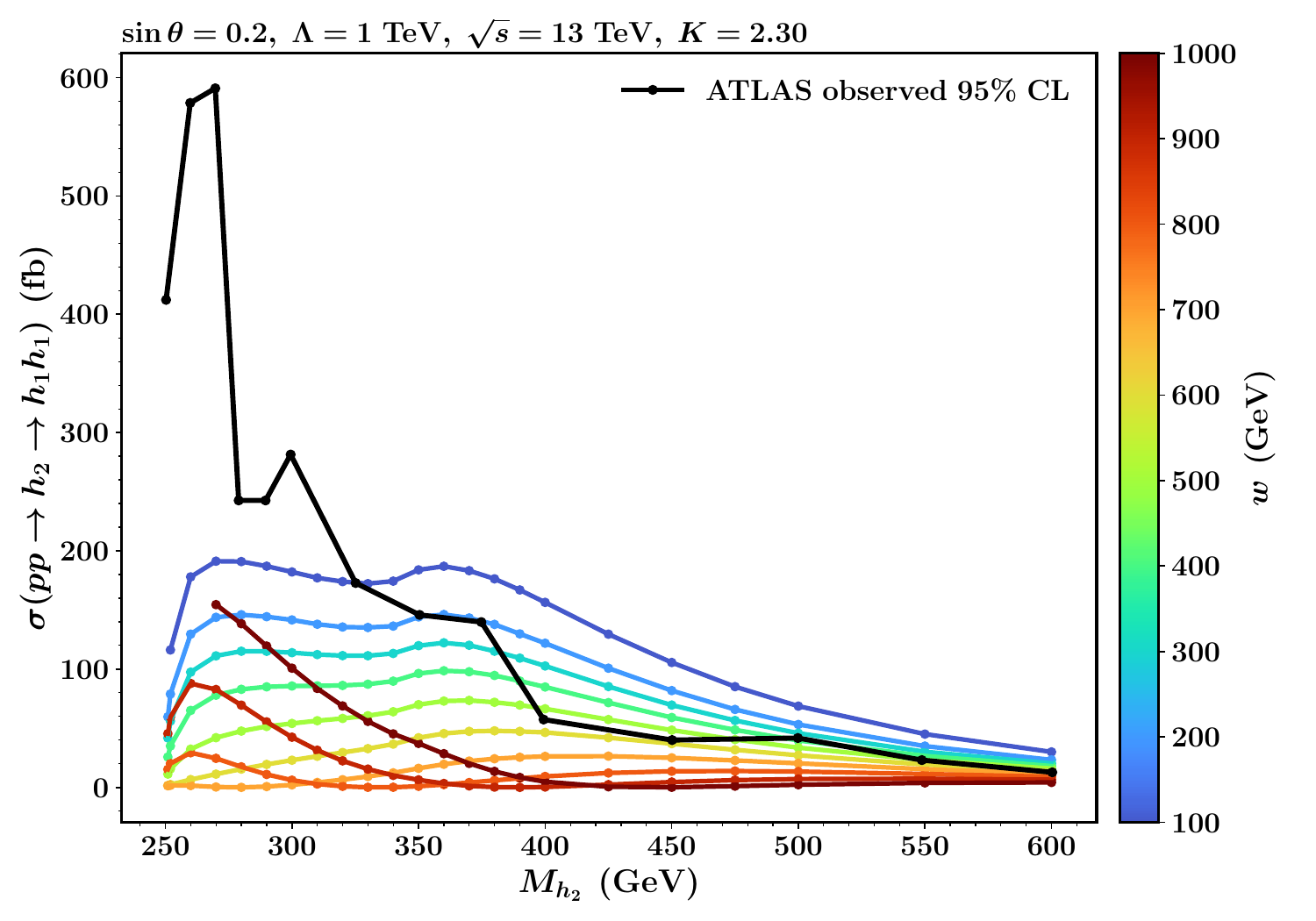}
        \caption{}
        \label{resonant1}
    \end{subfigure}\hfill
    \begin{subfigure}{0.48\textwidth}
        \centering
        \includegraphics[width=\linewidth]{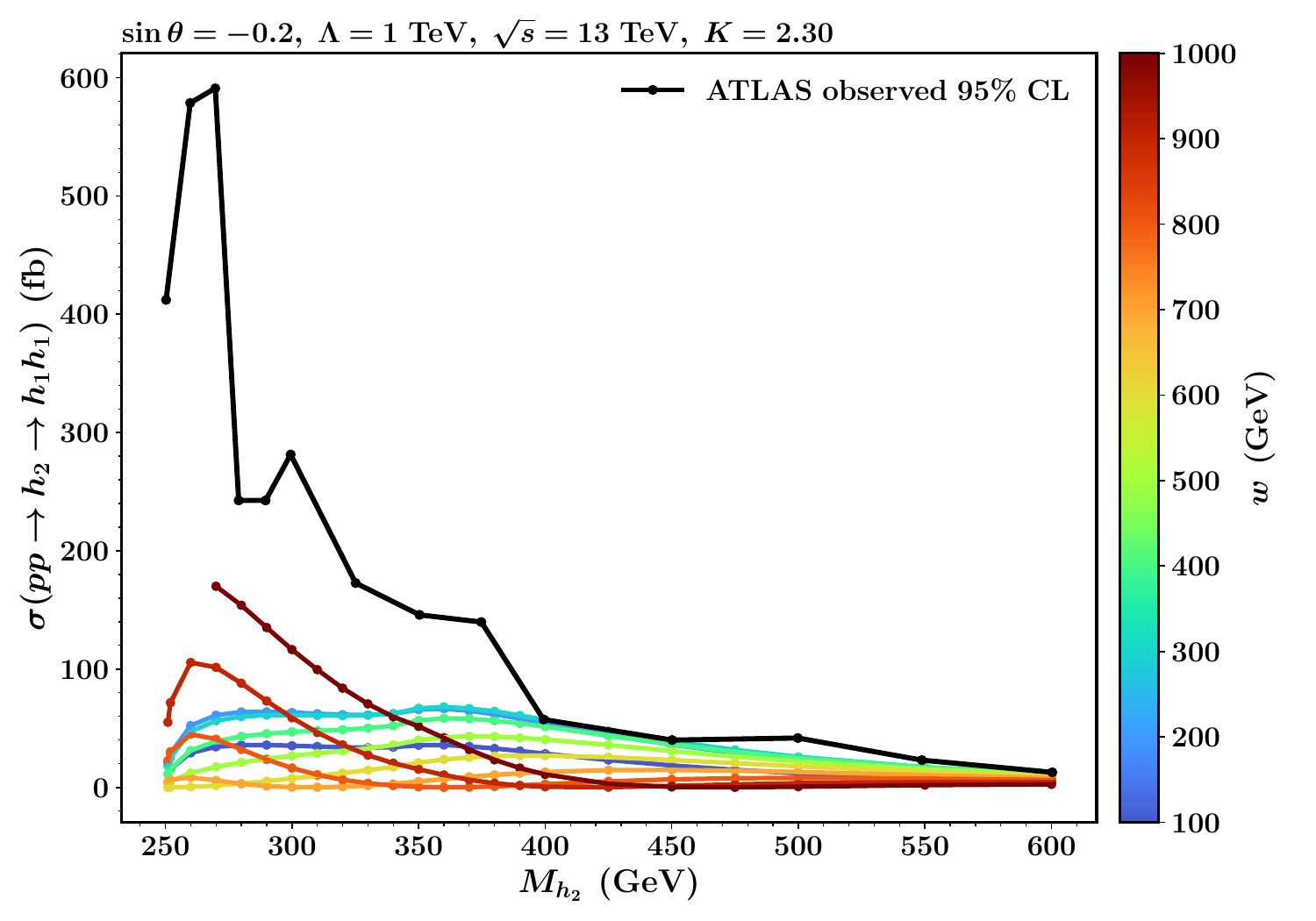}
        \caption{}
        \label{resonant2}
    \end{subfigure}
    \caption{Resonant di-Higgs production cross section
    $\sigma(pp \to h_2 \to h_1 h_1)$ as a function of $M_{h_2}$ for different values of $w$ (represented in color gradient.). The black curve represents the observed $95 \%$ CL upper limit on the resonant di-Higgs production set by ATLAS experiment.}
    \label{resonant}
\end{figure}

The ATLAS measurement of resonant di-Higgs production rate~\cite{Cheng:2025aev} at 13 TeV center of mass energy corresponding to an integrated luminosity of 139 $\text{fb}^{-1}$ also restrict the allowed parameter space of our model under consideration. The ATLAS collaboration  sets an observed upper limit on the di-Higgs rate as a function of the mass of the spin-0 resonance. In Figure \ref{resonant}, we plot the the resonant di-Higgs cross-section, $\sigma (pp \to h_2 \to h_1 h_1)$ as a function of the scalar mass $M_{h_2}$ for various $w$ assuming a $K$ factor of $2.3$~\cite{deFlorian:2014rta}. The cross-section as a function of $M_{h_2}$  in the range $250$ GeV to $600$ GeV for a fixed $w$ shows two local maxima with a minima in between. The first maxima corresponds to the resonance $M_{h_2} = 2 M_{h_1}$ while the second maxima corresponds to on-shell top quark contribution in the loop. It turns out that for $\sin\theta = 0.2$, singlet scalar VEV in the range $600<w<\Lambda $ is allowed for heavy scalar resonance of mass $> 250$ GeV. In the lower mass range of the scalar mass $250 \ \text{GeV} - 370 \ \text{GeV}$,  $w> 200 $GeV is allowed by the resonance di-Higgs search. This limits are further relaxed for $\sin \theta = -0.2$, almost all values of the dark-Higgs VEV in the range $100-\Lambda$ is allowed for scalar mass beyond $250$ GeV.

\section{Phase transitions}
\label{phase_transition}
In this section, we will discuss the phase transition dynamics in the context of the model described in Section~\ref{model} by incorporating one loop finite temperature correction to the tree-level potential given in Equation~\ref{v0}. Considering the fluctuations $h$, $s$ around the  background fields $\phi _1$ and $\phi _2$  one can write
\begin{equation}
	H= \frac{1}{\sqrt{2}}\begin{pmatrix}
		\chi_1 + i \chi _2 \\
		\phi_1 +h + i \chi _3
	\end{pmatrix} \qquad \phi = \frac{1}{\sqrt{2}} \left(\phi_2 + s + i \chi _4 \right).
    \label{fields}
\end{equation}
The field dependent mass-squared matrix of the scalars at zero temperature can be written as 
\begin{equation}
	m^2 (\phi_1, \phi_2)=	\begin{pmatrix}
		-\mu_h^2 + \dfrac{1}{2} \lambda_{hs} \phi_2^2 + 3 \lambda_h \phi_1^2 +  \dfrac{1}{4 \Lambda^2} \phi_2^4 & \lambda_{hs} \phi_1 \phi_2 + \frac{1}{\Lambda^2} \phi_1 \phi_2^3 \\
		\lambda_{hs} \phi_1 \phi_2 + \frac{1}{\Lambda^2} \phi_1 \phi_2^3 & -\mu_s^2 + \dfrac{1}{2} \lambda_{hs} \phi_1^2 + 3 \lambda_s \phi_2^2 + \dfrac{3}{2 \Lambda^2} \phi_1^2 \phi_2^2
	\end{pmatrix} \label{fieldmass}
\end{equation}
\noindent
Hence, the eigenvalues of the above matrix are:
\begin{equation}
	\begin{aligned}
		m_{h_1,h_2}^2(\phi_1,\phi_2) &= \frac{1}{8} \bigg\{ 
		-4\mu_h^2 - 4\mu_s^2 + 2(6\lambda_h + \lambda_{hs})\phi_1^2 
		+ 2\!\left(\lambda_{hs} + 6\lambda_s + \tfrac{3}{\Lambda^2}\phi_1^2\right)\phi_2^2 
		+ \tfrac{1}{\Lambda^2}\,\phi_2^4 \\
		& \pm \bigg[ 
		4\!\left(2\mu_h^2 - 2\mu_s^2 + (-6\lambda_h + \lambda_{hs})\phi_1^2\right)^2 
		+ 8\Big( 7\lambda_{hs}^2 \phi_1^2 - 6\!\left(2\lambda_s + \tfrac{1}{\Lambda^2}\phi_1^2\right)\!(-\mu_h^2 + \mu_s^2 + 3\lambda_h\phi_1^2)\\
		&\quad + \lambda_{hs}\!\left(-2\mu_h^2 + 2\mu_s^2 + 6(\lambda_h + \lambda_s)\phi_1^2 + \tfrac{3}{\Lambda^2}\phi_1^4\right) \Big)\phi_2^2  \\
		&\quad + 4\Big( (\lambda_{hs} - 6\lambda_s)^2 + \tfrac{1}{\Lambda^2}(25\lambda_{hs} + 36\lambda_s)\phi_1^2 
		+ \tfrac{9}{\Lambda^4}\phi_1^4 + \tfrac{2}{\Lambda^2}(-\mu_h^2 + \mu_s^2 + 3\lambda_h\phi_1^2) \Big)\phi_2^4 \\
		&\quad + \tfrac{4}{\Lambda^2}\left(\lambda_{hs} - 6\lambda_s + \tfrac{13}{\Lambda^2}\phi_1^2\right)\phi_2^6 
		+ \tfrac{1}{\Lambda^4}\phi_2^8 
		\bigg]^{1/2}
		\bigg\}  
	\end{aligned}
   \label{masseigen}
\end{equation}
which now receive contributions from the dimension six term.

The background field dependent masses for the other relevant fields, {\it e.g.}, $W^{\pm}, Z, \gamma_d , t$-quark and goldstone bosons ($\chi _i$'s) are given by:
\begin{equation}
	\begin{aligned}
		&m_W ^2 (\phi_1, \phi_2)= \frac{g_1^2}{4} \phi_1 ^2, \quad 
		m_Z ^2(\phi_1, \phi_2)= \frac{g_1^2 + g_2 ^2}{4} \phi_1^2 ,\quad 
		m_{\gamma _d}^2 = g_d^2 \phi_2^2 ,\\
		&m_{\chi_1}^2 = m_{\chi_2}^2 = m_{\chi_3}^2 = -\mu_h ^2 + \lambda_h \phi_1 ^2 + \frac{1}{2} \lambda_{hs} \phi_2 ^2 + \frac{1}{4 \Lambda^2} \phi_2 ^4 ,\\
		&m_{\chi_4}^2 = -\mu_s ^2 + \frac{1}{2} \lambda_{hs} \phi_1 ^2 + \lambda_s \phi_2 ^2 + \frac{1}{2\Lambda^2} \phi_1^2 \phi_2 ^2,\\
        &m_t^2 (\phi_1, \phi_2)= \frac{y_t^2}{2} \phi_1^2 , \quad 
	\end{aligned}
    \label{massspect}
\end{equation}
where $g_1, \ g_2 \ \text{and} \ g_d$ are the gauge coupling constants of the gauge groups $SU(2)_L$, $U(1)_Y$ and $U(1)_D$, respectively. 
The degrees of freedom ($n_i$'s) associated with each of these species are 
\begin{equation}
	n_{h_1}= n_{h_2} = n_{\chi_{1,2,3,4}}= 1, \quad n_W = 6,\quad n_Z = n_{\gamma _d} =3, \quad n_t =12 \label{ndof}
\end{equation}
We now introduce the Coleman-Weinberg (CW) correction \cite{Quiros:1999jp}, 
\begin{equation}
	\begin{aligned}
		V_{\text{CW}}(\phi_1 , \phi_2)= \frac{1}{64 \pi^2}\bigg\{ \sum_B n_B m_B^4 (\phi_1, \phi_2) \left[ \log(\frac{m_B^2 (\phi_1,\phi_2)}{Q^2}) -c_B \right]\\
		-\sum_F n_F m_F^4 (\phi_1, \phi_2) \left[ \log(\frac{m_F^2 (\phi_1,\phi_2)}{Q^2}) - \frac{3}{2} \right]\bigg\},
	\end{aligned} \label{vcw}
\end{equation}
where the sum in the first term runs over all the bosons ($B= h_1, h_2, \chi _i\text{'s}, W^{\pm}, Z, \gamma_d$) and the second one runs over all the fermions of the theory that couple to the higgs field. We have considered only the top quark contribution in the second term among all the SM fermions. Additionally, $Q$ is the renormalization scale which we will set to $800$ GeV without any loss of generality throughout our analysis. The value of the parameter $c_B$ in equation is $3/2$ ($5/6$) for scalar (vector) bosons, respectively. 

In order to ensure that the vacuum expectation values of the scalar fields and mass matrix at minima in presence of the CW term remain the same as that predicted at the tree level, 
we add additional counter terms to the above scalar potential  and use the following set of renormalization conditions. 
 
\begin{equation}
	\begin{aligned}
		&V_1 = V_{\text{CW}} + V_{\text{ct}}  \\
		&\left. \frac{\partial V_1}{\partial \phi_1}\right|_{\substack{\phi_1 = v_{_{\text{EW}}} \\ \phi_2 = w}}= \left. \frac{\partial V_1}{\partial \phi_2}\right|_{\substack{\phi_1 = v_{_{\text{EW}}} \\ \phi_2 = w}}= \left. \frac{\partial^2 V_1}{\partial \phi_1^2}\right|_{\substack{\phi_1 = v_{_{\text{EW}}} \\ \phi_2 = w}}= \left. \frac{\partial^2 V_1}{\partial \phi_2^2}\right|_{\substack{\phi_1 = v_{_{\text{EW}}} \\ \phi_2 = w}}= \left. \frac{\partial^2 V_1}{\partial \phi_1 \partial \phi_2}\right|_{\substack{\phi_1 = v_{_{\text{EW}}} \\ \phi_2 = w}}=   \left. \frac{\partial V_1}{\partial \phi_2 }\right|_{\substack{\phi_1 = 0 \\ \phi_2 = u}}=0
	\end{aligned}
     \label{renormcond}
\end{equation}
where $V_{\text{ct}}$ is given by
\begin{equation}
	V_{\text{ct}}(\phi_1, \phi_2)= \frac{1}{2} a\phi_1^2 + \frac{1}{4} b \phi_1^4 + \frac{1}{2} c \phi_2^2 + \frac{1}{4} d \phi_1^2 \phi_2^2 + \frac{1}{4} e\phi_2^4   + \frac{1}{8} f  \phi_1^2 \phi_2^4 \label{vct}
\end{equation}
Here $(v_{_{\text{EW}}},w)$ is the electroweak vacuum configuration corresponding to the tree-level scalar potential given in Equation~\ref{v0} and $(0, u)$ is the stationary field configuration in the $\phi_1 =0$ direction as discussed earlier. The six unknown coefficients ($a,b,c,d,e,f$) are determined by solving the following six algebraic equations.
\small
\begin{align}
	&\left. \frac{\partial V_{cw}}{\partial \phi_1}\right|_{\substack{\phi_1 = v_{_{\text{EW}}} \\ \phi_2 = w}}\hspace{-0.2cm}+ a v_{_{\text{EW}}} + b v_{_{\text{EW}}}^3 + \frac{d}{2} v_{_{\text{EW}}} w^2 + \frac{f}{4} v_{_{\text{EW}}} w^4 =0  \hspace*{0.3cm} \left. \frac{\partial^2 V_{cw}}{\partial \phi_1^2}\right|_{\substack{\phi_1 = v_{_{\text{EW}}} \\ \phi_2 = w}} \hspace{-0.2cm}+ a  + 3b v_{_{\text{EW}}}^2 + \frac{d}{2}  w^2 + \frac{f}{4}  w^4 =0 \nonumber\\
	&\left. \frac{\partial V_{cw}}{\partial \phi_2}\right|_{\substack{\phi_1 = v_{_{\text{EW}}} \\ \phi_2 = w}}\hspace{-0.2cm}+ c w + e w^3 + \frac{d}{2} v_{_{\text{EW}}}^2 w + \frac{f}{2} v_{_{\text{EW}}}^2 w^3 =0 \hspace*{0.3cm} \left. \frac{\partial^2 V_{cw}}{\partial \phi_2^2}\right|_{\substack{\phi_1 = v_{_{\text{EW}}} \\ \phi_2 = w}}\hspace{-0.2cm}+ c + 3 e w^2 + \frac{d}{2}  v_{_{\text{EW}}}^2 + \frac{3f}{2} v_{_{\text{EW}}}^2 w^2 =0 \nonumber\\
	&\left. \frac{\partial^2 V_{cw}}{\partial \phi_1 \partial \phi_2}\right|_{\substack{\phi_1 = v_{_{\text{EW}}} \\ \phi_2 = w}}\hspace{-0.2cm}+ d v_{_{\text{EW}}} w + f v_{_{\text{EW}}} w^3 =0 \hspace*{2.5cm} \left. \frac{\partial V_{cw}}{\partial \phi_2}\right|_{\substack{\phi_1 = 0 \\ \phi_2 = u}}+ c u + e u^3 =0 \label{cteqn}
\end{align}
\normalsize
When ($0,u$) is not a stationary point, which is the case in a very small region of parameter space (see Figure \ref{region}), we set $f=0$ and just and solve five equations for five unknowns. It is also important to note that the above choices of renormalization conditions is not unique. Alternative choices of renormalization conditions can have mild, if not significant, effect on the parameter space consistent with either the strong EWPT or the requirement that ($v_{_{\text{EW}}}, w$) is the electroweak vacuum at zero temperature \cite{Chen:2025ksr}.

Finally the finite temperature correction to the effective potential at one-loop is given by \cite{Quiros:1999jp},
\begin{equation}
	V_1^T (\phi_1, \phi_2, T)= \frac{T^4}{2\pi^2}\left[\sum_B n_B J_B \left(\frac{m_B^2(\phi_1, \phi_2)}{T^2} \right) +  n_t J_F \left(\frac{m_t^2(\phi_1, \phi_2)}{T^2} \right) \right] \label{v1T}
\end{equation}
where the thermal bosonic and fermionic function $J_B$ and $J_F$ are defined as :
\begin{equation}
	J_B(\theta) = \int_{0}^{\infty} y^2 \log(1- e^{- \sqrt{y^2 + \theta}}) dy \qquad 	J_F(\theta) = \int_{0}^{\infty} - y^2 \log(1+ e^{- \sqrt{y^2 + \theta}}) dy \label{jbjf}
\end{equation}
We have also taken care of the daisy resummation by redefining the field dependent masses of the scalars and longitudinal polarization of the gauge bosons as follows.
\begin{equation}
	m_i ^2 ( \phi_1, \phi_2, T)= m_i ^2 ( \phi_1, \phi_2) + \Pi _i (T) \label{ring1}
\end{equation}
where $i$ belongs to all the relevant bosonic degrees of freedom only.  The one-loop thermal self energy contributions ($\Pi _i(T)$)  are given by \cite{Carrington:1991hz, Hashino:2018zsi, Parwani:1991gq},
\begin{equation}
	\begin{aligned}
		& \Pi _{hh}= \Pi _{\chi_{1,2,3}}= \left(\frac{3 g_1 ^2}{16} + \frac{g_2 ^2}{16} + \frac{y_t^2}{4} + \frac{\lambda_h}{2} + \frac{\lambda_{hs}}{12} + \frac{1}{6\Lambda^2} w^2 \right) T^2 \ ,\\
		& \Pi _{ss}=  \Pi _{\chi_4}= \left(\frac{g_d ^2}{16} + \frac{\lambda_{hs}}{6} + \frac{\lambda_{s}}{3} + \frac{1}{6 \Lambda ^2} v^2 + \frac{1}{2 \Lambda^2} w^2 \right) T^2 \ , \qquad  \Pi _{hs}\approx 0  \ ,\\
		& \Pi _{A_{1L}}= \Pi_{A_{2L}}=\Pi_{A_{3L}}= \frac{11}{6} g_1^2 T^2, \qquad \Pi _{B_L}= \frac{11}{6}g_2^2 T^2, \qquad \Pi _{B{_{dL}}}= \frac{1}{3} g_d^2 T^2
	\end{aligned}
    \label{ring2}
\end{equation}
The field dependent squared masses of the scalars at finite temperature are the eigenvalues of the following thermally corrected hessian matrix.
\begin{equation}
	m^2(\phi_1, \phi_2) + \begin{pmatrix}
		\Pi _{hh} & 0 \\
		0 & \Pi _{ss}
	\end{pmatrix}
    \label{ring3}
\end{equation}
The mass matrix involving the longitudinal components of the gauge bosons at finite temperature in ($A_1$, $A_2$, $A_3$, $B$, $B_d$) basis is given by 

\begin{equation}
	\begin{pmatrix}
        \frac{1}{4} g_1^2 \phi_1 ^2 + \Pi _{A_{1L}} & 0 & 0 & 0 &0 \\
        0 &\frac{1}{4} g_1^2 \phi_1 ^2 + \Pi _{A_{2L}} & 0 & 0 &0 \\
		0 & 0 & \frac{1}{4} g_1^2 \phi_1 ^2 + \Pi_{A_{3L}} & -\frac{1}{4} g_1 g_2 \phi_1 ^2\vspace*{0.2cm}&0 \\ 
		0 & 0 & -\frac{1}{4} g_1 g_2 \phi_1 ^2 & \frac{1}{4} g_2^2 \phi_1 ^2 + \Pi _{B_L} &0\\
        0 & 0 & 0&0 & g_d^2 \phi_2^2 + \Pi_{B_{dL}}
	\end{pmatrix}
    \label{ring4}
\end{equation}
where we have neglected the kinetic mixing. It turns out that the photon develops a non-vanishing longitudinal component at finite temperature.

The complete one loop effective potential at finite temperature is given by :
\begin{equation}
    V_{eff}(\phi_1, \phi_2, T)= V_0(\phi_1, \phi_2) + V_{\text{CW}}(\phi_1, \phi_2) + V_{\text{ct}}(\phi_1, \phi_2) + V_1^T(\phi_1, \phi_2, T) \label{fullveff}
\end{equation}

 We are interested in studying the phase transition pattern and its strength for the above effective potential keeping in mind its implications for electroweak baryogenesis. As mentioned earlier, for successful electroweak baryogenesis in this scenario required baryon number violation is induced by sphaleron transition processes like in  the SM, while a SFOEWPT ensures departure from thermal equilibrium. During the electroweak phase transition, bubbles of the broken electroweak phase nucleate and expand within the surrounding electroweak-symmetric vacuum. The critical temperature ($T_c$) is defined as the temperature at which the electroweak symmetry preserving vacuum becomes degenerate with that corresponding to the broken-phase.

Any net baryon asymmetry generated at the critical temperature is efficiently washed out by sphaleron transitions in the symmetric phase. In contrast, it is preserved in the broken phase, where the sphaleron transition rate is exponentially suppressed as $\propto \exp\left(-\frac{E_{\text{sph}}(T)}{T}\right)$~\cite{Quiros:1999jp}. In particular, one requires $\frac{E_{\text{sph}}(T_c)}{T_c} \gtrsim 45$ in order to avoid washout of the generated baryon asymmetry in the broken phase. This condition is approximately satisfied if $\frac{\phi_c}{T_c} \gtrsim 1$, where $\phi_c$ denotes the discontinuity in the order parameter $\phi_1$ at the critical temperature, i.e., $\phi_c = \Delta \phi_1 (T_c)$. However, the precise lower bound on $\frac{\phi_c}{T_c}$ is subject to theoretical uncertainties and may vary in the range $0.6\text{--}1.4$~\cite{Patel:2011th, Fuyuto:2014yia}. In this analysis, we adopt the following conservative criterion for a strong first-order phase transition:
\[
\frac{\phi_c}{T_c} \geq 0.8 \, .
\]
\vspace{0.3cm}

\subsection{High temperature approximation}
\begin{figure}[h!]
    \centering
    \begin{subfigure}{0.49\textwidth}
        \centering
        \includegraphics[width=\linewidth]{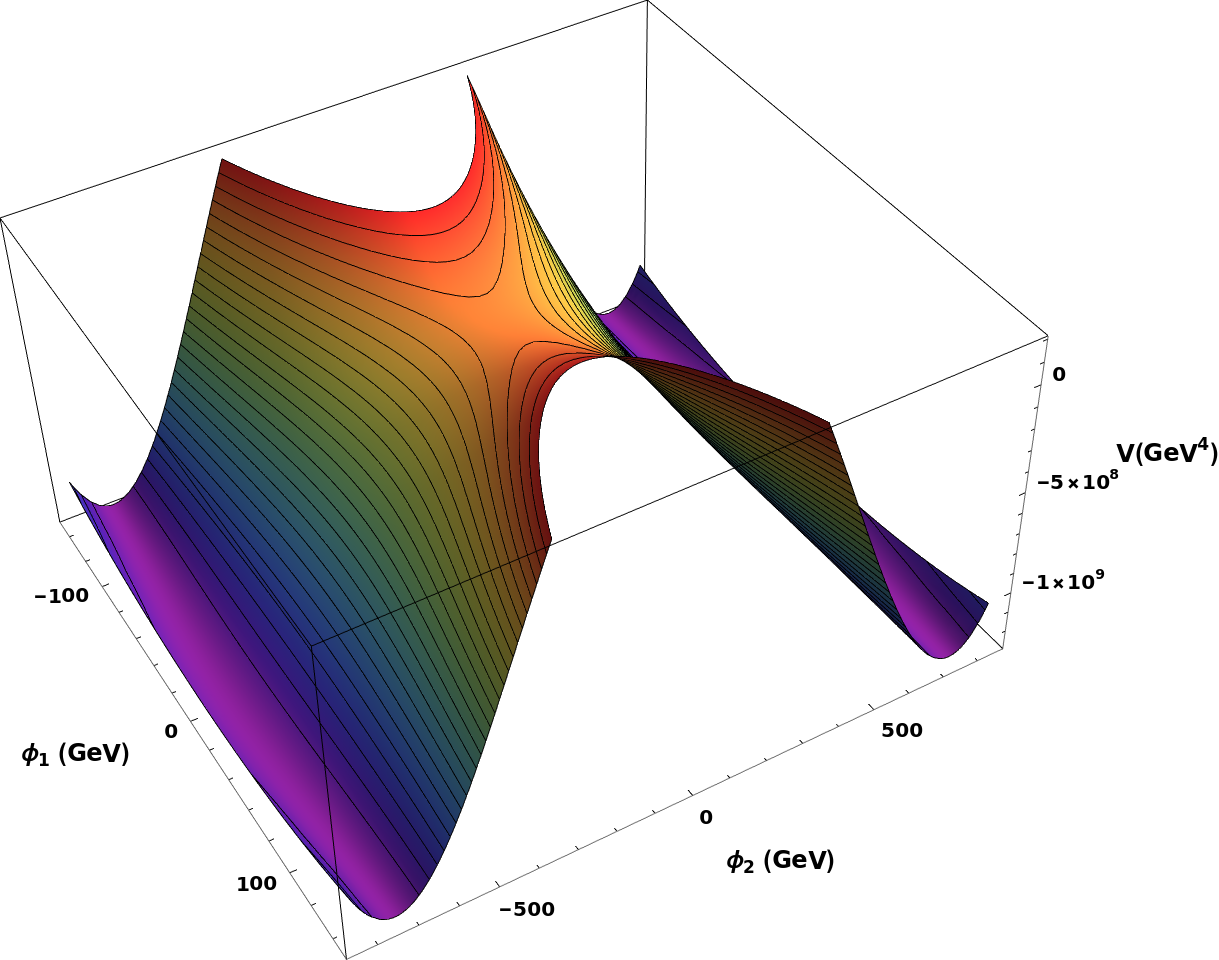}
        \caption{}
        \label{pot3d_1}
    \end{subfigure}\hfill
    \begin{subfigure}{0.48\textwidth}
        \centering
	\includegraphics[width = 0.95\linewidth]{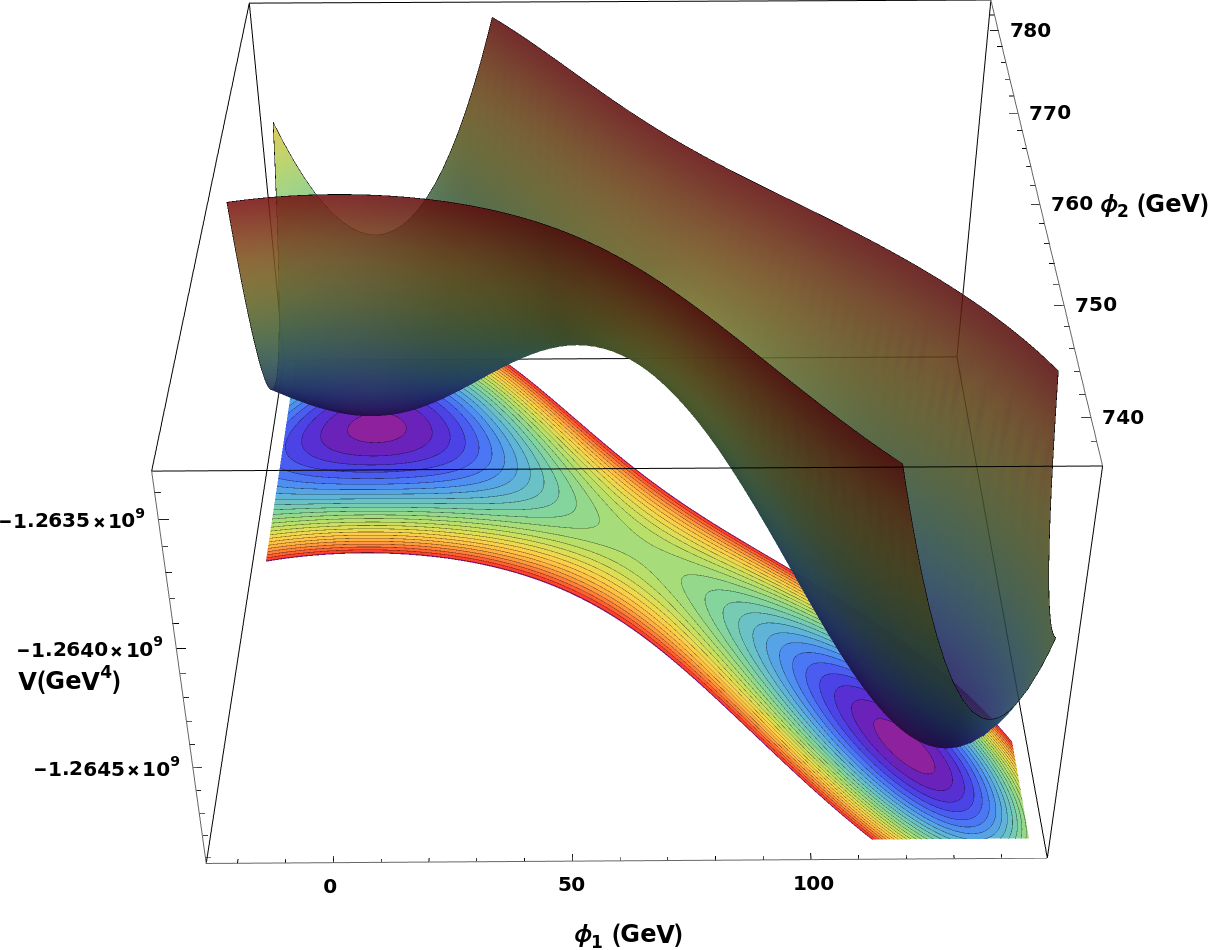}
	\caption{}
	\label{3dzoomed}
    \end{subfigure}
    \caption{3D plot of the one loop finite temperature corrected effective potential at high temperature approximation (see Equation~\ref{vht}) in the $\phi_1-\phi_2$ plane at $T_c= 106.2$ GeV. This plot is obtained neglecting the CW and daisy correction. Here the color gradient representing the value of the potential. The figure in the right (b) represents appropriate scaled version of figure in the left (a). }
    \label{pot3d}
\end{figure}

In this subsection, we try to understand the phase transition pattern of the effective potential at high temperature approximation ($T> m$) neglecting the CW and daisy corrections. In this approximation, with leading order temperature dependent terms, the effective potential has the following form\footnote{Strictly speaking, this approximation is not justified over the entire space of the background field values. Here, we use the high temperature approximation for the illustrative purpose only. The results presented in the following sections are based on exact numerical simulation and do not rely on this approximation. }

\begin{align}
	V_{eff}(\phi_1 , \phi_2, T) \approx &-\frac{1}{2} (\mu_h^2 -c_h T^2) \phi_1^2 + \frac{\lambda_h}{4} \phi_1^4 
	-\frac{1}{2} (\mu_s^2 -c_s T^2) \phi_2^2 \nonumber \\
	&+ \left( \frac{\lambda_{s}}{4} + \frac{1}{24 \Lambda^2} T^2 \right) \phi_2^4 
	+ \left(\frac{\lambda_{hs}}{4} + \frac{1}{12 \Lambda^2} T^2 \right) \phi_1^2 \phi_2^2 
	+ \frac{1}{8 \Lambda^2} \phi_1^2 \phi_2^4 \label{vht}
\end{align}
where, \begin{equation}
	c_h = \frac{1}{48} \left( 9 g_1^2 + 3 g_2^2 + 2 (6 y_t^2 + 12 \lambda_h + 2 \lambda_{hs} ) \right), \qquad
	c_s = \frac{1}{12} \left( 2 \lambda_{hs} + 4 \lambda_s + 3 g_d^2 \right). \label{chcs}
\end{equation}
In order to  illustrate that the above potential gives rise to a barrier that facilitates a FOEWPT (even in the absence of $T\phi_i^3$ term in the effective potential), we choose a representative benchmark point (BP)
 $M_{h_2} =250$ GeV, $\sin \theta =0.15$, $w = 725.7$ GeV, $M_{\gamma_d} = 60$ GeV and $\Lambda = 800$ GeV. This corresponds to a critical temperature $T_c= 106.2$ GeV  and the effective potential displays degenerate ground state configurations ($0,u(T_c)$) and ($v(T_c),w(T_c)$) (see Figure~\ref{pot3d}).

 In terms of numerical values at this benchmark point, the effective potential can be written as 
\begin{align}
&V_{eff}(\phi_1, \phi_2, T\sim 106 \ \text{GeV}) \approx \  \frac{\Tilde{\mu}_h^2}{2}\,\phi_1^2
+ \frac{\Tilde{\lambda}_h}{4}\,\phi_1^4
+\frac{\Tilde{\mu}_s^2}{2}\,\phi_2^2
+\frac{\Tilde{\lambda}_{hs}}{4}\,\phi_1^2 \phi_2^2 
+\frac{\Tilde{\lambda}_s}{4}\,\phi_2^4
+ \frac{1}{8\Lambda^2}\phi_1^2 \phi_2^4
	\label{vhtnumb} \\
    &\text{with } \ \ \Tilde{\mu}_h^2= 9.36\times 10^4, \ \Tilde{\lambda}_h= 1.38\times 10^{-1}, \ \Tilde{\mu}_s^2 = -8.4 \times 10^3, \ \Tilde{\lambda}_{hs}= - 7.8\times10^{-1}, \Tilde{\lambda}_s= 1.4\times 10^{-2},
\end{align}
and the values of the coefficients $c_h$ and $c_s$ are 
\begin{equation}
	c_h = 0.338 \qquad c_s = -0.125. \label{chcsnumb}
\end{equation}
Near the origin ($0,0$) the leading terms are quadratic in $\phi_1$ and $\phi_2$. Since the overall coefficient of $\phi_1^2$ and $\phi_2^2$ is positive and negative, respectively, at $T=T_c$ this corresponds to a saddle point of the scalar potential. 

Along the $\phi_1 =0$ direction the coefficient of $\phi_2^2$ is negative and $\phi_2^4$ is positive at critical temperature. Therefore one expects an extrema at ($0, u(T_c)$). In particular, this should corresponds to a minima along the $\phi_1=0$ direction. The behaviour of the potential around this point in the orthogonal direction is determined by the value of $u(T_c)$. Because $u(T_c)$ determines the overall coefficient of $\phi_1^2$ around this point which in this case turns out to be positive, suggesting ($0,u(T_c)$) is actually a local minima. 
The potential along $\phi_2 =0$ direction always increases as long as $\mu_h^2$ is negative. Therefore, one doesn't encounter any further local extrema along $\phi_2 =0$ direction in this approximation.

Finally, due to the presence of the $\phi_1^2 \phi_2^2$ term with negative coefficient, gives rise to the possibility of additional local minima at ($v(T_c),w(T_c)$). This is because for certain field values of $\phi_2$ at critical temperature, i.e. $w(T_c)$, the overall coefficient of the $\phi_1^2$ term becomes negative. The particular value of $w(T_c) $ for which this happens depends on the relative magnitude of the coefficients of $\phi_1^2$, $\phi_1^2 \phi_2^2$ and $\phi_1^2 \phi_2^4$. The overall sign of the coefficient of the $\phi_1^2 \phi_2^2$ term is determined by the value of $\lambda_{hs}$ as $\frac{T_c^2}{\Lambda^2}$ is a small number (see Equation~\ref{vht}).  In fact, it is crucial to have $\lambda_{hs}$ negative in order to get a local minima at ($v(T_c), w(T_c)$) for a wide range of parameter space in this scenario\footnote{Actually the relative sign and the magnitude among the coefficients of $\phi_1^2$, $\phi_1^2 \phi_2^2$ and $\phi_1^2 \phi_2^4$ play a crucial role to have a local minima at ($v(T_c), w(T_c)$).  }. As we will see later $\lambda_{hs} \sim -1$ will facilitate a SFOEWPT.

 As discussed in \cite{Carena:2019une} the large $\lambda_{hs}$ helps to enhance the thermal barrier via the $\phi_1^3$ term in the singlet scalar extension of the SM with spontaneous $\mathcal{Z}_2$ breaking or in theories with spontaneous $U(1)_D$ breaking \cite{Hashino:2018zsi}. In both of these cases $\lambda_{hs}$ is a function of the independent parameters, namely, the scalar mixing angle ($\sin\theta$), the VEV of the singlet scalar ($w$) along with the masses of the scalars. In fact, the magnitude as well as sign of $\lambda_{hs}$ are strongly correlated with $\sin\theta$ in these models. For higher $w$ one needs higher $\sin\theta$ value to achieve the same magnitude of $\lambda_{hs}$ in usual singlet scalar extension models. 
    However, in presence of the dimension six operator as considered in this work, this correlation is weaken. The same magnitude of $\lambda_{hs}$ can be achieved even for smaller $\sin\theta$ assuming large $w$ (alternatively for lower value of the cut-off scale $\Lambda$). This is due to the fact that $\lambda_{hs}$ receives additional contribution ($\sim -\frac{w^2}{\Lambda^2}$) coming from the dimension-six term. This gives a greater flexibility in terms of the allowed region of parameter space consistent with SFOEWPT.
    
The sign and magnitude of $\lambda_{hs}$ also has important implications for symmetry restoration at high temperature. This dependency enters via $c_h$ and $c_s$, sign of which crucially control the symmetry restoration at high temperature. Since $c_h$ is always positive even for  $\lambda_{hs}\sim -1$ 
    the overall coefficient of $\phi_1^2$ term in Equation \ref{vht} at high temperature  becomes positive. Which implies at high temperature electroweak symmetry is  restored. On the other hand, the high temperature restoration of the $U(1)_D$ symmetry depends on the interplay between $\lambda_{hs}$ and dark gauge coupling ($g_d$). The former one is mostly negative whenever the requirement of SFOEWPT is met.
    It turns out that $c_s$ is negative due to $\lambda_s, g_d < |\lambda_{hs}|$. In this context, it is important to note the role of dark-photon mass in the restoration of the $U(1)_D$ symmetry at high temperature. For low mass dark-photon the value of $g_d$ is small. However, as dark photon mass increases, the dark gauge coupling $g_d$ also increases for a fixed $w$. Hence, $c_s$ can in principle become positive which in effect helps to restore the $U(1)_D$ symmetry at high temperature.

\subsection{Exact numerical results}
We have implemented the full one-loop finite temperature corrected effective potential, including CW and daisy corrections described by Equation~\ref{fullveff} in {\tt CosmoTransitions} \cite{Wainwright:2011kj}. For the finite temperature correction, full numerical integration has been implemented using spline interpolation. The infrared divergences for the zero Matsubara modes of the bosonic degrees of freedom have been taken care of by leading order resummation and redefining the masses including the thermal correction (Parwani Method~\cite{Parwani:1991gq}) as discussed in this section. To obtain the viable parameter space consistent with the requirement of SFOEWPT and  its dependence on various parameters, we have performed dedicated parameter space scan using {\tt CosmoTransitions}.

\begin{figure}[h!]
	\centering
	\begin{subfigure}[b]{0.48\textwidth}
		\centering
		\includegraphics[width = \textwidth]{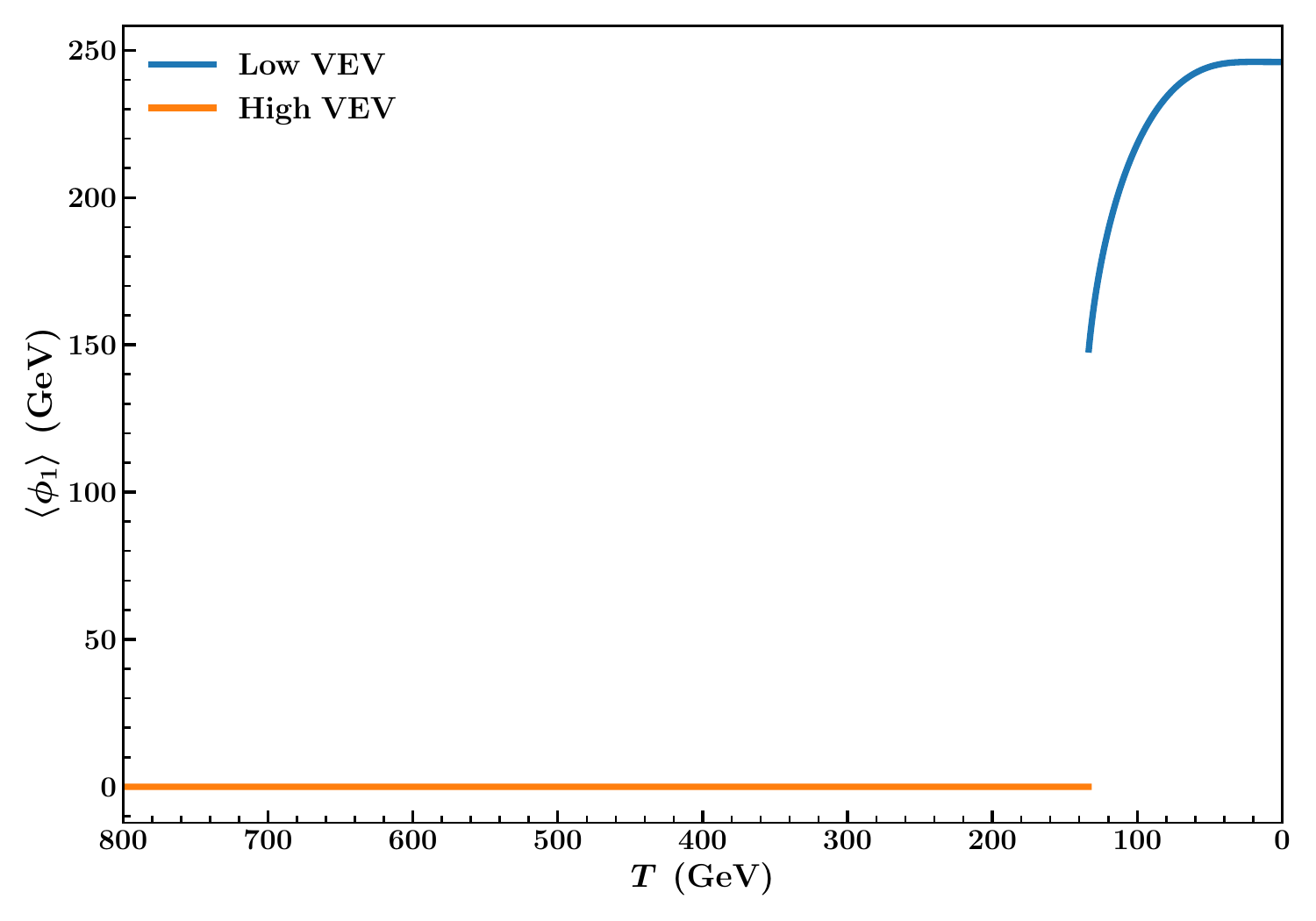}
		\caption{}
		\label{fig9}
	\end{subfigure}
	\hfill
	\begin{subfigure}[b]{0.48\textwidth}
		\centering
		\includegraphics[width= \textwidth]{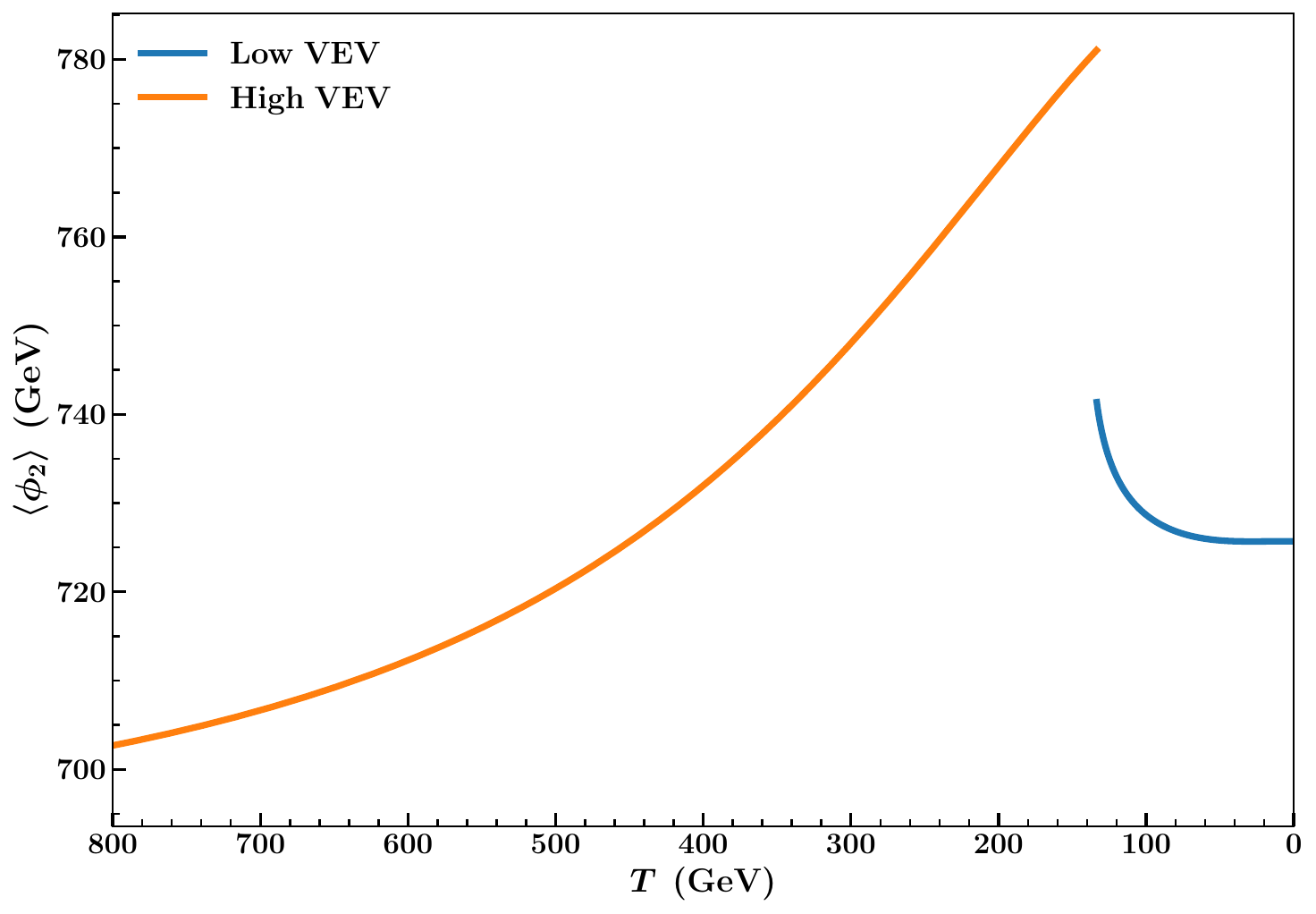}
		\caption{}
		\label{fig10}
	\end{subfigure}
	\caption{ Thermal evolution of (a) $\phi_1$ and (b) $\phi_2$ VEVs for $M_{h_2} = 250$ GeV, $\sin \theta =0.15 $, $w=725.7$ GeV, $M_{\gamma_d}= 60 $ GeV and $\Lambda = 800$ GeV. The phase transition proceeds from $(0,781~\text{GeV}) \to (147.7~\text{GeV}, 741.6~\text{GeV})$ at $T = 133.6 $ GeV}
    \label{vevevo}
\end{figure}

In Figure \ref{vevevo}, we illustrate the evolution of the  VEVs of of the doublet and the singlet scalar fields with temperature for the previously mentioned benchmark point ($M_{h_2} = 250$ GeV, $\sin \theta= 0.15$ , $w=725.7$ GeV, $M_{\gamma_d}= 60$ GeV, $\Lambda = 800$ GeV ) . It shows that the $\phi_1$ VEV ($\langle \phi_1 \rangle$) has a discontinuity of $\Delta \phi_1= 147.7$ GeV at the critical temperature $T=T_c = 133.6$ GeV leading to a FOEWPT of strength $\frac{\Delta\phi_1}{T_c} = \frac{\phi_c}{T_c} = 1.1$.

\begin{figure}[h!]
	\centering
	\begin{subfigure}[b]{0.48\textwidth}
		\centering
		\includegraphics[width = \textwidth]{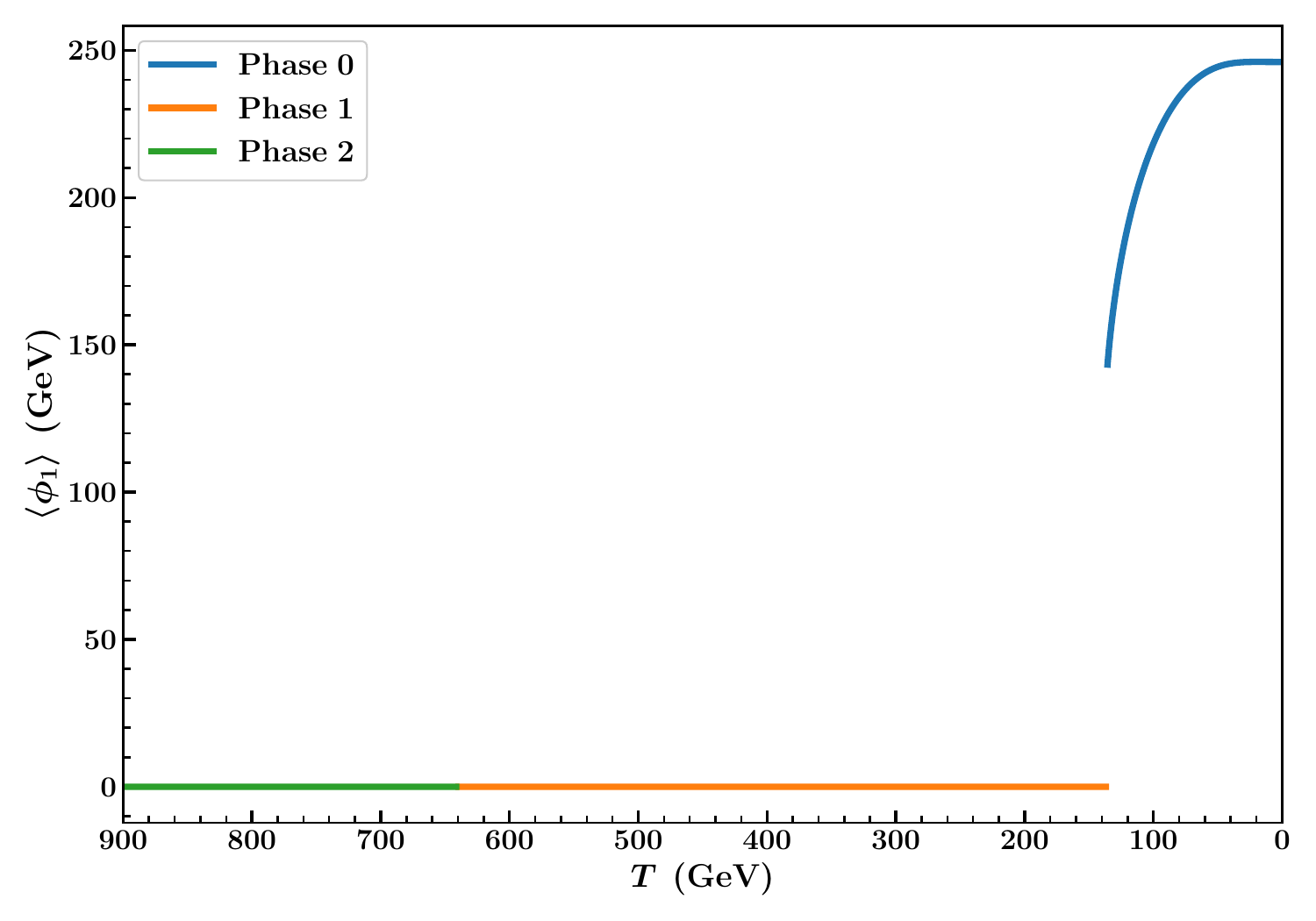}
		\caption{}
		\label{fig9_1}
	\end{subfigure}
	\hfill
	\begin{subfigure}[b]{0.48\textwidth}
		\centering
		\includegraphics[width= \textwidth]{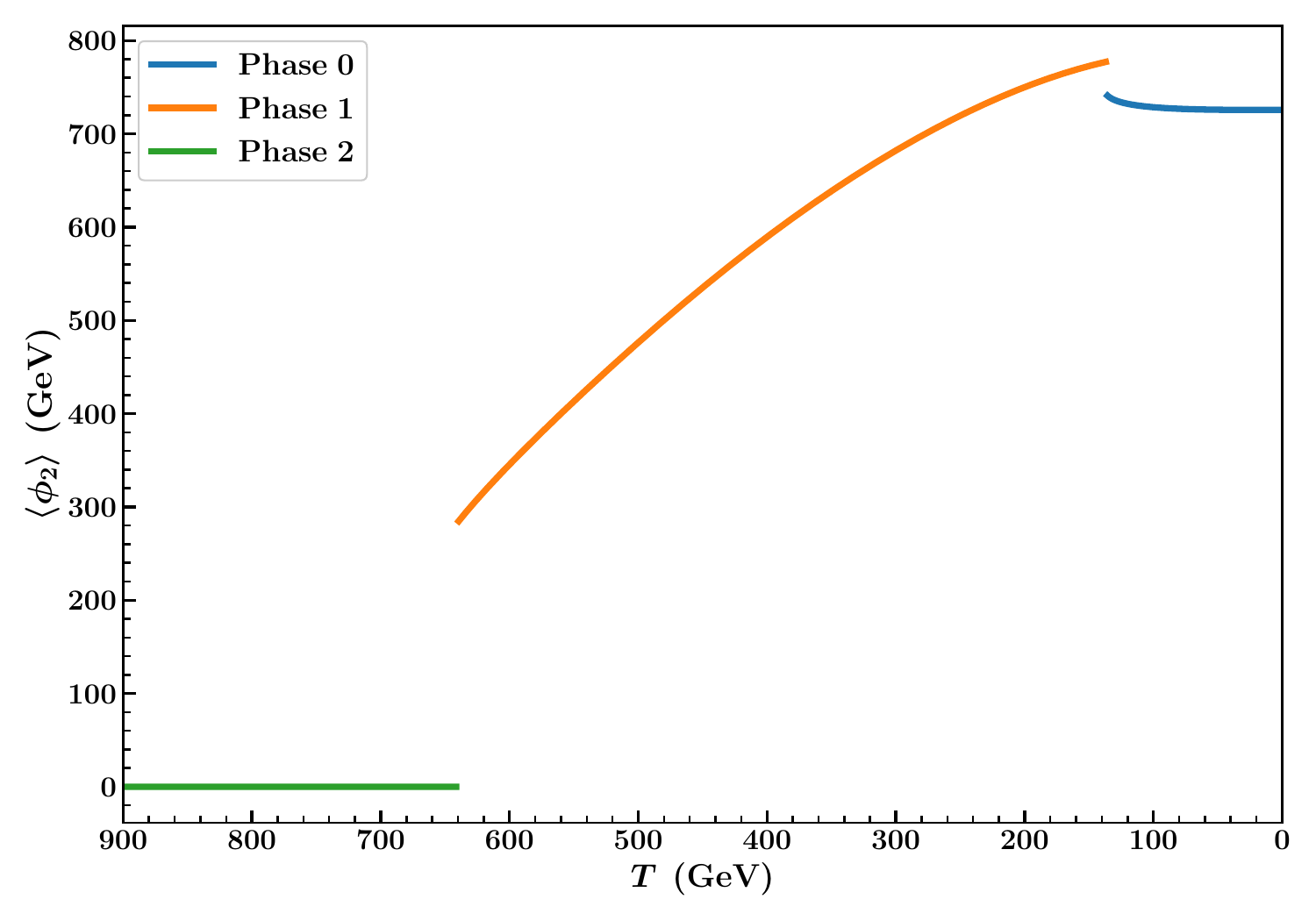}
		\caption{}
		\label{fig10_1}
	\end{subfigure}
	\caption{Thermal evolution of (a) $\phi_1$ and (b) $\phi_2$ VEVs for $M_{h_2} = 250$ GeV, $\sin \theta =0.15 $, $w=725.7$ GeV, $M_{\gamma_d}= 400 $ GeV and $\Lambda = 800$ GeV. The phase transition proceeds from $(0,0) \to (0, 284.7~\text{GeV})$ at $T= 639.4$ GeV, then from  $(0, 777.4~\text{GeV}) \to (142.3~\text{GeV}, 741.1~\text{GeV})$ at $T= 135.76$ GeV}
    \label{vevevo_1}
\end{figure}

As discussed previously that for low dark photon mass $M_{\gamma_d} = 60$ GeV, the value of $c_s$ in Equation~\ref{vht} was negative resulting in non restoration of $U(1)_D$ symmetry even at high temperature as can be seen from Figure~\ref{fig10}. However, for high dark photon mass the $\phi_2$ VEV at high temperature can in principle become zero, eventually restoring the $U(1)_D$ symmetry. To illustrate this, we plot the evolution of $\phi_1$ and $\phi_2$ VEVs with temperature for the same set of parameters $M_{h_2}, w, \sin\theta$ but now we take $M_{\gamma_d}= 400 $ GeV, in Figure~\ref{vevevo_1}. As a result of heavy dark photon mass the $U(1)_D$ symmetry is restored at high temperature (see Figure~\ref{fig10_1}). Moreover, the resulting phase transition is strongly first-order with a PT pattern $(0,0)\to (0,u(T)) \to (v(T), w(T))$. Such phase transitions might be useful in the context of dark matter relic abundance calculation as claimed in~\cite{Huang:2026qdc, Mahapatra:2026fyv} and GW signal. However, the strength of the EWPT remains almost unaffected or weakened for higher dark photon mass. Therefore, for the rest of our analysis we work with a low mass dark photon, e.g., $M_{\gamma_d}= 60$ GeV as a representative choice, consistent with existing low energy observations.

\begin{figure}[h!]
	\centering
	\begin{subfigure}[b]{0.48\textwidth}
		\centering
		\includegraphics[width= \textwidth]{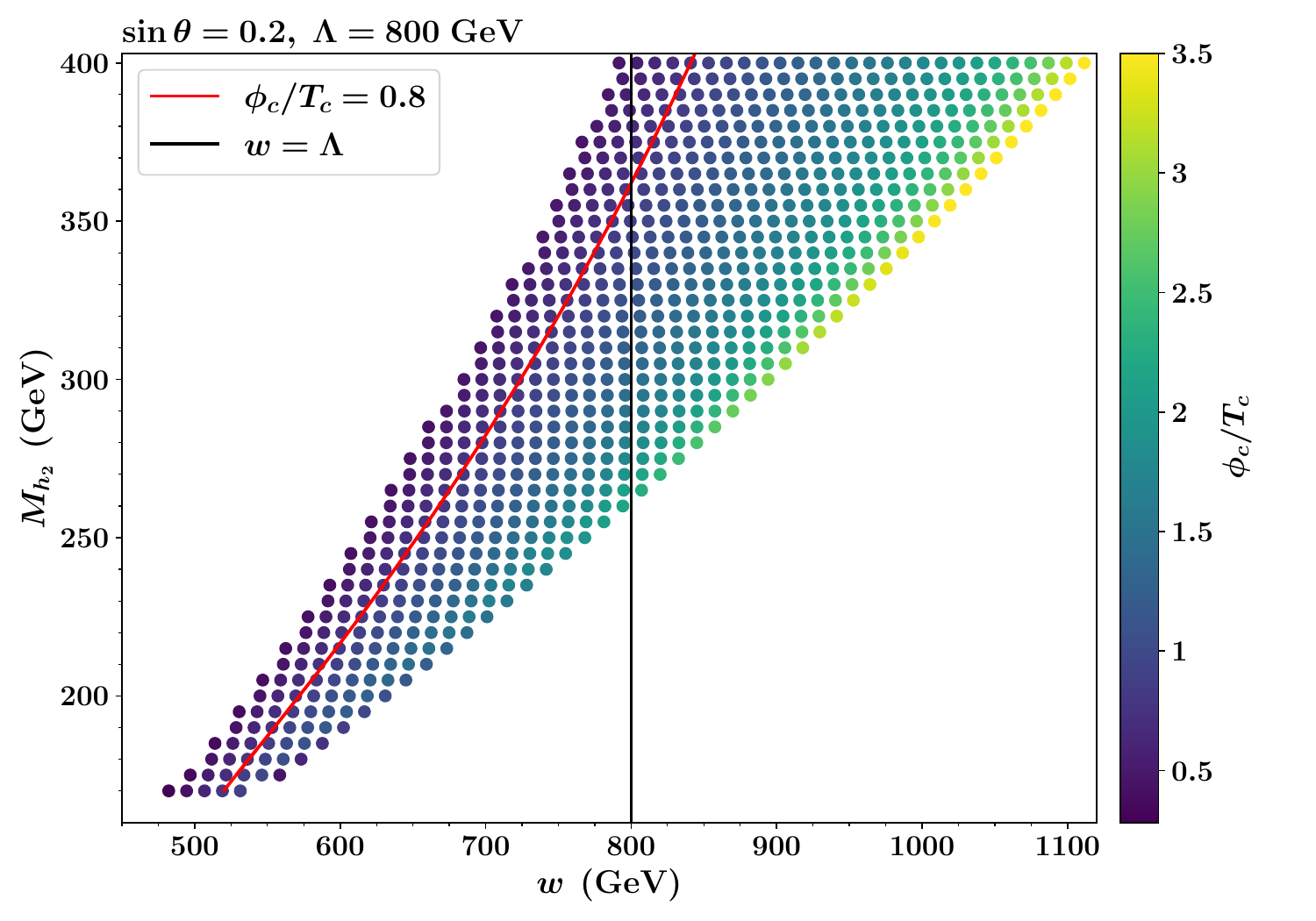}
		\caption{}
		\label{fig15}
	\end{subfigure}
	\hfill
	\begin{subfigure}[b]{0.48\textwidth}
		\centering
		\includegraphics[width=\textwidth]{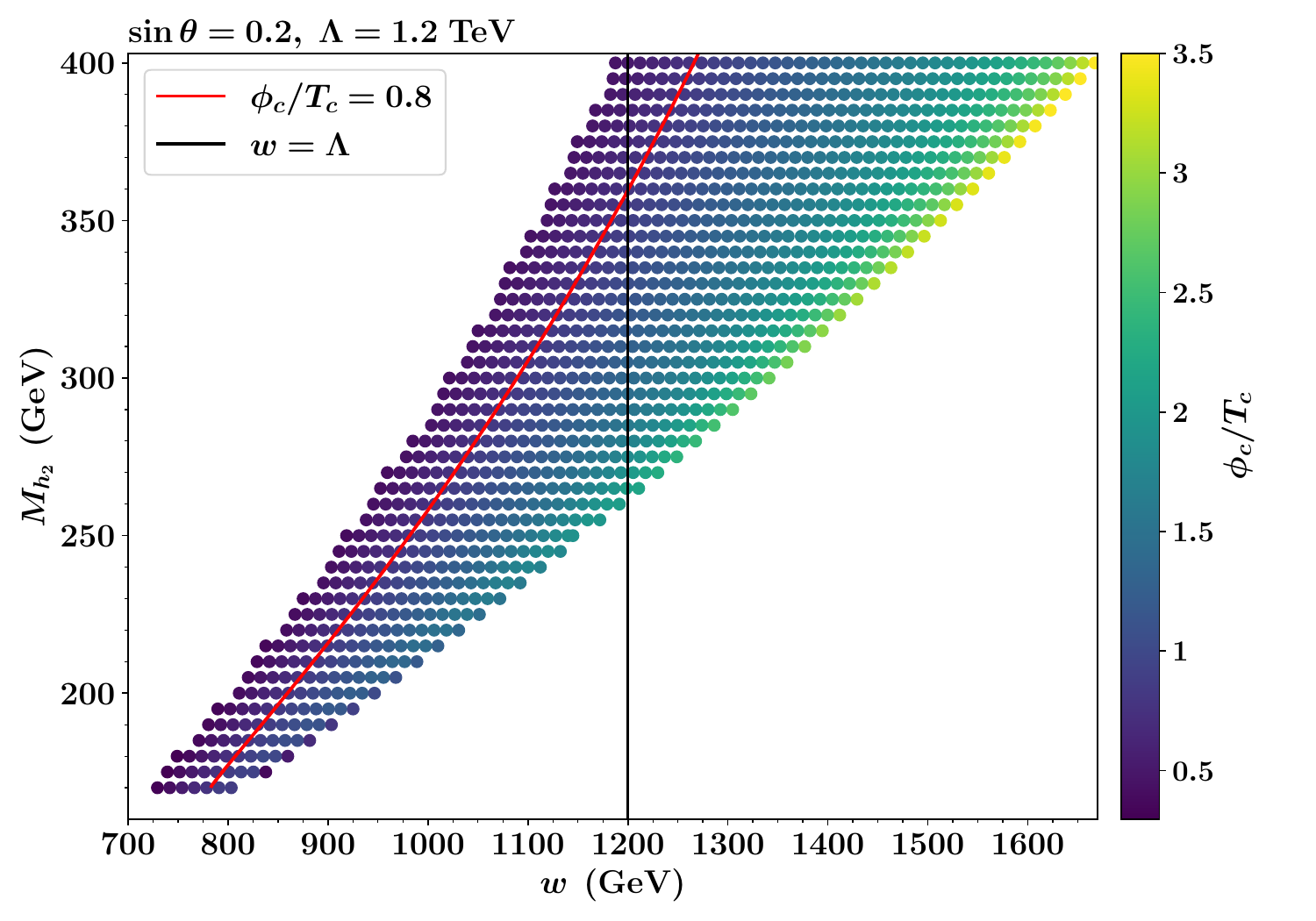}
		\caption{}
		\label{fig16}
	\end{subfigure}
\vspace{0.3cm} 

\begin{subfigure}[b]{0.48\textwidth}
	\centering
	\includegraphics[width=\textwidth]{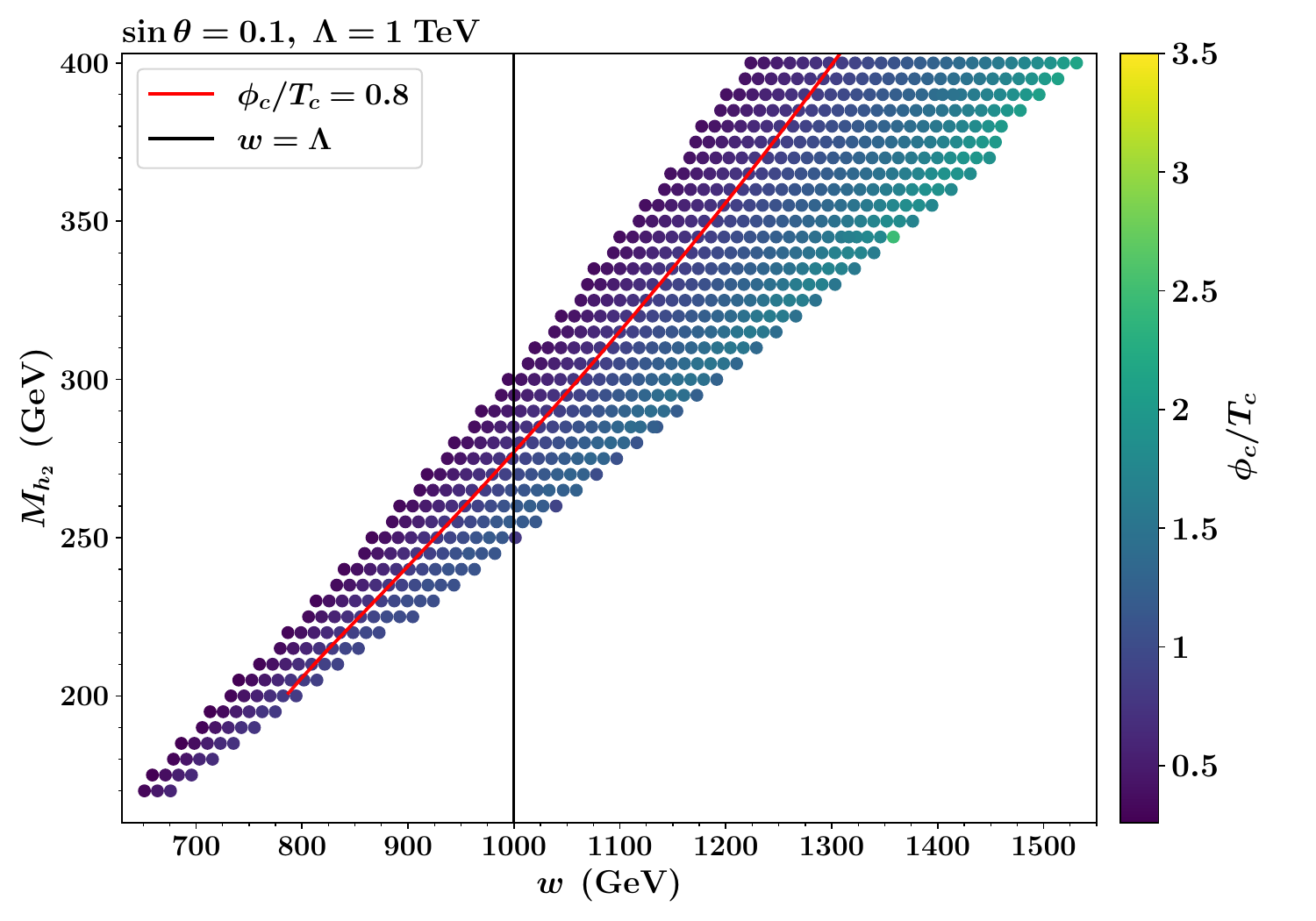}
	\caption{}
	\label{fig19}
\end{subfigure}
\hfill
\begin{subfigure}[b]{0.48\textwidth}
	\centering
	\includegraphics[width=\textwidth]{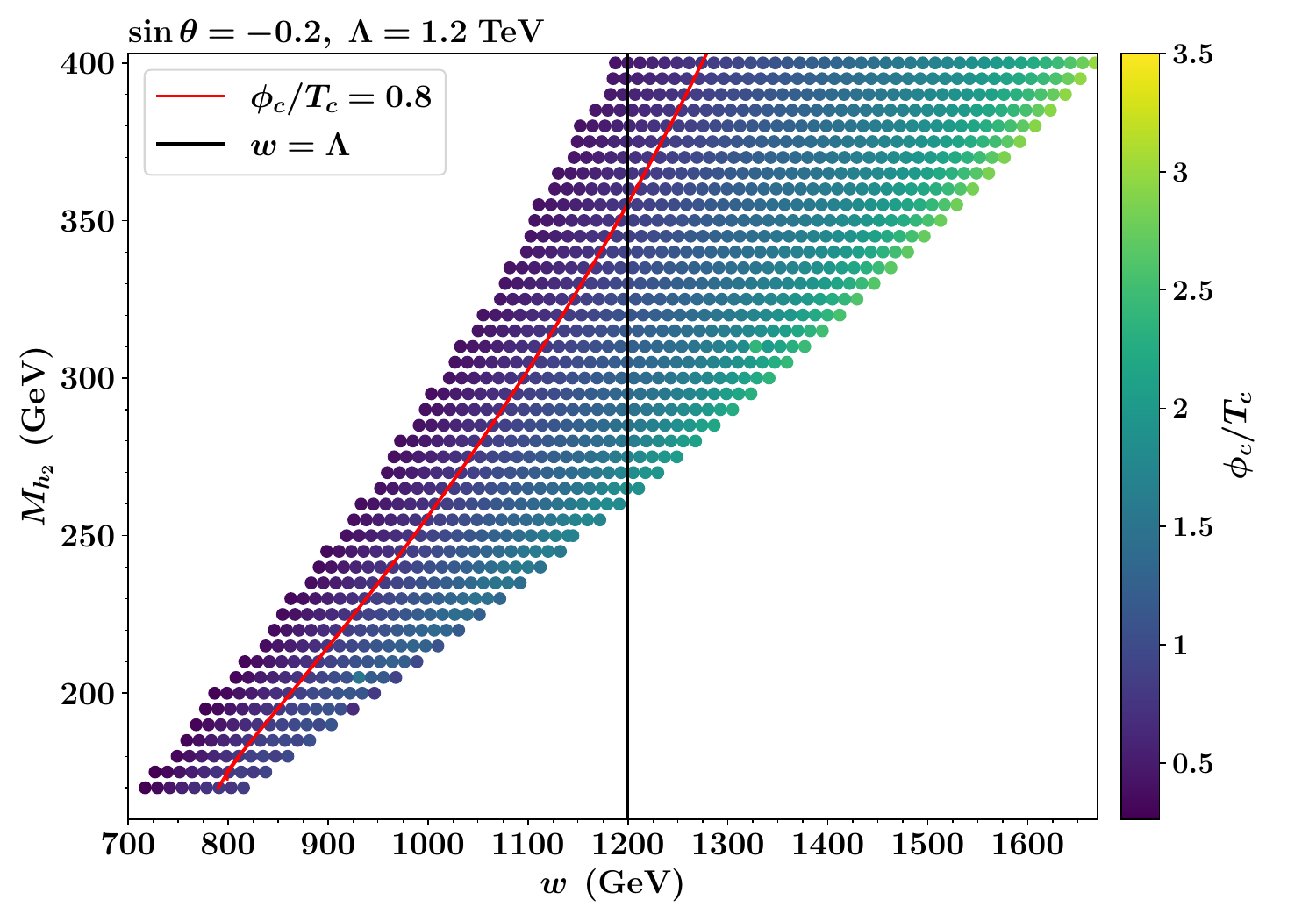}
	\caption{}
	\label{fig20}
\end{subfigure}

\caption{Strength of the EWPT ($\frac{\phi_c}{T_c}$) in the $w - M_{h_2}$ plane for different choices of $\sin\theta$ and $\Lambda$. The color bar on the right of the plots represent the value of $\frac{\phi_c}{T_c}$. The points on the left of the red line corresponds to $\frac{\phi_c}{T_c}< 0.8$. The black line corresponds to the equation $w= \Lambda$.}
\label{fig:Ms_tanb_combined}
\label{Ms-tanb}
\end{figure}

 In Figure~\ref{Ms-tanb}, we display the region of parameter space in the  $M_{h_2}-w$ plane consistent with the requirement of first order EWPT for various choices of scalar mixing angle and the cut-off scale. The value of $\frac{\phi_c}{T_c}$ has been shown in color gradient. The colored region which gives first order EWPT is bounded from the left by the requirement of SFOEWPT $\frac{\phi_c}{T_c}\geq 0.8$ and on the right by the requirement that $(v_{_{\text{EW}}},w)$ is the global minima of the tree-level potential (Equation~\ref{bbcond1} and Equation~\ref{bbcond2}) and  the potential is bounded from below (Equation \ref{bbcond}). The region on the right of the verticle black line which corresponds to $w>\Lambda$ is excluded by the requirement of validity of the EFT approach.

\begin{figure}[h!]
    \centering
	\begin{subfigure}[b]{0.48\textwidth}
		\centering
		\includegraphics[width= \textwidth]{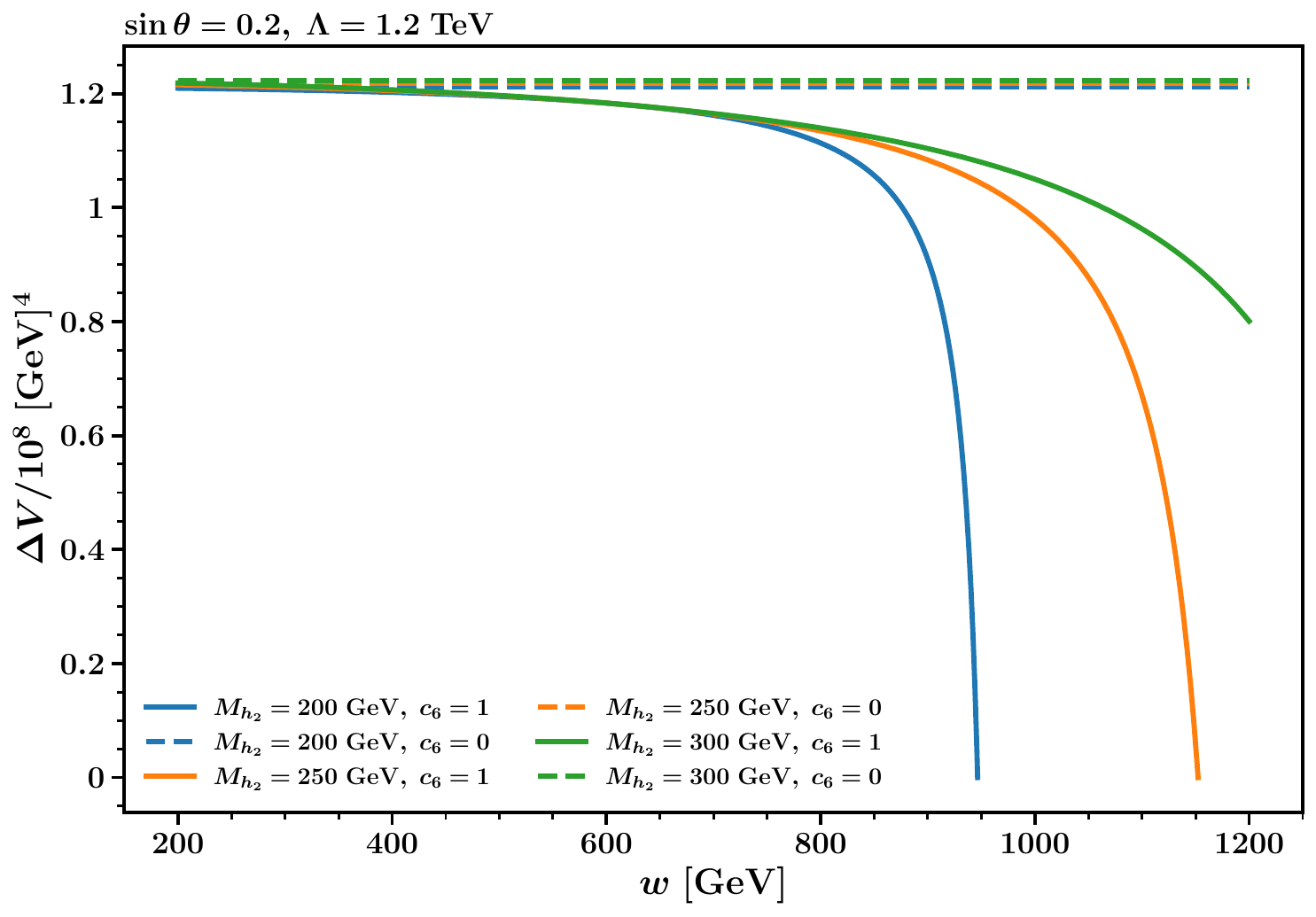}
		\caption{}
		\label{deltaV_Mh2}
	\end{subfigure}
	\hfill
	\begin{subfigure}[b]{0.48\textwidth}
		\centering
		\includegraphics[width=\textwidth]{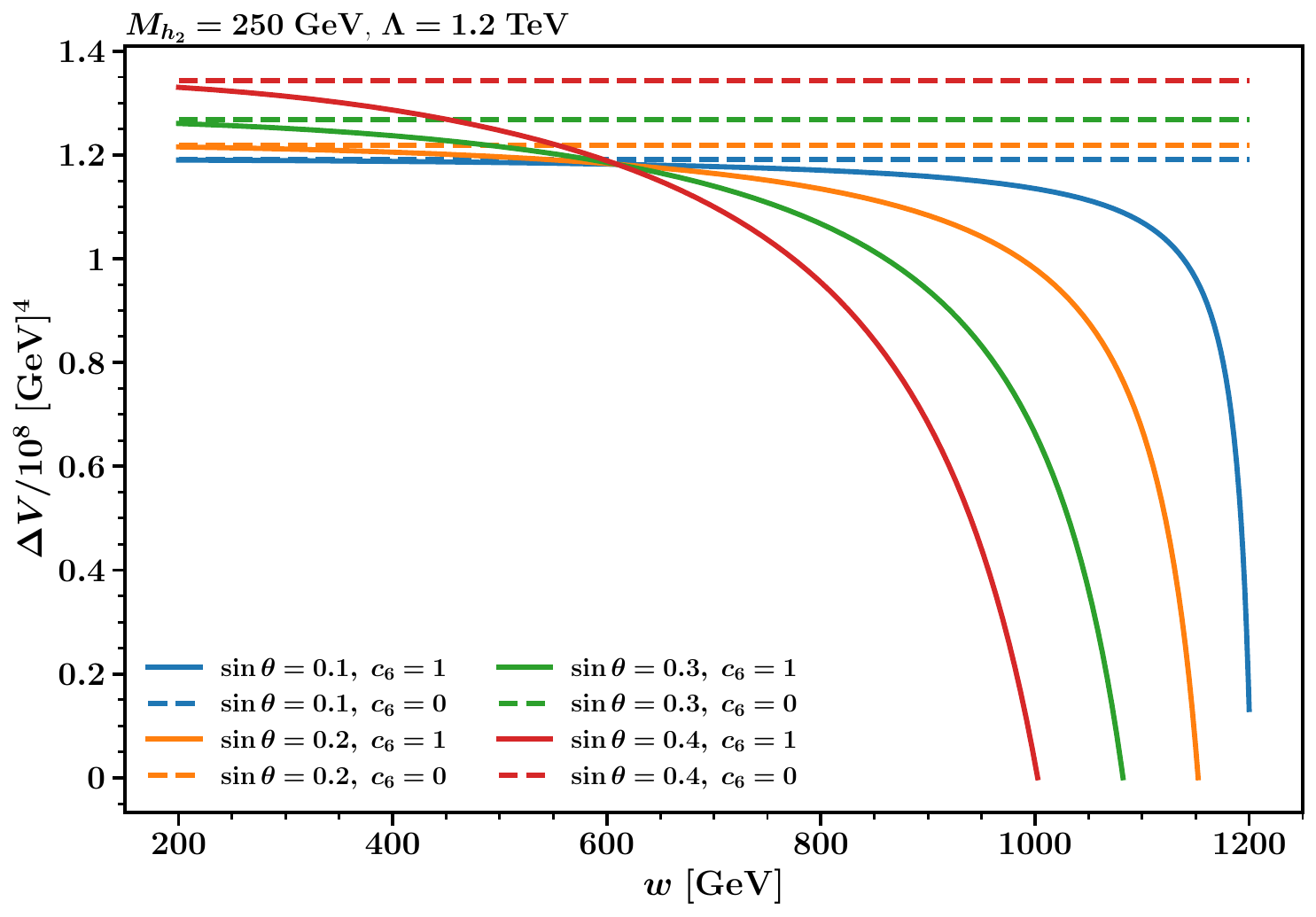}
		\caption{}
		\label{deltaV_sintheta}
	\end{subfigure}
	\vspace{0.3cm} 
	
\begin{subfigure}[b]{0.48\textwidth}
		\centering
		\includegraphics[width=\textwidth]{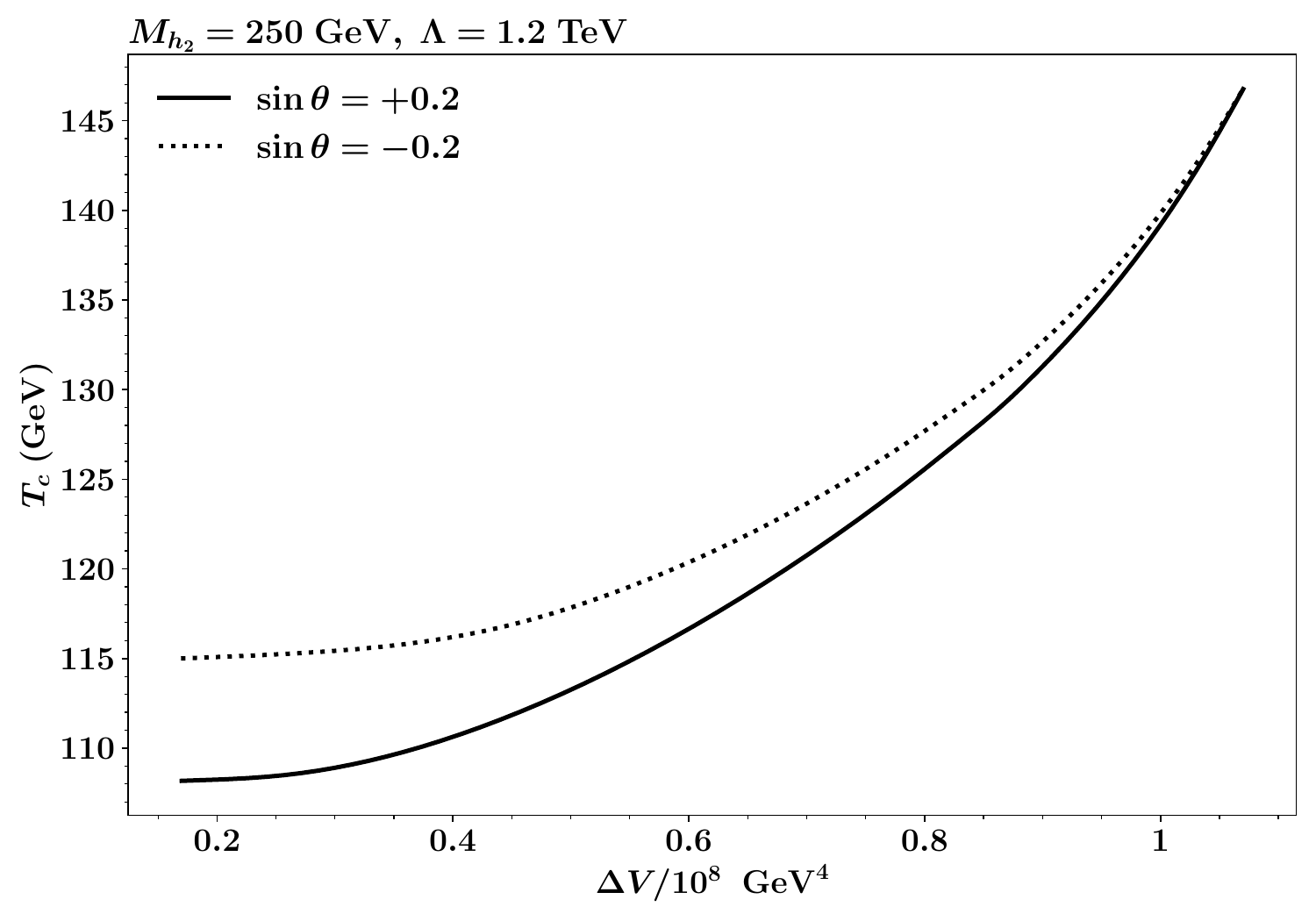}
		\caption{}
		\label{potdiff_lambdahs}
	\end{subfigure}
    \hfill
    \begin{subfigure}[b]{0.48\textwidth}
		\centering
		\includegraphics[width=\textwidth]{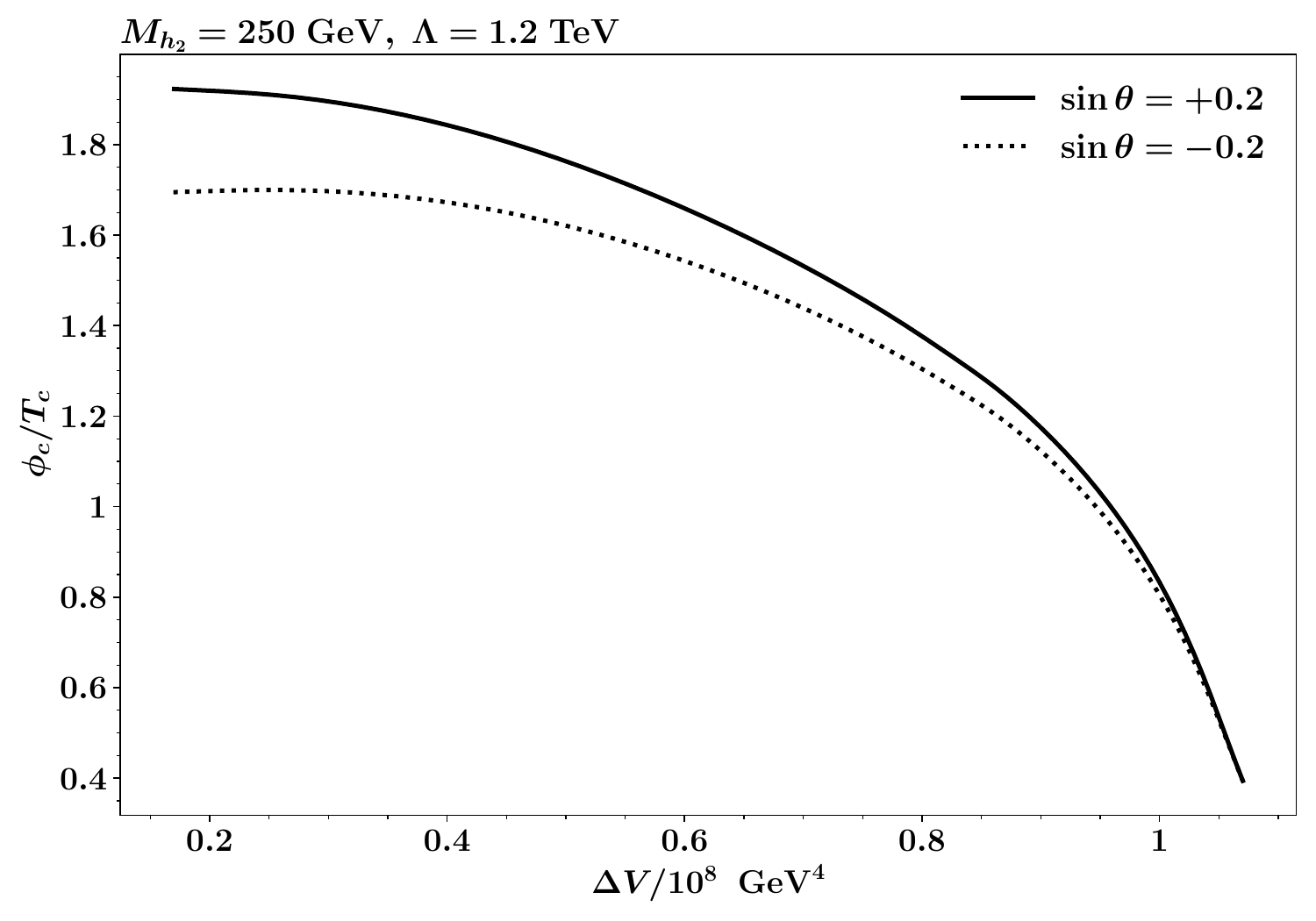}
		\caption{}
		\label{fig23}
	\end{subfigure}
    \vspace{0.3cm} 
	\centering
	\begin{subfigure}[b]{0.48\textwidth}
		\centering
		\includegraphics[width= \textwidth]{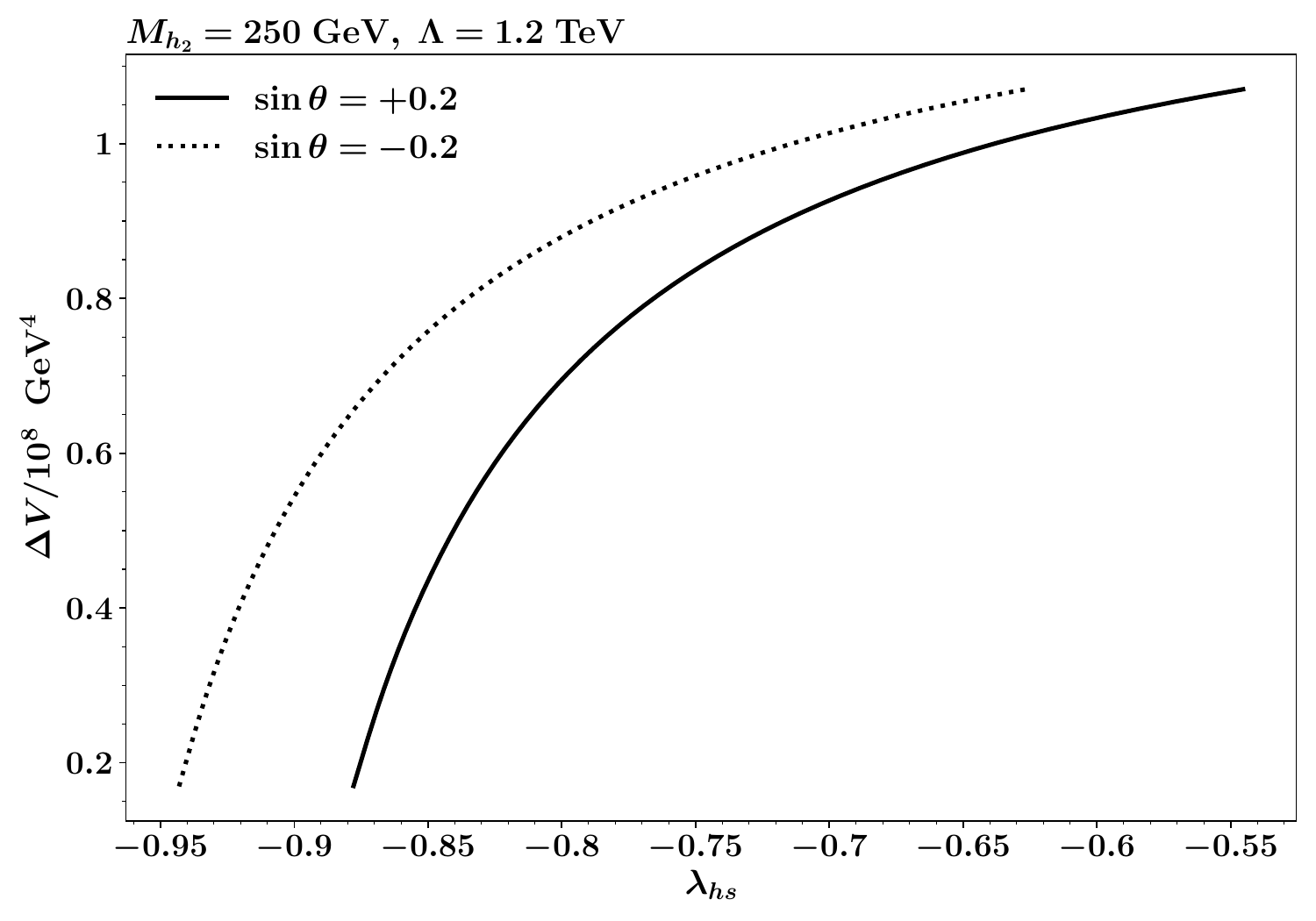}
		\caption{}
		\label{deltaV_lambdahs}
	\end{subfigure}
	\hfill
	\begin{subfigure}[b]{0.48\textwidth}
		\centering
		\includegraphics[width=\textwidth]{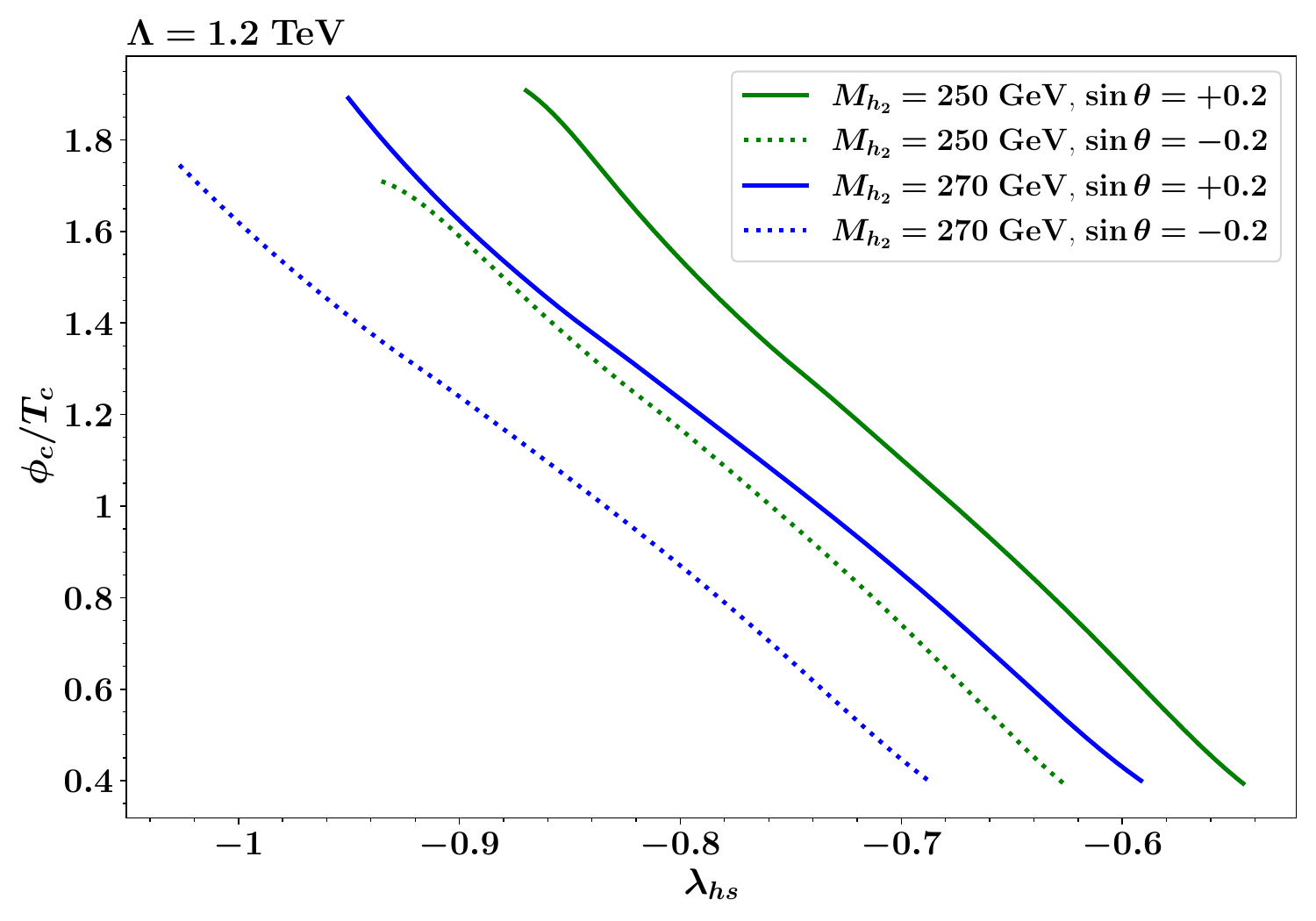}
		\caption{}
		\label{phictc_lambdahs}
	\end{subfigure}

	\caption{Tree-level potential difference, $\Delta V = V_0(0,u) - V_0(v_{_{\text{EW}}},w)$ as a function of $w$ (a) for fixed $\sin\theta$ and varying $M_{h_2}$, and (b) for fixed $M_{h_2}$ and varying $\sin\theta$. Here, solid (dashed) lines correspond to $c_6=1$ ($c_6=0$). Panel (c) and (d) show the dependence of $T_c$ and $\phi_c/T_c$, respectively on $\Delta V$. Panel (e) and (f) display the variation of $\Delta V $ and $\frac{\phi_c}{T_c}$, respectively with $\lambda_{hs}$. In sub-figures (c)–(f), solid (dotted) lines denote $\sin\theta = +0.2$ ($-0.2$). The cut-off scale is assumed to be $1.2$ TeV to obtain these plots.}
	\label{fig:lambda_hs}
\end{figure}

One can see that, for a given value of $M_{h_2}$, the ratio $\phi_c/T_c$ increases with increase in $w$ within its allowed range. This behavior can be traced back to the fact that, as $w$ increases, the zero-temperature tree-level potential difference ($\Delta V = V_0(0, u)- V_0(v_{_{\text{EW}}},w ) $) between the extremum $(0,u)$, which preserves electroweak (EW) symmetry, and the true vacuum $(v_{_{\text{EW}}}, w)$, corresponding to the broken EW phase, gradually decreases. Eventually, beyond a certain value of $w$, the point $(0,u)$ becomes the global minimum of the tree-level potential. This behavior can be understood from a careful inspection of Figure~\ref{region} and clearly from Figure~\ref{deltaV_Mh2} and ~\ref{deltaV_sintheta}. It is commonly understood that a smaller tree-level potential difference $\Delta V$ leads to a lower critical temperature~\cite{Huang:2014ifa, Harman:2015gif, Dorsch:2017nza, Carena:2019une}, which in turn enhances the strength of the EW phase transition. We explicitly illustrate this with the help of Figure~\ref{potdiff_lambdahs} and ~\ref{fig23}, respectively.

This feature can also be understood in terms of the lagrangian parameter $\lambda_{hs}$.
As $w$ increases, $\lambda_{hs}$ attains sizable negative values ($|\lambda_{hs}|\sim 1$) (see Equation~\ref{lambdahs} and the corresponding plot in Figure~\ref{lambda_hs_w_200_300} in the appendix). The tree-level potential difference ($\Delta V$) decreases as $\lambda_{hs}$ attains more and more negative values, resulting in an enhanced $\frac{\phi_c}{T_c}$. These findings are also supported by Figure~\ref{deltaV_lambdahs} and \ref{phictc_lambdahs}.
Therefore, it is clear that negative $\lambda_{hs}$ favors SFOEWPT in our scenario. Figures~\ref{potdiff_lambdahs}--\ref{phictc_lambdahs} further indicate that the potential difference is larger, as a consequence $\frac{\phi_c}{T_c}$ is smaller, for negative $\sin\theta$ compared to the corresponding positive value of same magnitude for a fixed $\lambda_{hs}$. The fact that $\lambda_{hs}$ and $\sin\theta$ are not strongly correlated is a characteristic feature of the particular model framework considered in this work.

\begin{figure}[h!]
	\centering
	\begin{subfigure}[b]{0.48\textwidth}
		\centering
		\includegraphics[width=\textwidth]{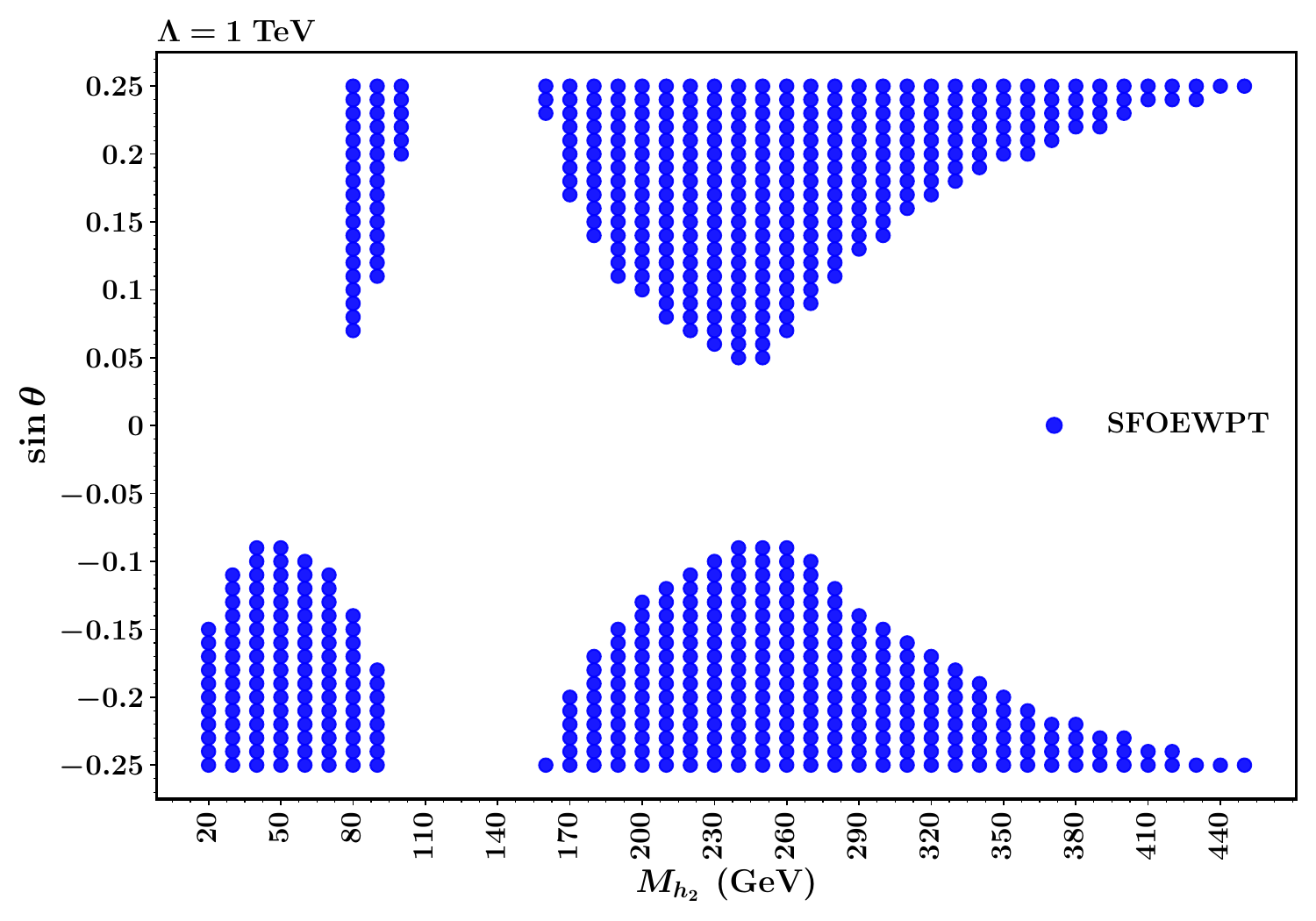}
		\caption{}
		\label{fig:scan1000}
	\end{subfigure}
    \begin{subfigure}[b]{0.48\textwidth}
		\centering
		\includegraphics[width=\textwidth]{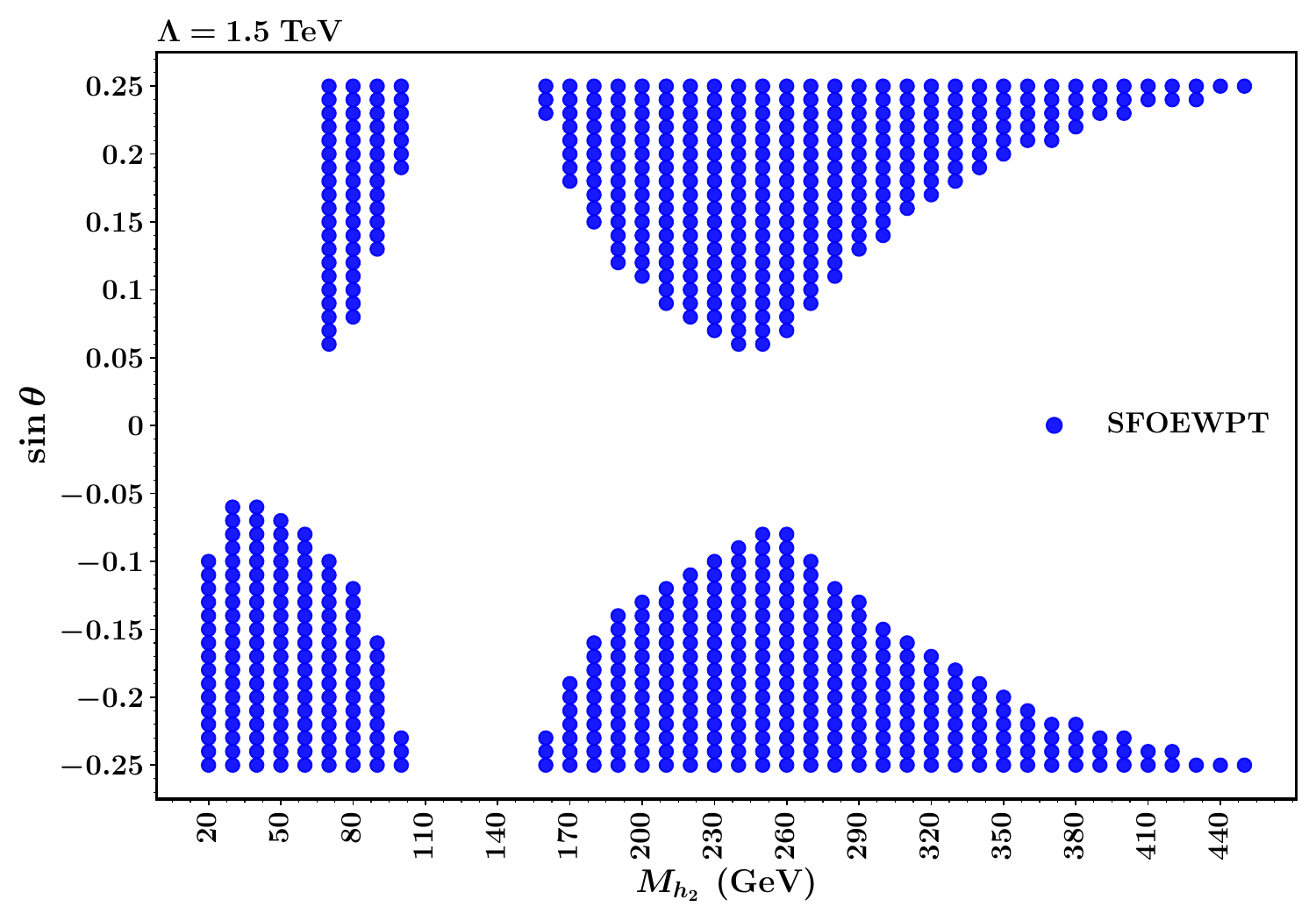}
		\caption{}
		\label{fig:scan1500}
	\end{subfigure}
	\hfill
	\caption{SFOEWPT allowed parameter space ($\frac{\phi_c}{T_c}\geq 0.8$) in the $M_{h_2}$--$\sin\theta$ plane for two different choices of the cutoff scale (a) $\Lambda = 1$ TeV (left), (b) $\Lambda = 1.5$ TeV (right) assuming full one loop finite temperature corrected effective potential described by Equation~\ref{fullveff}.}
	\label{fig:sfoewpt_scan}
\end{figure}

In light of the above discussion, we now present the region of parameter space in the $M_{h_2}-\sin\theta$ plane consistent with SFOEWPT in Figure~\ref{fig:sfoewpt_scan}. The parameter space can be divided into four regions. In the absence of the effective operator, it is well known that only for $M_{h_2} \ll M_{h_1}$ does one obtain an SFOEWPT-compatible parameter space with sizeable $\sin\theta$~\cite{Carena:2019une}. For higher values of $M_{h_2}$, it is difficult to accommodate a strong FOEWPT without assuming large $\sin\theta$, which is already ruled out by the Higgs signal strength measurements. This is in contrast with the present scenario, where, in the presence of the effective operator, the correlation between $\lambda_{hs}$ and $\sin\theta$ is weakened, and a sizeable $\lambda_{hs}$ can be achieved even for small $\sin\theta$. One can see that there exists a wide range of SFOEWPT-compatible parameter space in the region $M_{h_2} > M_{h_1}$ for both positive and negative $\sin\theta$, as it is now facilitated by large $w$ rather than $\sin\theta$. The slight asymmetry between the allowed regions corresponding to positive and negative $\sin\theta$ for $M_{h_2} > M_{h_1}$ can be traced back to Figure~\ref{potdiff_lambdahs}, which suggests that the tree-level potential difference $\Delta V$ is smaller for positive $\sin\theta$ for a fixed $\lambda_{hs}$.

In the left half of Figure~\ref{fig:sfoewpt_scan} (for $M_{h_2} < M_{h_1}$), there exist two competing phase transition patterns: $(0,0)\rightarrow (v(T),w(T))$ or $(0,u(T)) \rightarrow (v(T),w(T))$. The former is more likely when $(0,u)$ is not an extremum of the tree-level potential. This occurs when $\mu_s^2$ is negative. In fact, for very low $M_{h_2}$ and positive $\sin\theta$, $\mu_s^2$ is mostly negative (see Figure~\ref{mus2}). Transitions of the type $(0,0)\rightarrow (v(T),w(T))$ are less likely as the tree-level potential difference between $V_0(0,0)$ and $V_0(v,w)$ is large enough to rule out strong or even first-order phase transitions\footnote{This can be understood by looking at the bottom left corner of Figure~\ref{reg1}, which suggests that the allowed green shaded region (with $\mu_s^2 <0$) is well separated from the $V_0(v_{_{\text{EW}}},w)=V_0(0,0)$ (yellow) line by the requirement $\lambda_s>0$.}. For higher values of $M_{h_2}$, $\mu_s^2$ becomes positive again (see Figure~\ref{mus2_w_40_80} in appendix), and the phase transition proceeds via $(0,u(T)) \rightarrow (v(T),w(T))$, as usual. 
For negative $\sin\theta$ and very small $M_{h_2}$, the phase transition pattern is similar to that in the singlet scalar extension of the SM with spontaneous $\mathcal{Z}_2$ breaking, as the contribution from the dimension-six operator is suppressed by $\frac{w^2}{\Lambda^2}$. As the value of $M_{h_2}$ increases, the phase transition proceeds via the usual pattern for both positive and negative $\sin\theta$.

    There are no SFOEWPT-compatible parameter space points in the $M_{h_2}-\sin\theta$ plane in the degenerate ($M_{h_1} \sim M_{h_2}$) and zero-mixing limit ($\sin\theta \to 0$). Even though in this limit $(v,w)$ still remains the global minimum of the tree-level potential, the tree-level potential difference $\Delta V$ becomes independent of the singlet scalar VEV $w$ (see Equation~\ref{gmlimit}). Therefore, one can not minimize this difference further by increasing $w$ and consequently it does not help in lowering the critical temperature $T_c$. There are also no parameter space points consistent with SFOEWPT in the extreme right region of the plots in Figure~\ref{fig:sfoewpt_scan}, as it requires the value of $w$ to be greater than the cut-off scale $\Lambda$.

\begin{figure}[h!]
	\centering
	\begin{subfigure}[b]{0.48\textwidth}
		\centering
		\includegraphics[width=\textwidth]{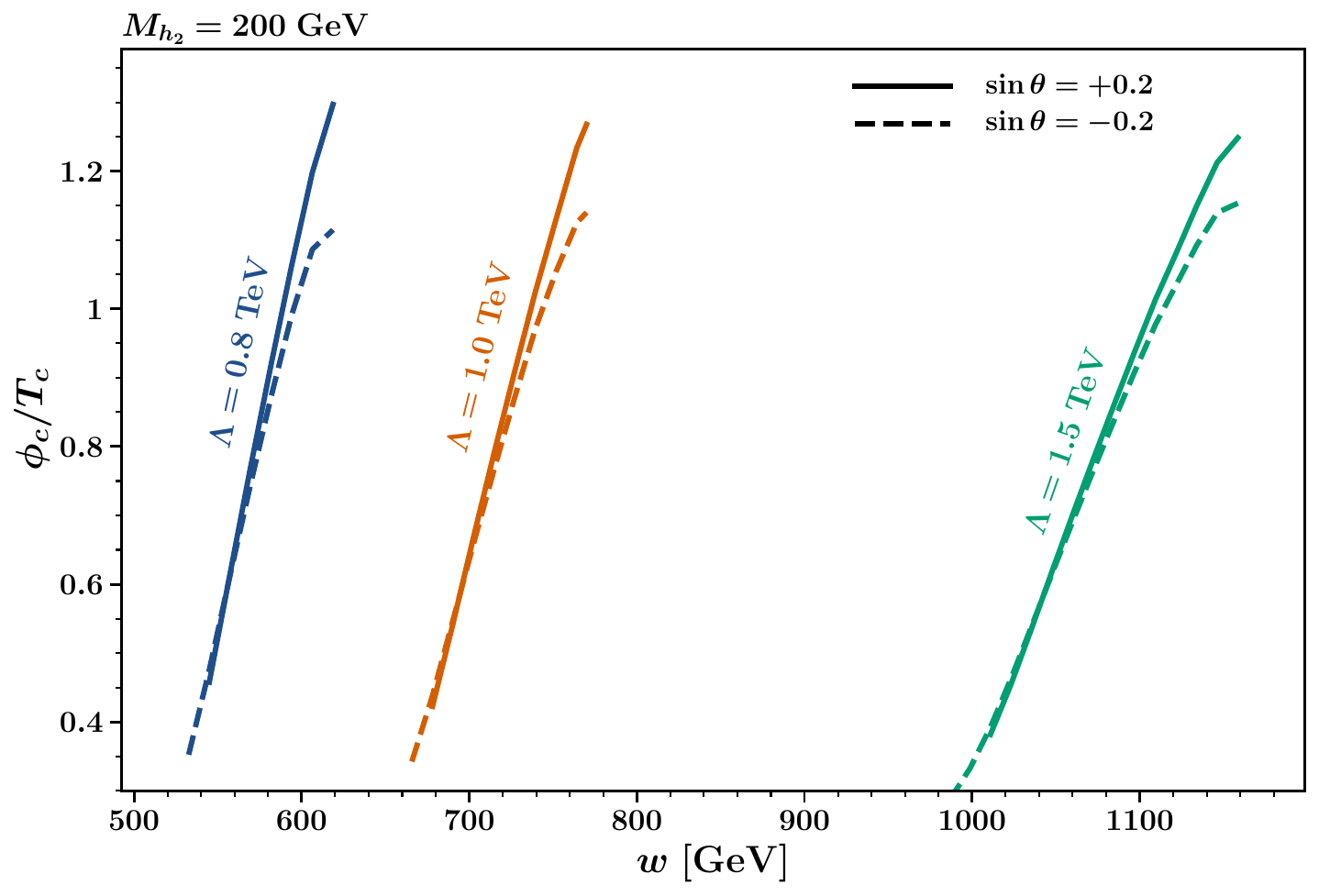}
		\caption{}
		\label{w_Lambda_200}
	\end{subfigure}
    \begin{subfigure}[b]{0.48\textwidth}
		\centering
		\includegraphics[width=\textwidth]{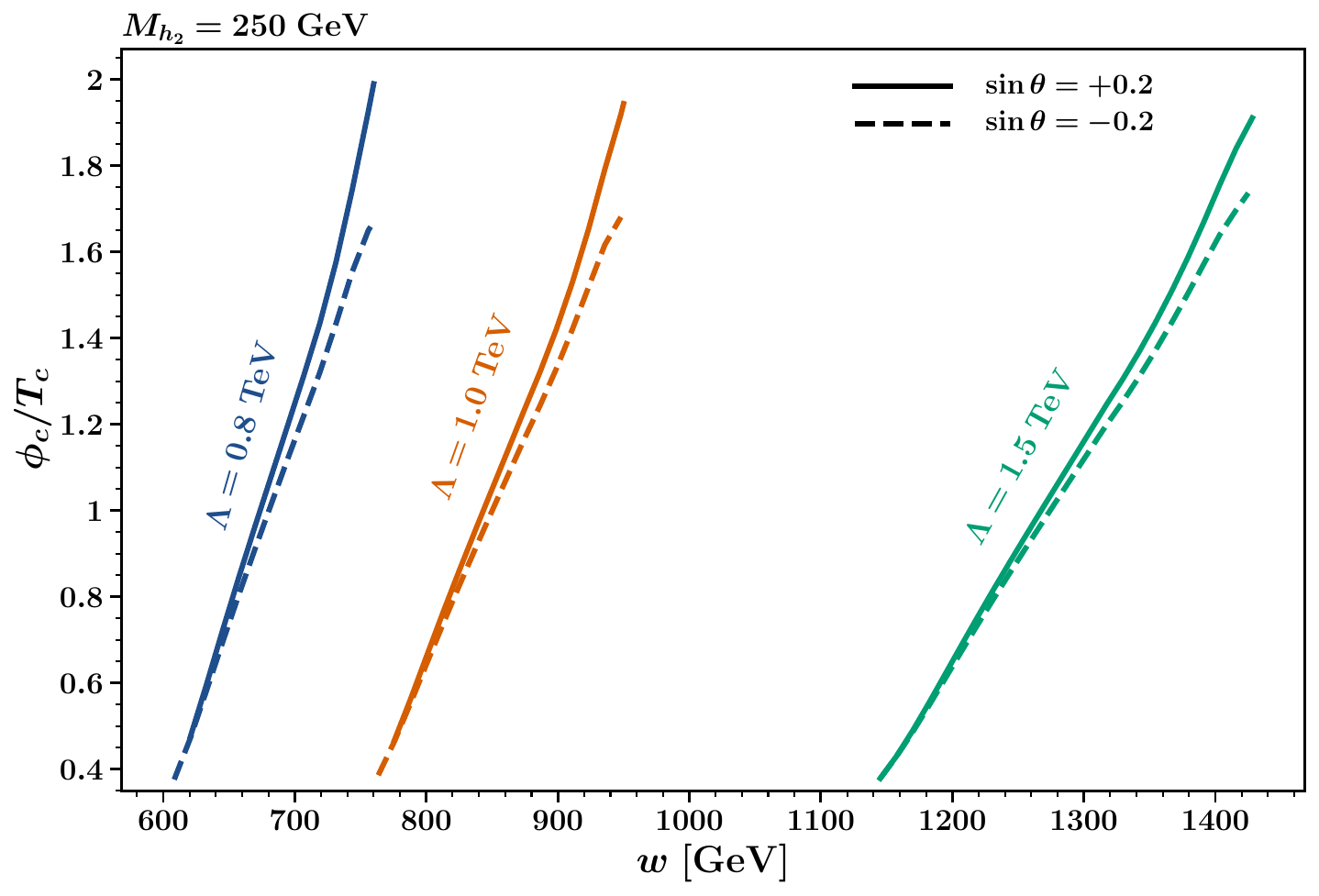}
		\caption{}
		\label{w_Lambda_250}
	\end{subfigure}
	\hfill
	\caption{Variation of $\frac{\phi_c}{T_c}$ with $w$ for three different choices of the cut-off scale (800 GeV, 1 TeV and 1.5 TeV) assuming (a) $M_{h_2}= 200$ GeV  and (b) $M_{h_2}= 250$ GeV. The solid and dashed lines correspond to $\sin\theta= +0.2$ and $-0.2$, respectively. }
	\label{woverlambda}
\end{figure}

     In Table~\ref{tab:ptpoints}, we present few representative benchmark points $M_{h_2},\ \sin\theta$ and corresponding minimum values of singlet scalar VEV ($w$) which lead to $\phi_c/T_c \geq 0.8$ assuming the cut-off scale $\Lambda=1$ TeV. Figure~\ref{woverlambda} illustrates the dependency of $\phi_c/T_c$ on the singlet scalar VEV $w$ for three different choices of the cut-off scale (800 GeV, 1 TeV, and 1.5 TeV). These plots also suggest that the ratio $w/\Lambda$ instead of the absolute value of the cut-off scale is a relevant factor in the context of EWPT.

\begin{table}[h!]
\centering
\small
\setlength{\tabcolsep}{6pt}
\begin{tabular}{cccc}
\toprule
$M_{h_2}$ [GeV] & $\sin\theta$ & $w$ [GeV] & $\phi_c/T_c$ \\
\midrule
\multirow{2}{*}{170} & $+0.2$ & 650 & 0.80 \\
                     & $-0.2$ & 660 & 0.80 \\
\midrule
\multirow{2}{*}{180} & $+0.2$ & 672 & 0.80 \\
                     & $-0.2$ & 676 & 0.81 \\
\midrule
\multirow{2}{*}{200} & $+0.2$ & 716 & 0.80 \\
                     & $-0.2$ & 719 & 0.80 \\
\midrule
\multirow{2}{*}{220} & $+0.2$ & 758 & 0.80 \\
                     & $-0.2$ & 762 & 0.81 \\
\midrule
\multirow{2}{*}{240} & $+0.2$ & 798 & 0.80 \\
                     & $-0.2$ & 802 & 0.80 \\
\midrule
\multirow{2}{*}{260} & $+0.2$ & 838 & 0.82 \\
                     & $-0.2$ & 841 & 0.80 \\
\midrule
\multirow{2}{*}{280} & $+0.2$ & 873 & 0.81 \\
                     & $-0.2$ & 878 & 0.80 \\
\midrule
\multirow{2}{*}{300} & $+0.2$ & 907 & 0.80 \\
                     & $-0.2$ & 913 & 0.80 \\
\midrule
\multirow{2}{*}{320} & $+0.2$ & 940 & 0.81 \\
                     & $-0.2$ & 947 & 0.81 \\
\midrule
\multirow{2}{*}{340} & $+0.2$ & 971 & 0.81 \\
                     & $-0.2$ & 977 & 0.80 \\
\midrule
\multirow{2}{*}{350} & $+0.2$ & 985 & 0.80 \\
                     & $-0.2$ & 993 & 0.81 \\
\bottomrule
\end{tabular}
\caption{Representative benchmark points $M_{h_2},\ \sin\theta$ and corresponding minimum values of singlet scalar VEV ($w$) that yield $\phi_c/T_c \geq 0.8$ assuming the cut-off scale $\Lambda=1$ TeV.}
\label{tab:ptpoints}
\end{table}

Finally, we conclude this section with the following remark. Even though we have considered only the dimension-six term of the form $|H|^2|\phi|^4$ a complete analysis in the context of the electroweak phase transition and its signatures at GW detectors and at the LHC requires a comprehensive study of all possible dimension-six operators allowed by the all possible symmetries of the theory. In principle, the UV theories that can potentially generate the above operator at low energy after integrating out heavy degrees of freedom can also give rise to other dimension-six operators, such as $|H|^4|\phi| ^2, |H|^6, |\phi|^6 $ with Wilson coefficients of the same order ($\sim \frac{1}{\Lambda^2}$) as above. For example, the inclusion of the operator $|H|^4|\phi| ^2$, in addition to the existing one, modifies Equations~\ref{lambdah}, \ref{lambdas}, \ref{lambdahs} (see Appendix) in the following way (assuming the Wilson coefficients to be of the same order)

\begin{align}
	\lambda_h &= \frac{M_{h_1}^2 \cos^2\theta + M_{h_2}^2 \sin^2\theta}{2 v_{_{\text{EW}}}^2} - \frac{w^2}{2 \Lambda^2},\label{lambdah_1} \\
	\lambda_s &= \frac{ M_{h_1}^2 \sin^2\theta + M_{h_2}^2 \cos^2\theta}{2 w^2} - \frac{v_{_{\text{EW}}}^2}{2 \Lambda^2} ,\label{lambdas_1} \\
	\lambda_{hs} &=- \frac{v_{_{\text{EW}}}^2}{ \Lambda^2} -\frac{w^2}{\Lambda^2} + \frac{\left(-M_{h_1}^2 + M_{h_2}^2\right) \cos\theta \sin\theta}{v_{_{\text{EW}}} w} \label{lambdahs_1}
\end{align}
One can see from Equation~\ref{lambdah_1} that the quartic Higgs coupling becomes negative for certain values of the ratio $\frac{w}{\Lambda}$ depending on the value of $\sin\theta$. The smaller the $\sin\theta$, the more stringent the upper limit becomes on the ratio $\frac{w}{\Lambda}$. However, a negative quartic Higgs coupling at tree level is perfectly acceptable if one includes all possible dimension-six operators in the scalar potential.
 The additional advantage one has in the presence of all possible dimension-six operators is both $\lambda_{h}$ and $\lambda_s$ are allowed to become negative. This eventually helps to create a barrier to facilitate FOEWPT, wider range of allowed parameter space, and at the same time helps to relax the upper bound on the cut-off scale, compared to that in the Higgs SMEFT scenario \cite{Postma:2020toi, Wagner:2023vqw}. Moreover, $\lambda_{hs}$ now receives an additional contribution of order $- \frac{v_{_{\text{EW}}}^2}{ \Lambda^2}$ (Equation~\ref{lambdahs_1}) which can induce SFOEWPT even at lower $w$ compared to the case where only the $|H|^2|\phi| ^4$ term is present. A complete analysis of EWPT including the full list of dimension-six operators in the singlet scalar extension framework will be taken up in a future work.

%

\section{Nucleation}
\label{nucleat}

During a strong first-order cosmological phase transition, the Universe evolves via thermal
tunneling from a metastable false vacuum to the true vacuum via the nucleation of
critical bubbles. The bubble nucleation rate per unit volume at temperature $T$ is given by~\cite{Linde:1981zj}
\begin{equation}
\Gamma(T) \simeq T^4
\left(\frac{S_3(T)}{2\pi T}\right)^{3/2}
\exp\!\left[-\frac{S_3(T)}{T}\right],
\end{equation}
where $S_3(T)$ denotes the three-dimensional Euclidean action of the critical bubble and expressed as
\begin{equation}
S_3 = 4\pi \int_0^\infty r^2 dr \left[ \frac{1}{2}\sum_{i} \left( \frac{d\phi_i}{dr} \right)^2 + V_{eff}(\phi_1, \phi_2, T) \right]
\end{equation}

\begin{figure}[h!]
   \begin{subfigure}[b]{0.48\textwidth}
		\centering
	\includegraphics[width = \textwidth]{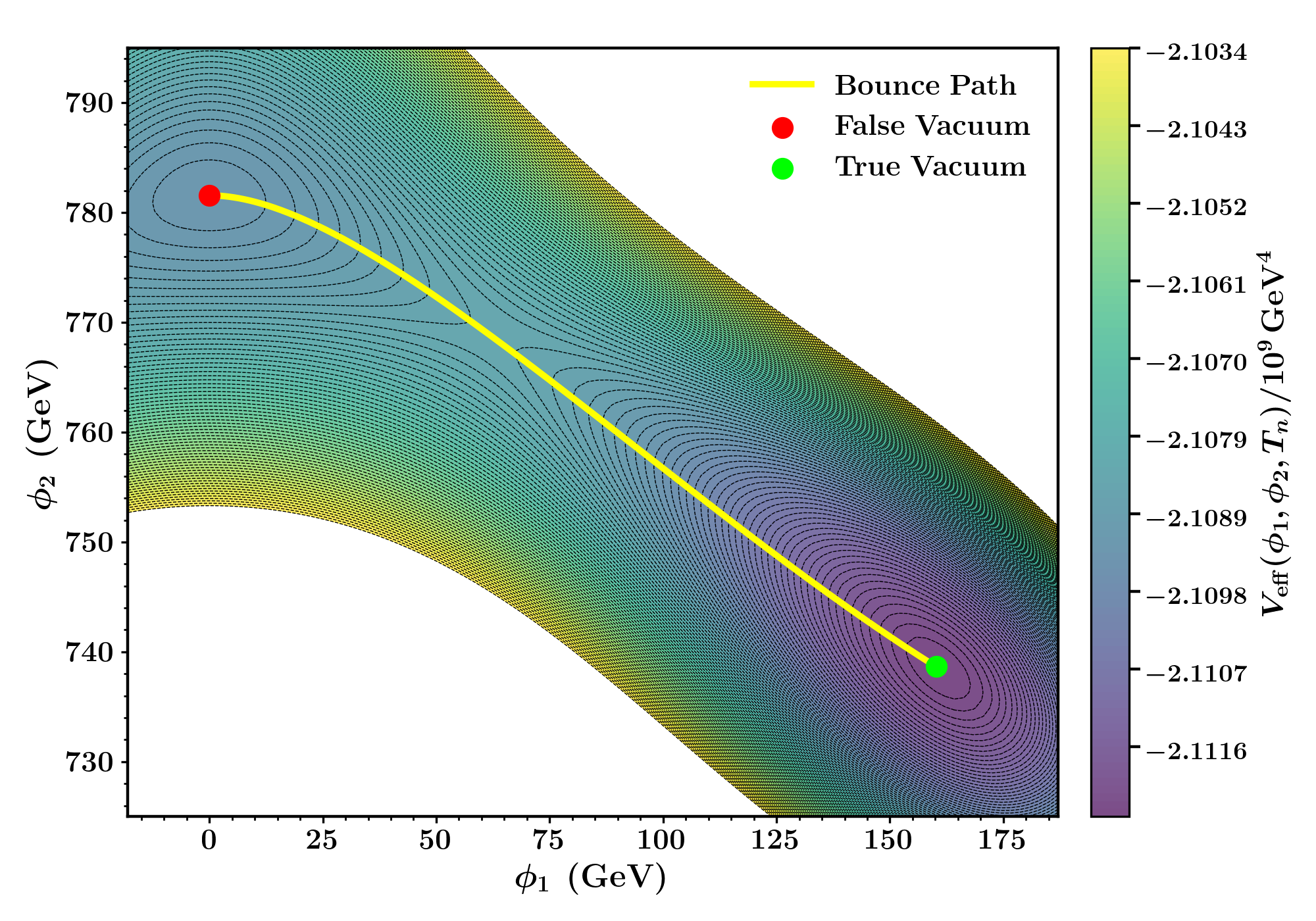}
		\caption{}
		\label{bounce}
	\end{subfigure}
	\hfill
\begin{subfigure}[b]{0.48\textwidth}
		\centering
		\includegraphics[width=\textwidth]{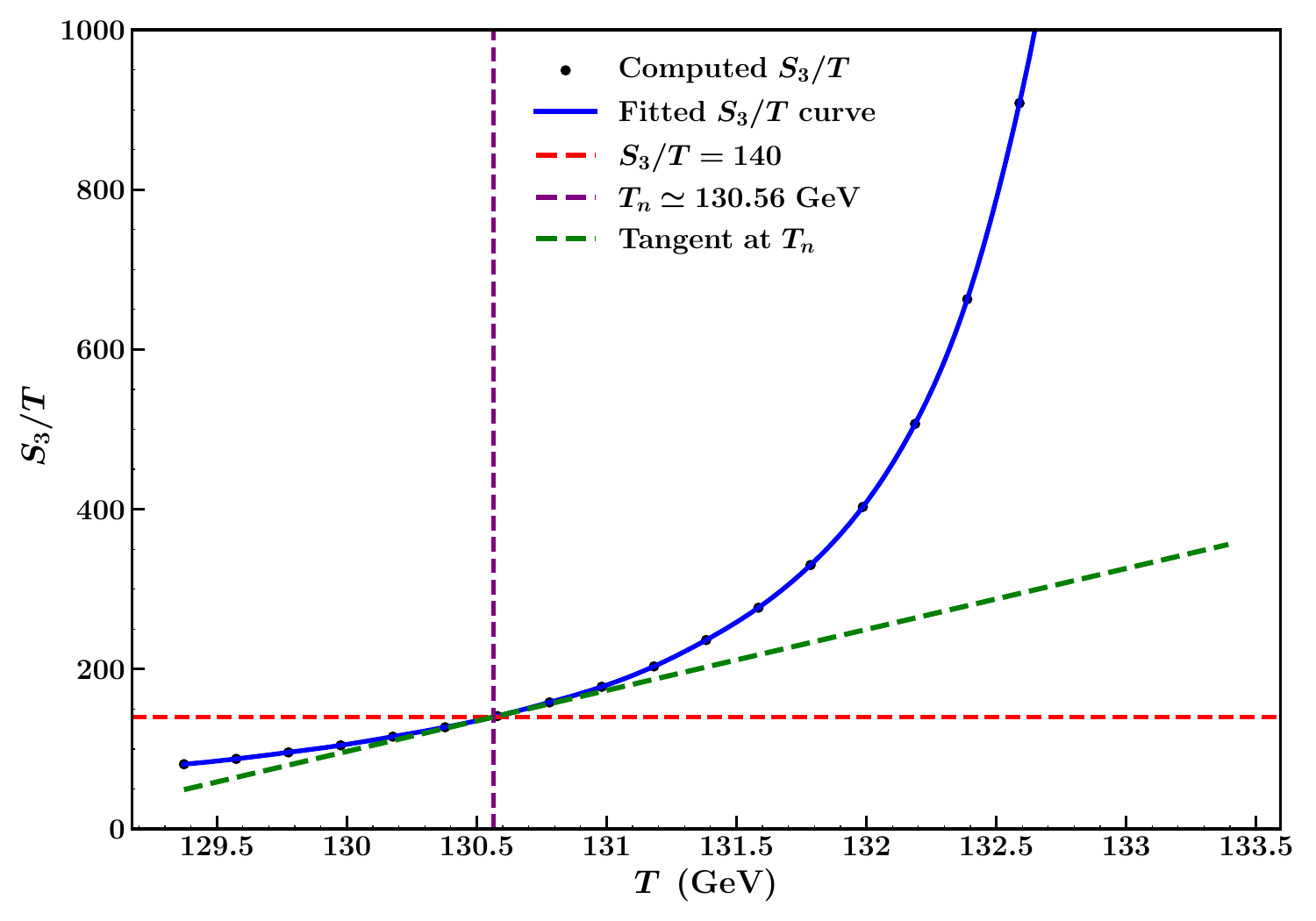}
	\caption{}
	\label{S3byT}
	\end{subfigure}
    \caption{(a) Bounce solution between false and true vacuum at the nucleation temperature. (b) Evolution of $S_3/T$ with $T$. The numerical values of $S_3/T$ computed using \texttt{CosmoTrasitions} are shown by black dotted points. The green dashed line representing the slope of the fitted curve (blue) at $T=T_n$ determines the inverse phase transition duration $\beta$ at $T = T_n$. Both plots are for $M_{h_2} = 250$ GeV, $\sin \theta =0.15 $, $w=725.7$ GeV, $M_{\gamma_d}= 60 $ GeV and $\Lambda = 800$ GeV.}
    \label{nuc}
\end{figure}

The specific field configuration $\left( \phi_1(r), \phi_2(r) \right)$ that minimizes this action is called the ``bounce'' solution. This is obtained by solving the Euler-Lagrange equation, which takes the form of a non-linear differential equation with a ``friction" term:
\begin{equation}
\frac{d^2\phi_i}{dr^2} + \frac{2}{r} \frac{d\phi_i}{dr} = \frac{\partial V_{eff}(\phi_1, \phi_2, T)}{\partial \phi_i}
\end{equation}
satisfying the boundary conditions:
\begin{equation}
\left. \frac{d\phi_i}{dr} \right|_{r=0} = 0 \ , \qquad
 \lim_{r \to \infty} \phi_i(r) = 0
\end{equation}
The nucleation temperature ($T_n$), is defined as the temperature where the probability of nucleating at least one bubble per Hubble volume is order unity:
\begin{equation}
\int_{T_n}^{T_c} \frac{dT}{T} \frac{\Gamma(T)}{H(T)^4} \sim 1
\qquad \Longrightarrow \qquad
\frac{S_3(T_n)}{T_n} \simeq \mathcal{O}(140),
\end{equation}
with $H(T)$ the Hubble expansion rate.  In models with two scalar fields ($\phi_1, \phi_2$), like the one considered here, the bounce is a trajectory in the 2-dimensional field space that extremizes the coupled equations i.e. the solution to the coupled Euclidean equations for both fields. However, in our scenario the dominant field excursion takes place along the Higgs direction, as has been shown in Figure~\ref{bounce}. 
The evolution of the $O(3)$-invariant Euclidean action ($S_3$) against temperature is displayed in Figure~\ref{S3byT}. In particular we numerically evaluate the $\frac{S_3}{T}$  using \texttt{CosmoTransitions} \cite{Wainwright:2011kj}\footnote{To determine the nucleation parameters, we compute the ratio $S_3/T$ on a descending temperature grid beginning at $T_c$ using \texttt{CosmoTransitions}. To mitigate numerical instabilities inherent in any bounce solver package, we reconstruct the discrete data using a smooth functional fit. The nucleation temperature $T_n$ is then identified by the condition $S_3(T_n)/T_n = 140$, and the inverse phase transition duration $\beta$ is extracted by evaluating the temperature derivative of the fitted curve at $T = T_n$.}.

\begin{figure}[h!]
	\centering
	\includegraphics[width = 0.4\textwidth]{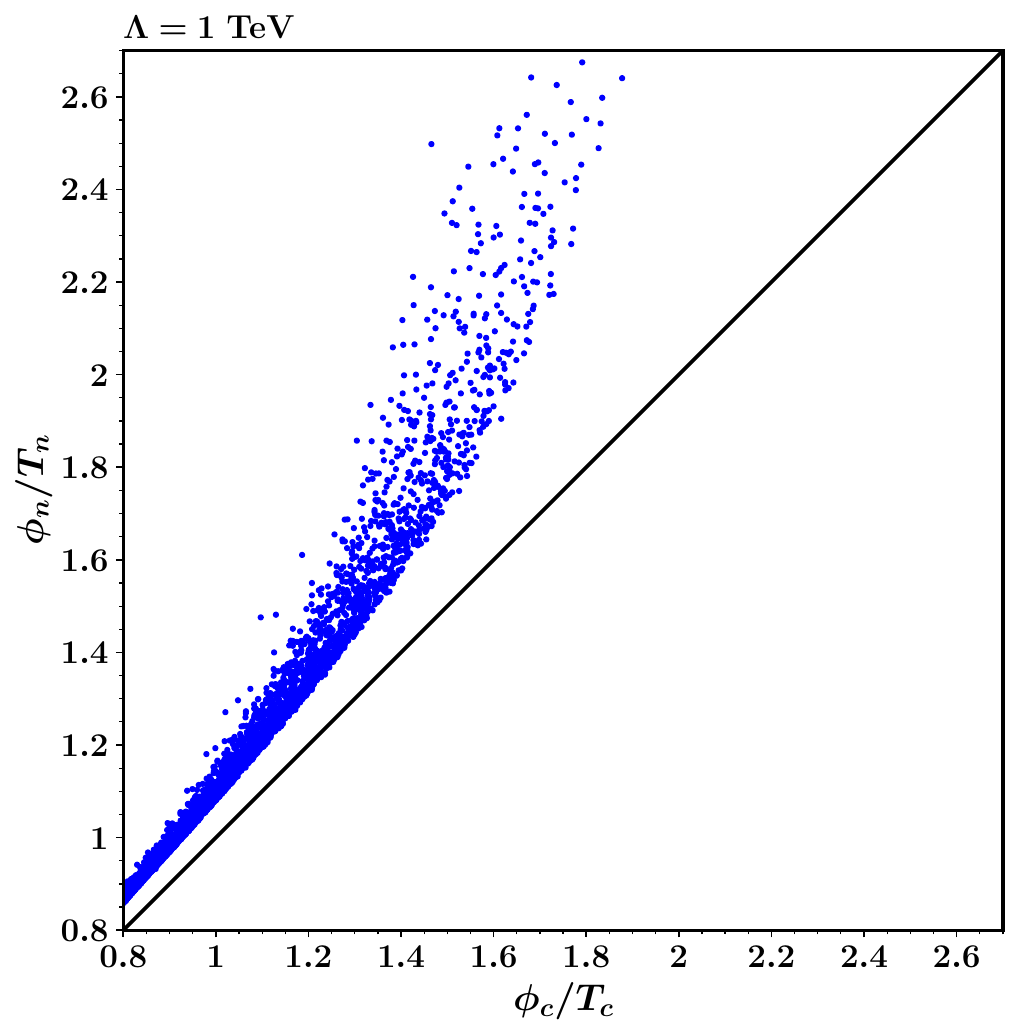}
	\caption{Possible $\frac{\phi_n}{T_n}$ values for a given $\frac{\phi_c}{T_c}$ for various $M_{h_2}, \sin\theta$ and $w$ values that predict SFOEWPT, assuming $\Lambda=1$ TeV ( see Figure~\ref{fig:scan1000}).}
	\label{phictc_phintn}
\end{figure}

In order to achieve a sustainable electroweak baryogenesis during bubble nucleation  one needs $\frac{\phi_c}{T_c}\gtrsim 0.6 - 1.4$,  as discussed in Section~\ref{phase_transition}. We have set $\frac{\phi_c}{T_c}\geq 0.8$, which in turn implies a larger value for  $\frac{\phi_n}{T_n}$ , where $\phi_n$ is the discontinuity of the order parameter $\phi_1$ at nucleation temperature. This is also illustrated in Figure~\ref{phictc_phintn} where we display $\frac{\phi_n}{T_n}$ outcomes for a given $\frac{\phi_c}{T_c}$ for various choices of $M_{h_2}, \sin\theta$ and $w$ that correspond to SFOEWPT.  While $\phi_c/T_c$ is often taken to be the standard measure of the transition strength, the growth of $\phi_n/T_n$ above the diagonal line illustrates that the nucleation typically starts only after significant supercooling.

Once nucleated, bubbles of the true vacuum expand due to the pressure difference between
the two phases and eventually collide and percolate, completing the phase transition.
The highly out-of-equilibrium dynamics associated with bubble expansion and collisions
source a stochastic background of gravitational waves.
The key parameters which determine the strength of the gravitational wave signal are the following:
\begin{equation}
\alpha \equiv \frac{\Delta \rho}{\rho_{\rm rad}},
\end{equation}
where $\Delta \rho$ is the released latent heat and $\rho_{\rm rad}$ is the radiation
energy density evaluated at $T_n$.
The characteristic time scale of the transition is described by the inverse duration
parameter
\begin{equation}
\frac{\beta}{H_n}
= T_n \left.\frac{d}{dT}\left(\frac{S_3}{T}\right)\right|_{T=T_n},
\end{equation}
where $H_n \equiv H(T_n)$.

\section{Gravitational wave}
\label{gravwav}

\begin{figure}[h!]
   \begin{subfigure}[b]{0.46\textwidth}
		\centering
		\includegraphics[width=\textwidth]{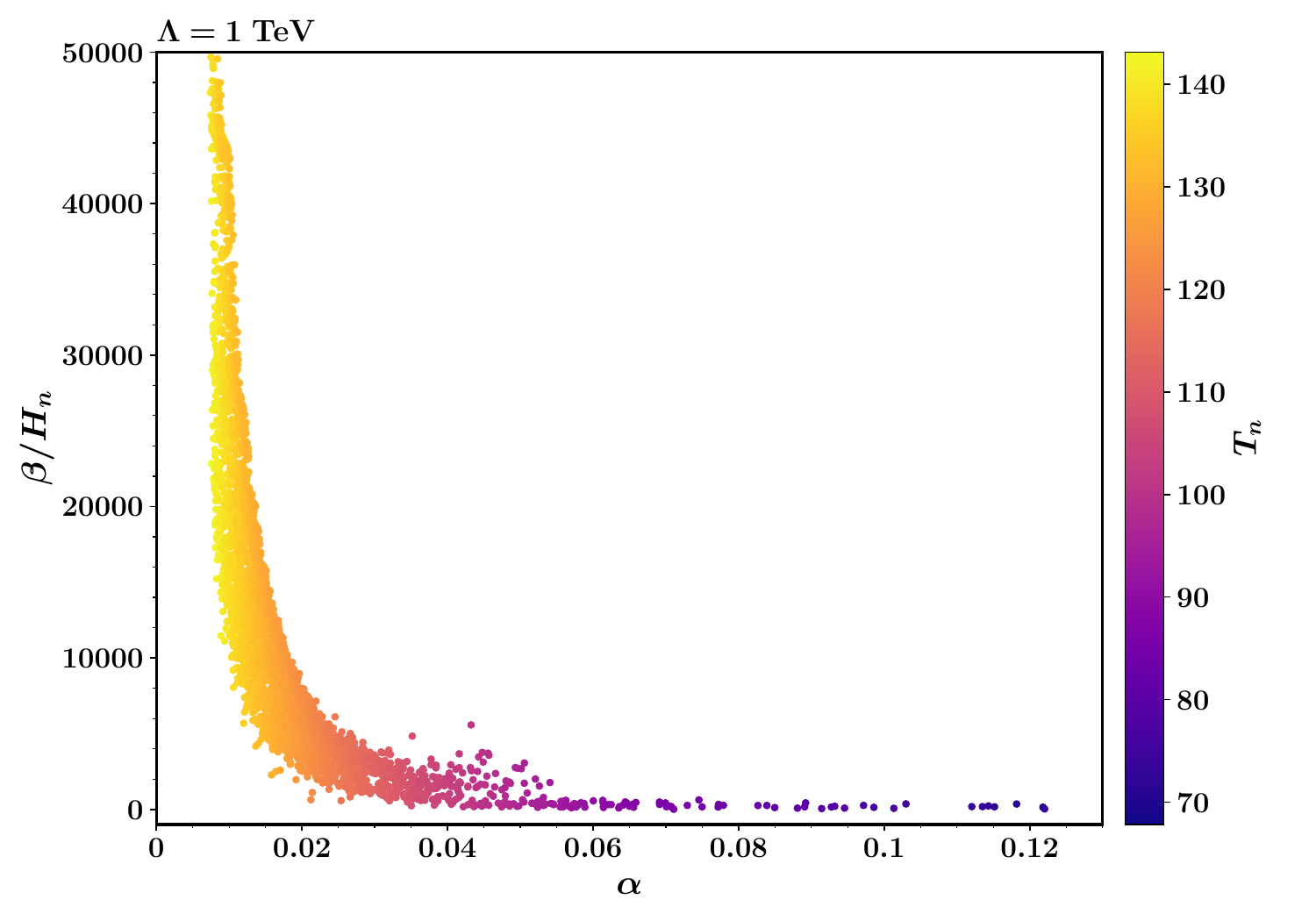}
		\caption{}
		\label{alphabeta}
	\end{subfigure}
	\hfill
\begin{subfigure}[b]{0.46\textwidth}
		\centering
		\includegraphics[width=\textwidth]{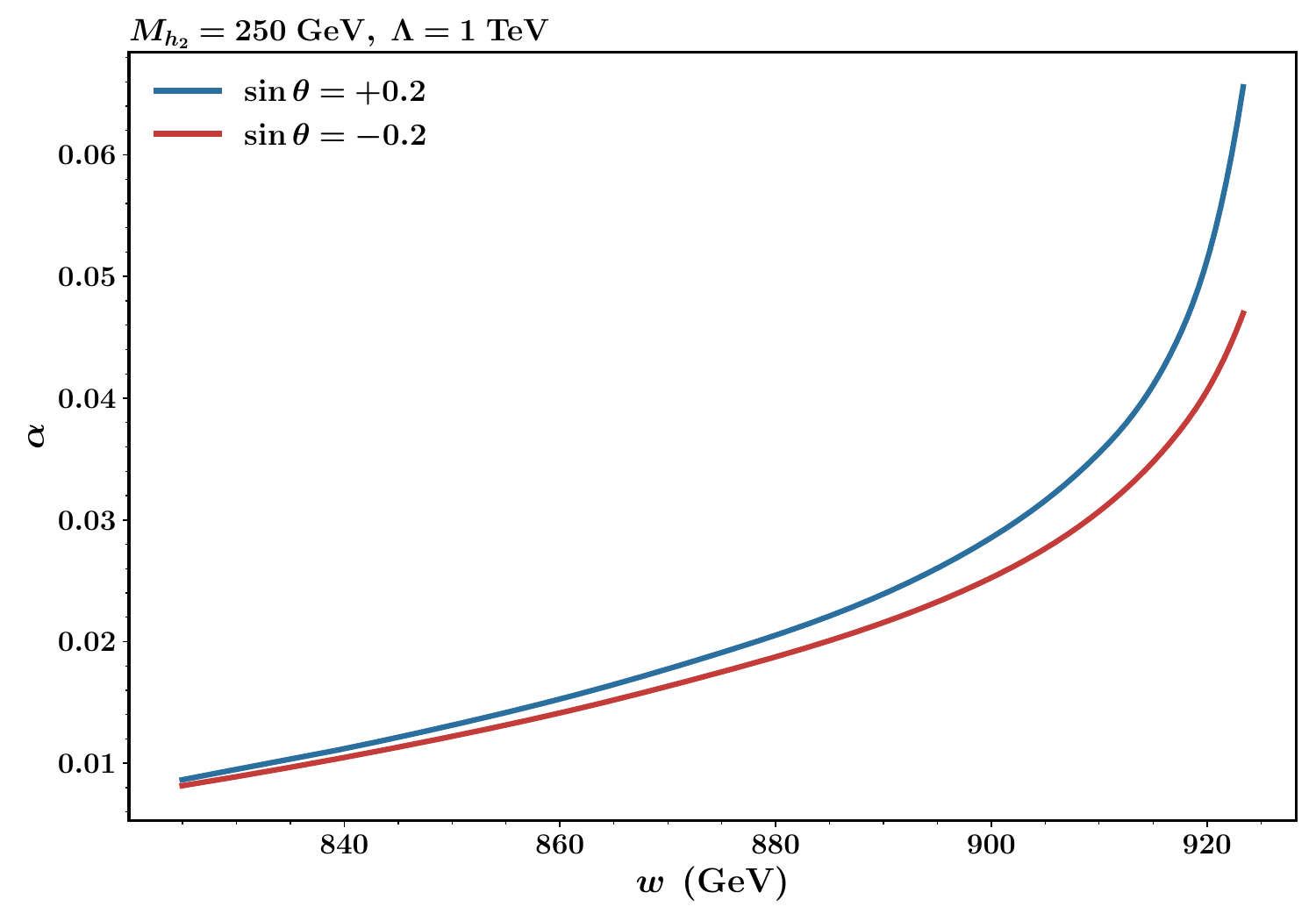}
		\caption{}
		\label{alpha}
	\end{subfigure}
	\caption{The variations of the various GW signal quantifiers. The left panel: $\alpha$ vs $\beta/H_n$ with color gradient as nucleation temperature $T_n$, for the SFOEWPT compatible parameter space points in Figure~\ref{fig:scan1000}. The right panel : variations of $\alpha$ is shown against the VEV of the singlet-like scalar for an illustrative value $M_{h_2} = 250$ GeV. The blue (red) curve corresponds to the scalar mixing angle $\sin \theta = 0.2~(-0.2)$. Both of these assume $\Lambda =1$ TeV. }
\end{figure}

The gravitational wave provides a robust probe of first-order electroweak phase transitions in the early Universe, with promising prospects for
detection at future space-based interferometers. The strength of the GW signals are quantified by
$\alpha$, $\beta/H_n$, and the bubble wall velocity.
A large $\alpha$ and a smaller $\beta$ would typically result in a detectable signal over the background. From Figure~\ref{alphabeta}, we can see
that large $\alpha$ and small $\beta/H_n$ are only possible for supercooled transitions that are characterized by small nucleation temperatures. The increase in 
$\alpha$ with the increase in the VEV of the singlet scalar ($w$) for a fixed cut-off scale $\Lambda$ can be inferred from Figure~\ref{alpha}. There is a mild variation of this dependency due to the {\it sign} of the scalar mixing angle $\sin\theta$.

The resulting stochastic background is conventionally characterized by the present-day
energy density spectrum,
\begin{equation}
\Omega_{\rm GW}(f) \equiv
\frac{1}{\rho_c}\frac{d\rho_{_{\rm GW}}}{d\ln f},
\end{equation}

The stochastic gravitational wave (GW) generated during a first-order phase transition receives contributions primarily from three sources: collisions of the bubble wall, sound waves in the surrounding plasma, and magneto-hydrodynamic (MHD) turbulence. Most of the energy released during the transition is pumped into the surrounding plasma as heat and kinetic energy. Numerical simulations indicate that bubble collisions primarily create fluid perturbations in the form of sound waves which act as a persistent and potent source of GWs until the expansion of the Universe (Hubble expansion) eventually dampens them. The total present-day GW energy density spectrum, in terms of the dimensionless Hubble parameter $h = \frac{H_0}{100}~\rm{km}~ \rm{s}^{-1}~\rm{Mpc}^{-1}$ (where $H_0$ is the current Hubble constant) \cite{DES:2017txv}, can therefore be written as~\cite{Caprini:2015zlo}
\begin{equation}\label{eq:GWTotal}
\Omega_{\rm GW} h^2
\simeq
\Omega_{\rm col} h^2
+
\Omega_{\rm sw} h^2
+
\Omega_{\rm tur} h^2,
\end{equation}
 As a result, the shape and amplitude of the GW spectrum provide a direct observational window into the particle-physics dynamics underlying the phase transition.

 The energy-momentum tensor, dominated by scalar fields, acts as the fundamental source for gravitational waves. The corresponding energy density spectrum as a function of frequency $f$ can be estimated when the bubble walls expand pushing the surrounding plasma and eventually collide at near-relativistic speeds and is given by~\cite{Jinno:2016vai}
\begin{equation}\label{eq:GWcoldetails}
\Omega_{\rm col} h^2 = 1.67 \times 10^{-5}
\left(\frac{\beta}{H_n}\right)^{-2}
\left( \frac{\kappa_c \alpha_n}{1 + \alpha_n} \right)^2
\left( \frac{100}{g_{\ast}} \right)^{1/3}
\left( \frac{0.11 v_w^3}{0.42 + v_w^2} \right)
\frac{3.8 \left( f/f_{\rm col} \right)^{2.8}}{1 + 2.8 \left( f/f_{\rm col} \right)^{3.8}},
\end{equation}
where $v_w$ indicates the velocity of the bubble wall, $\alpha_n$ signifies the strength of the phase transition at the nucleation temperature, and $\kappa_c$ is a factor quantifying the conversion of the vacuum energy into the kinetic energy of the expanding bubble walls. This efficiency factor can be parametrized as \cite{Kamionkowski:1993fg}
\begin{equation}\label{eq:kcfac}
\kappa_c =
\frac{0.715\,\alpha_n + \frac{4}{27}\sqrt{\frac{3\alpha_n}{2}}}
{1 + 0.715\,\alpha_n}.
\end{equation}

For the collision term, the frequency of the resulting GW spectrum peaks at 
\begin{equation}\label{eq:PF1}
f_{\rm col} = 16.5 \times 10^{-6} \left( \frac{f_{\ast}}{\beta} \right) \left( \frac{\beta}{H_n} \right) \left( \frac{T_n}{100 \, {\rm GeV}} \right) \left( \frac{g_{\ast}}{100} \right)^{1/6} \, {\rm Hz},
\end{equation}
where $f_\ast/\beta$ is a fitted function and is given by~\cite{Jinno:2016vai}
\begin{equation}\label{eq:fastbybetadetails}
\frac{f_{\ast}}{\beta} = \frac{0.62}{1.8 - 0.1 v_w + v^2_w}.
\end{equation}

Assuming the bubbles reach relativistic speeds at the time of collision, we consider $v_w = 1$~\cite{Kamionkowski:1993fg,Espinosa:2010hh} in the subsequent discussion. 

A continuous, low-frequency ``hum" typically arises from the sound waves generated in the plasma during the expansion of the bubble after nucleation~\cite{Hindmarsh:2013xza,Hindmarsh:2016lnk,Hindmarsh:2017gnf}. The corresponding GW energy density spectrum is
\begin{equation}\label{eq:GWswpart}
\Omega_{\rm sw} h^2 = 2.65 \times 10^{-6}\; \Upsilon(\tau_{\rm sw}) \left(  \frac{\beta}{H_n} \right)^{-1} v_w \left( \frac{\kappa_{\rm sw} \alpha_n}{1 + \alpha_n} \right)^2 \left( \frac{100}{g_{\ast}} \right)^{1/3} \left( \frac{f}{f_{\rm sw}} \right)^3 \left[ \frac{7}{4 + 3 \left( f/f_{\rm sw} \right)^2} \right]^{7/2},
\end{equation}
The efficiency of converting the kinetic energy of the plasma (sound waves) into gravitational waves is characterized by a dimensionless parameter, $\kappa_{\rm sw}$ ~\cite{Kamionkowski:1993fg,Espinosa:2010hh}:
\begin{equation}\label{eq:kappasw}
\kappa_{\rm sw} =\frac{\sqrt{\alpha_n}}{0.135+\sqrt{0.98+\alpha_n}}.
\end{equation}
The factor $\Upsilon(\tau_{\rm sw})$ accounts for the finite lifetime of the sound wave period and is defined as
\begin{equation}\label{eq:swtimepart}
\Upsilon(\tau_{\rm sw}) = 1 - \frac{1}{\sqrt{1+2 \tau_{\rm sw} H_{\ast}}},
\end{equation}
where $\tau_{\rm sw}$ denotes the sound wave lifetime. Following Ref.~\cite{Hindmarsh:2017gnf}, we approximate $\tau_{\rm sw} \approx R_n/\overline{U}_f$, with the mean bubble separation $R_n = (8\pi)^{1/3} v_w \beta_n^{-1}$ and the root-mean-squared fluid velocity $\overline{U}_f = \sqrt{3\kappa_{\rm sw}\alpha_n/4}$.
The redshifted peak frequency associated with the sound wave contribution is
\begin{equation}\label{eq:PF2}
f_{\rm sw} = 1.9 \times 10^{-5} \left( \frac{1}{v_w} \right) \left( \frac{\beta}{H_n} \right) \left( \frac{T_n}{100 \, {\rm GeV}} \right) \left( \frac{g_{\ast}}{100} \right)^{1/6} \, {\rm Hz}.
\end{equation}

Finally, the strong magnetic fields coupled to the ionized plasma creates turbulent motions during the bubble expansion, often called magnetohydrodynamic (MHD) turbulence, and can be another potential source for GWs~\cite{Caprini:2009yp}. The contribution to the GW energy density spectrum from such MHD turbulence is:
\begin{equation}\label{eq:GWturpart}
\Omega_{\rm tur} h^2 = 3.35 \times 10^{-4} \left( \frac{\beta}{H_n} \right)^{-1} v_w \left( \frac{\kappa_{\rm tur} \alpha_n}{1 + \alpha_n} \right)^{3/2} \left( \frac{100}{g_{\ast}}\right)^{1/3} \left[ \frac{\left( f/f_{\rm tur} \right)^3}{\left[ 1 + \left( f/f_{\rm tur} \right) \right]^{11/3} \left( 1 + \frac{8 \pi f}{h_{\ast}} \right)} \right],
\end{equation}
where $h_{\ast} = 16.5\times 10^{-6} \left( \frac{T_n}{100 \, {\rm GeV}} \right) \left( \frac{g_{\ast}}{100} \right)^{1/6} \, {\rm Hz}$ is the inverse Hubble time. The peak frequency of the MHD turbulence term appears as
\begin{equation}\label{eq:PF3}
f_{\rm tur} = 2.7 \times 10^{-5} \frac{1}{v_w} \left( \frac{\beta}{H_n} \right) \left( \frac{T_n}{100 \, {\rm GeV}} \right) \left( \frac{g_{\ast}}{100} \right)^{1/6} \, {\rm Hz}.
\end{equation}
The turbulence efficiency factor is taken to be $\kappa_{\rm tur} = \epsilon \kappa_{\rm sw}$, where $\epsilon$ parameterizes the fraction of bulk kinetic energy converted into turbulent motion. Following previous studies~\cite{Zhou:2022mlz}, we adopt $\kappa_{\rm tur} \simeq 0.1\,\kappa_{\rm sw}$ in our numerical analysis.
\begin{figure}[h!]
	\centering
	\includegraphics[width = 0.55\textwidth, height=6cm]{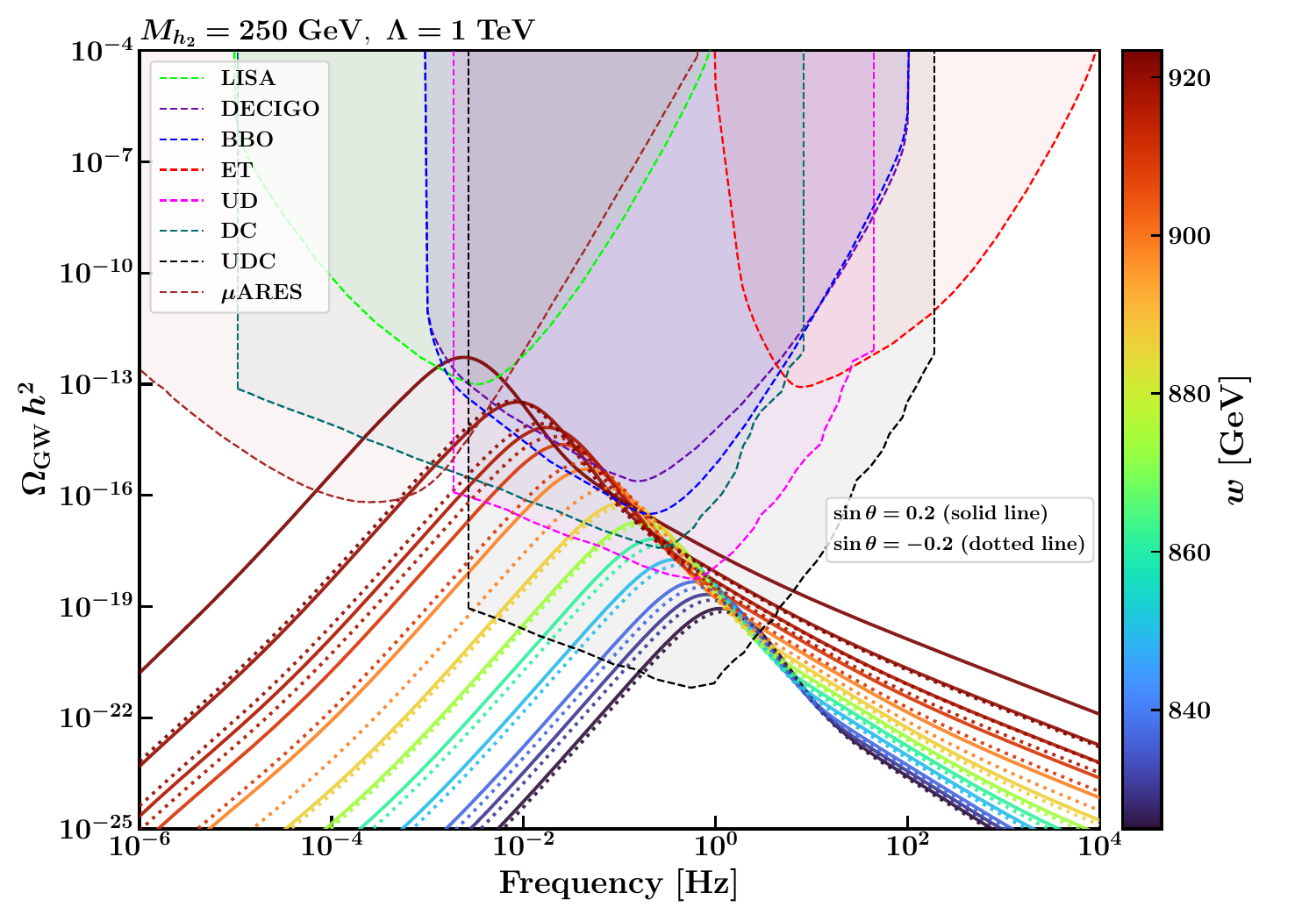}
	\caption{Variations of the gravitational wave energy densities with the frequency for different values of $w$ for $M_{h_2}= 250$ GeV and $\Lambda= 1$ TeV. The light shaded regions represent the expected sensitivity reach of the next generation space-based and ground-based detectors like, LISA~\cite{LISA:2017pwj}, BBO~\cite{Yagi:2011wg}, DECIGO~\cite{Nakayama:2009ce}, Einstein Telescope (ET)~\cite{Punturo:2010zz}, $\mu$Ares~\cite{Sesana:2019vho},  ultimate-DECIGO (UD), DECIGO-corr (DC) and ultimate-DECIGO-corr (UDC)~\cite{Nakayama:2009ce}. The solid and dotted lines are for two distinct values of $\sin \theta = 0.2~\rm{and}~-0.2$, respectively.}
	\label{gw_vs_w}
\end{figure}

To illustrate the gravitational wave signals inherent to this model, we have estimated the spectral density of stochastic GW generated for $M_{h_2} = 250$ GeV and a fixed new physics scale $\Lambda = 1$ TeV with the singlet VEV $w$ varying. The numerical results for some discrete values of $w$ are synthesized in Figure~\ref{gw_vs_w}, which displays the spectral density across various frequency ranges for two values of $\sin \theta = \pm 0.2$. We have considered the standard ``runaway" scenario in which bubbles of a new vacuum state expand with nearly unrestrained speed in the absence of friction of the plasmic medium. The peak frequencies of the cumulative contributions of collision, sound, and MHD terms are within reach of the sensitivities of LISA~\cite{LISA:2017pwj}, BBO~\cite{Yagi:2011wg}, DECIGO~\cite{Nakayama:2009ce}, ultimate-DECIGO (UD), DECIGO-corr (DC) and ultimate-DECIGO-corr (UDC)~\cite{Nakayama:2009ce} as can be seen in Figure~\ref{gw_vs_w}. A few general comments on the patterns of the generated GW spectra for this scenario are as follows. The maximum energy density shifts towards a lower frequency and higher amplitude for larger values of the singlet VEV ($w$), which means greater chance of observing a GW signal over background noise. The pattern can be understood from Figure~\ref{alpha} which shows the strength of the transition increases with the increase of the singlet VEV. Another observation would be the peak amplitude is marginally less for the negative values of the $\sin\theta$ compared to its positive counterpart.

\begin{figure}[h!]
   \begin{subfigure}[b]{0.48\textwidth}
		\centering
	\includegraphics[width = \textwidth]{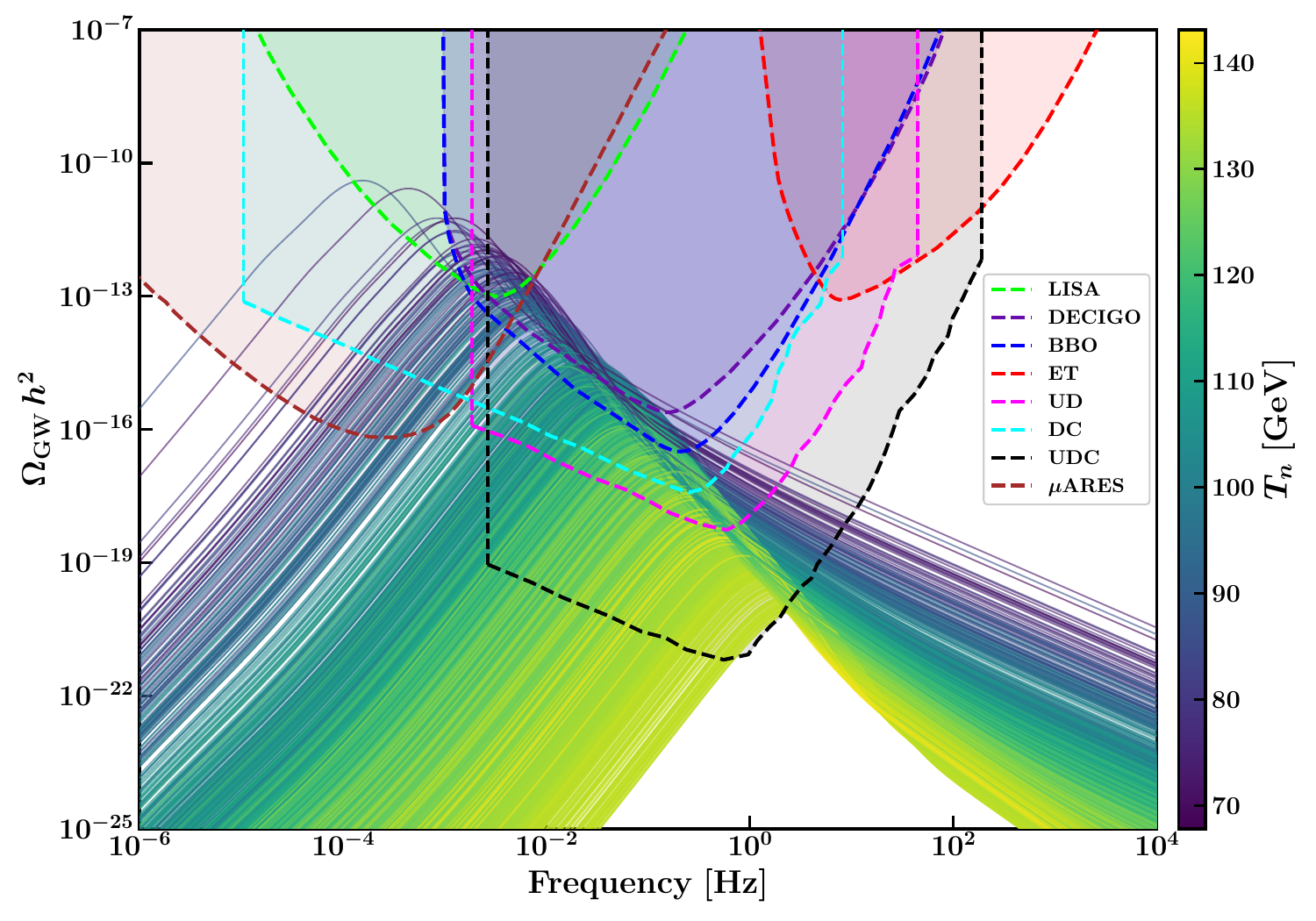}
	\caption{\footnotesize{}\normalsize}
	\label{fig25}
	\end{subfigure}
	\hfill
\begin{subfigure}[b]{0.48\textwidth}
		\centering
		\includegraphics[width=\textwidth]{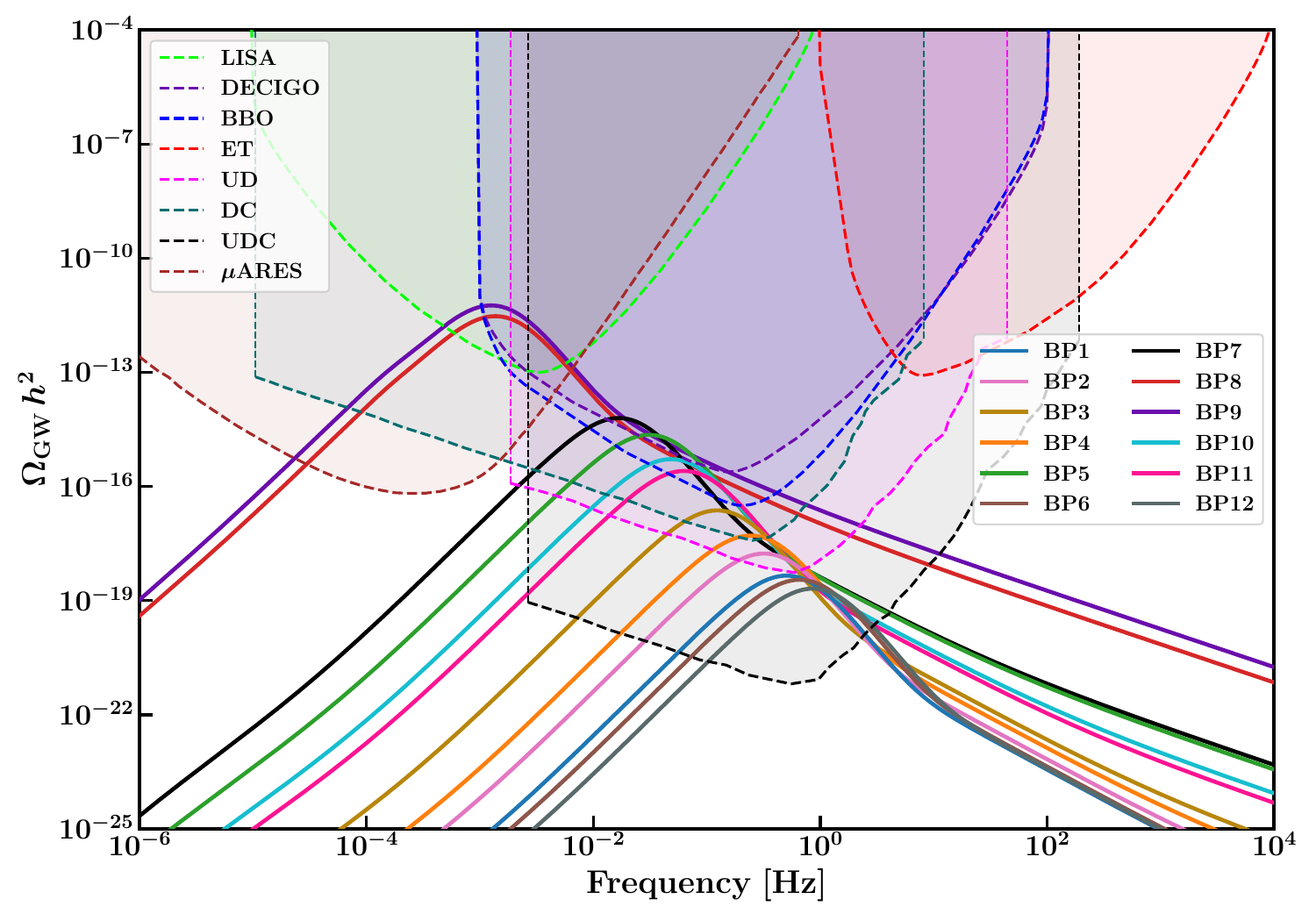}
		\caption{}
		\label{bp}
	\end{subfigure}
    
	\caption{The detectable GW signal coming from the cumulative contributions of the SW, bubble collision and MHD turbulence, as a function of frequency. (a) The left panel shows the GW signals for all those SFOEWPT compatible parameter space points in the $M_{h_2}-\sin\theta$ plane ( see Figure~\ref{fig:scan1000}) have been plotted with color gradient as nucleation temperature. (b) The right panel displays the same but for certain selected benchmark points tabulated in Table~\ref{tab:benchmark_points}. A fixed cut-off scale $\Lambda=1$ TeV has been considered for both of these figures.}
    \label{bp_1}
	
\end{figure}
In Figure~\ref{fig25}, we display the total GW signal coming from all three contributing sources generated during phase transitions, for all those parameter space points in the $M_{h_2}-\sin \theta$ plane satisfying SFOEWPT (see Figure~\ref{fig:scan1000}). The color gradient indicates the nucleation temperatures. The highest peak corresponds to the lowest nucleation temperature and is in accordance with the supercooled phase of the Universe, signifying a larger ($T_c - T_n$), which in turn increases $\alpha$, the strength of the transition.

\begin{table}[h!]
\small
\centering
\renewcommand{\arraystretch}{1.3}
\begin{tabular}{c c c c c c c c c c}
\hline\hline
BP &
$M_{h_2}$ [GeV] &
$\sin\theta$ &
$w$ [GeV] &
$T_c$ [GeV] &
$\phi_c/T_c$ &
$T_n$ [GeV] &
$\phi_n/T_n$ &
$\alpha$ &
$\beta/H$
\\
\hline
BP1  & 170 &  0.20     & 655.0 & 143.1 & 0.84 & 141.3 & 0.93 & 0.009 & 18771 \\
BP2  & 180 &  0.17     & 708.3 & 140.5 & 0.92 & 137.9 & 1.04 & 0.011 & 12211 \\
BP3  & 200 & $-0.19$   & 778.7 & 136.0 & 1.06 & 130.8 & 1.24 & 0.015 & 4976  \\
BP4  & 220 &  0.18     & 803.9 & 135.3 & 1.07 & 132.3 & 1.19 & 0.014 & 9668  \\
BP5  & 240 &  0.20     & 890.2 & 120.1 & 1.56 & 100.1 & 2.13 & 0.045 & 1635  \\
BP6  & 240 & $-0.20$   & 816.4 & 138.2 & 0.91 & 136.8 & 0.98 & 0.010 & 25325 \\
BP7  & 250 & $-0.19$   & 928.4 & 122.8 & 1.47 & 104.8 & 1.98 & 0.040 & 843   \\
BP8  & 250 & $-0.23$   & 907.2 & 116.7 & 1.66 &  76.7 & 2.90 & 0.103 & 94.1  \\
BP9  & 260 &  0.20     & 956.0 & 113.8 & 1.75 &  67.8 & 3.34 & 0.145 & 97.8  \\
BP10 & 280 &  0.20     & 975.4 & 121.4 & 1.50 & 112.0 & 1.80 & 0.031 & 2279  \\
BP11 & 280 & $-0.20$   & 975.4 & 124.2 & 1.41 & 116.1 & 1.66 & 0.027 & 2853  \\
BP12 & 300 &  0.20     & 926.2 & 135.9 & 0.93 & 134.8 & 1.00 & 0.010 & 33558 \\
\hline\hline
\end{tabular}

\caption{The predicted values of $T_c, \frac{\phi_c}{T_c}, T_n, \frac{\phi_n}{T_n}, \alpha,$ and $\frac{\beta}{H}$ for different representative benchmark points assuming $\Lambda = 1~\mathrm{TeV}$.}
\label{tab:benchmark_points}
\end{table}
\noindent
The GW signals for a set of twelve selected BPs tabulated in Table~\ref{tab:benchmark_points} are also depicted in Figure~\ref{bp}. 
The signals comprise all three said gravitating sources of that epoch. The peak frequencies vary by three orders of magnitudes. It should be noted that a wide range of interferometer based experiments can, in principle, detect such stochastic GW signals. In particular, LISA ~\cite{LISA:2017pwj}, the most imminent among all, should be able to see the gravitational waves predicted in this model for certain BPs in the mid-frequency region. Even energy density peaks as small as $\sim 10^{-19}$ can be probed in $\mu$Hz range interferometer experiments, like $\mu$ARES \cite{Sesana:2019vho}.

\section{Multi-scalar production at the LHC}
\label{multisc}
As mentioned, one of the profound implications of the additional scalar field and dimension-six operator involving Higgs and singlet scalar field described in Section~\ref{model}, is the modifications of the scalar potential, in particular, the latter one modifies the Higgs portal coupling and various triple and quartic scalar couplings (see Appendix~\ref{appendix1}, and~\ref{trip}).
 The detection of stochastic gravitation wave signals in any future experiments is definitely a novel way to gain knowledge on the structure of the scalar potential and its evolution in the early Universe. However, one can also explore the deviation of various trilinear and quartic scalar couplings which in turn  can probe the phase transition dynamics, by investigating multi-scalar production at the LHC. More specifically, we look for di-scalar bosons ($h_1h_1$, $h_1 h_2$ and $h_2 h_2$) and triple-Higgs ($h_1 h_1 h_1$) production at the LHC. For the numerical estimation of multi-scalar production cross-sections, we implement the Lagrangian of the model described in Section~\ref{model} in \texttt{FeynRules v2.3}~\cite{Alloul:2013bka, Christensen:2008py} coupled with NLOCT~\cite{Degrande:2014vpa} terms generated using \texttt{Feynarts v3.12}~\cite{Hahn:2000kx} in order to handle the one-loop diagrams and counter terms. The Universal Feynrule Output (\texttt{UFO})~\cite{Degrande:2011ua, deAquino:2011ub} of the Feynrules is then interfaced with \texorpdfstring{\texttt{MadGraph5\_aMC@NLO}}{MadGraph5_aMC@NLO}~\cite{Alwall:2014hca} to generate multi-scalar production in $p p$ collision at one-loop level. The renormalization and the factoraization scales have been set at the default dynamical scale of \texorpdfstring{\texttt{MadGraph5\_aMC@NLO}}{MadGraph5_aMC@NLO}.

\subsection{Di-scalar bosons production}
\begin{figure}[H]
	\centering
	\begin{subfigure}[b]{0.24\textwidth}
		\centering
		\includegraphics[width = \textwidth]{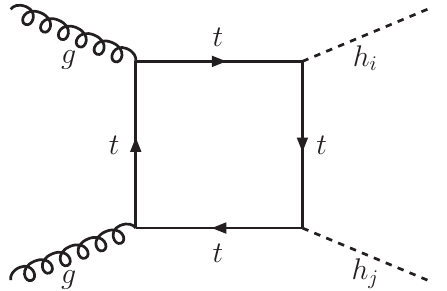}
		\caption{}
		\label{feyn1}
	\end{subfigure}
	\hfill
    \begin{subfigure}[b]{0.24\textwidth}
		\centering
		\includegraphics[width = \textwidth]{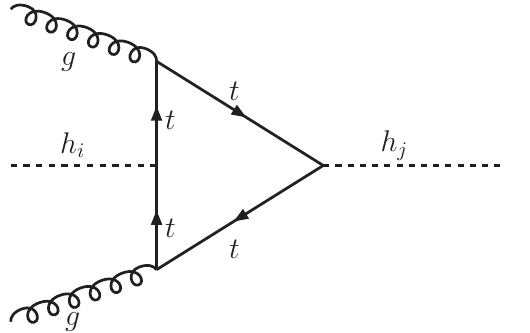}
		\caption{}
		\label{feyn2}
	\end{subfigure}
    \hfill
	\begin{subfigure}[b]{0.24\textwidth}
		\centering
		\includegraphics[width= \textwidth]{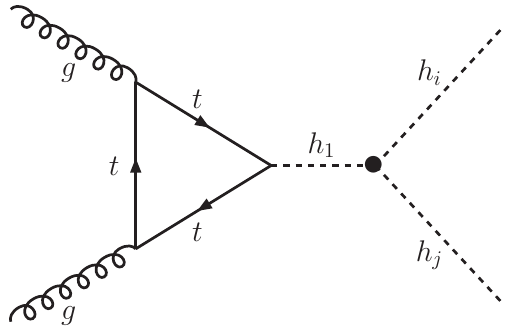}
		\caption{}
		\label{feyn3}
	\end{subfigure}
    \hfill
    	\begin{subfigure}[b]{0.24\textwidth}
		\centering
		\includegraphics[width= \textwidth]{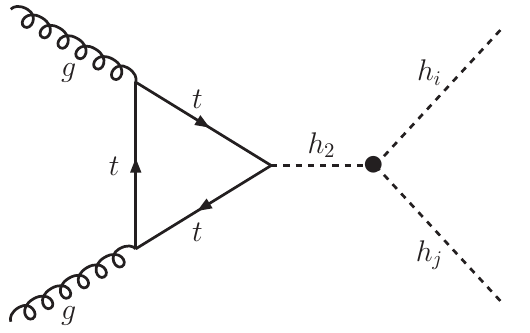}
		\caption{}
		\label{feyn4}
	\end{subfigure}
	\caption{Representative Feynman diagrams for $pp\to h_i h_j$ process. Here, the black dot in the above diagrams represent various modified triple scalar couplings in presence of the effective operator. }
    \label{hihj}
\end{figure}

Di-scalar bosons pair production ($h_i h_j$ with $i,j= 1,2$) at the LHC proceeds primarily via gluon-gluon fusion (ggF) process due to large flux of the initial state gluons in $pp$ collisions at the LHC center of mass energies. The corresponding Feynman diagrams for this process include the box and triangle diagrams, as shown in Figure~\ref{hihj}. Although, the presence of the dimension-six operator does not introduce to any new Feynman diagrams, its effects enter through modifications of the $h_i-h_j-h_k$ vertex factors (Appendix~\ref{trip}). The different Feynman diagrams in Figure~\ref{hihj} can interfere with each other in a constructive or destructive manner depending on the overall sign of these triple scalar couplings ($\lambda_{h_i h_j h_k}$). Moreover, these couplings are also sensitive to the ratio $\frac{w}{\Lambda}$, as can be seen from equation~\ref{lambda211}. In Figure~\ref{lambda211_w}, (of Appendix~\ref{trip}) we present the variation of $\lambda_{h_2 h_1 h_1}$ coupling as a function of the singlet scalar VEV, which is one of the crucial parameters in the context of FOEWPT as demonstrated in this manuscript. This plot also suggests that $\lambda_{h_2 h_1 h_1}$ has a zero  at certain values of $w$, depending on $M_{h_2}$ and $\sin\theta$. Therefore, to illustrate the intricate role that $w$ plays in the context of di-Higgs cross-sections from collider perspective, we work with three different choices of $w$, namely $w_1, w_2$ and $w_3$, with $w_1 > w_2 > w_3$. These three choices can be motivated as follows. The first choice, $w_1$ leads to SFOEWPT (as can be seen in Table~\ref{tab:ptpoints}), while the choice, i.e., $w_2$ leads to $\lambda_{h_2 h_1 h_1}=0$. The third choice $w_3$, corresponds to comparable value of $\lambda_{h_2 h_1 h_1}$ coupling as that predicted by $w_1$, but with opposite {\it sign}. 

\begin{figure}[h!]
   \begin{subfigure}[b]{0.48\textwidth}
		\centering
	\includegraphics[width = \textwidth]{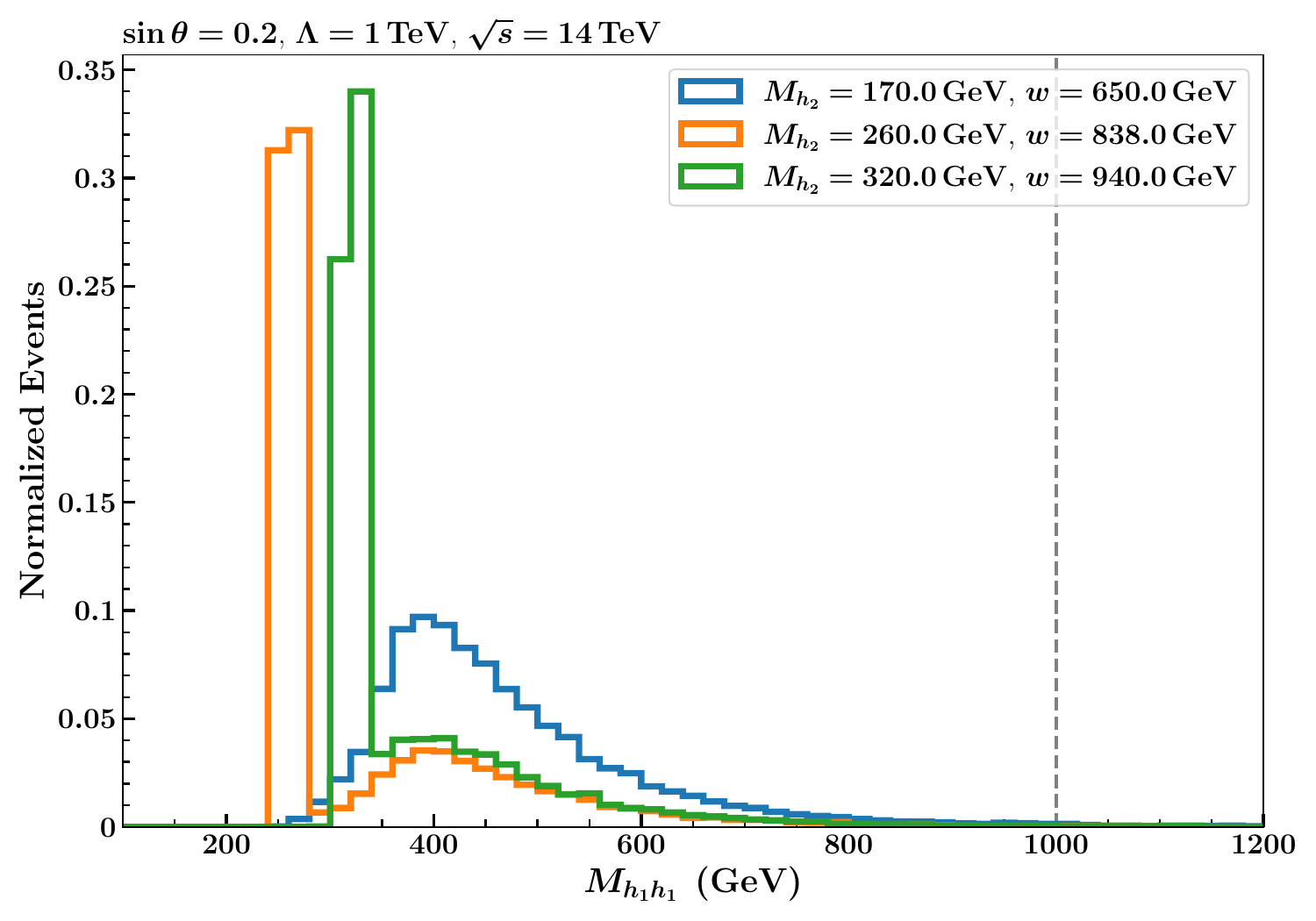}
	\caption{}
	\label{mh1h1_highw}
	\end{subfigure}
	\hfill
\begin{subfigure}[b]{0.48\textwidth}
		\centering
		\includegraphics[width=\textwidth]{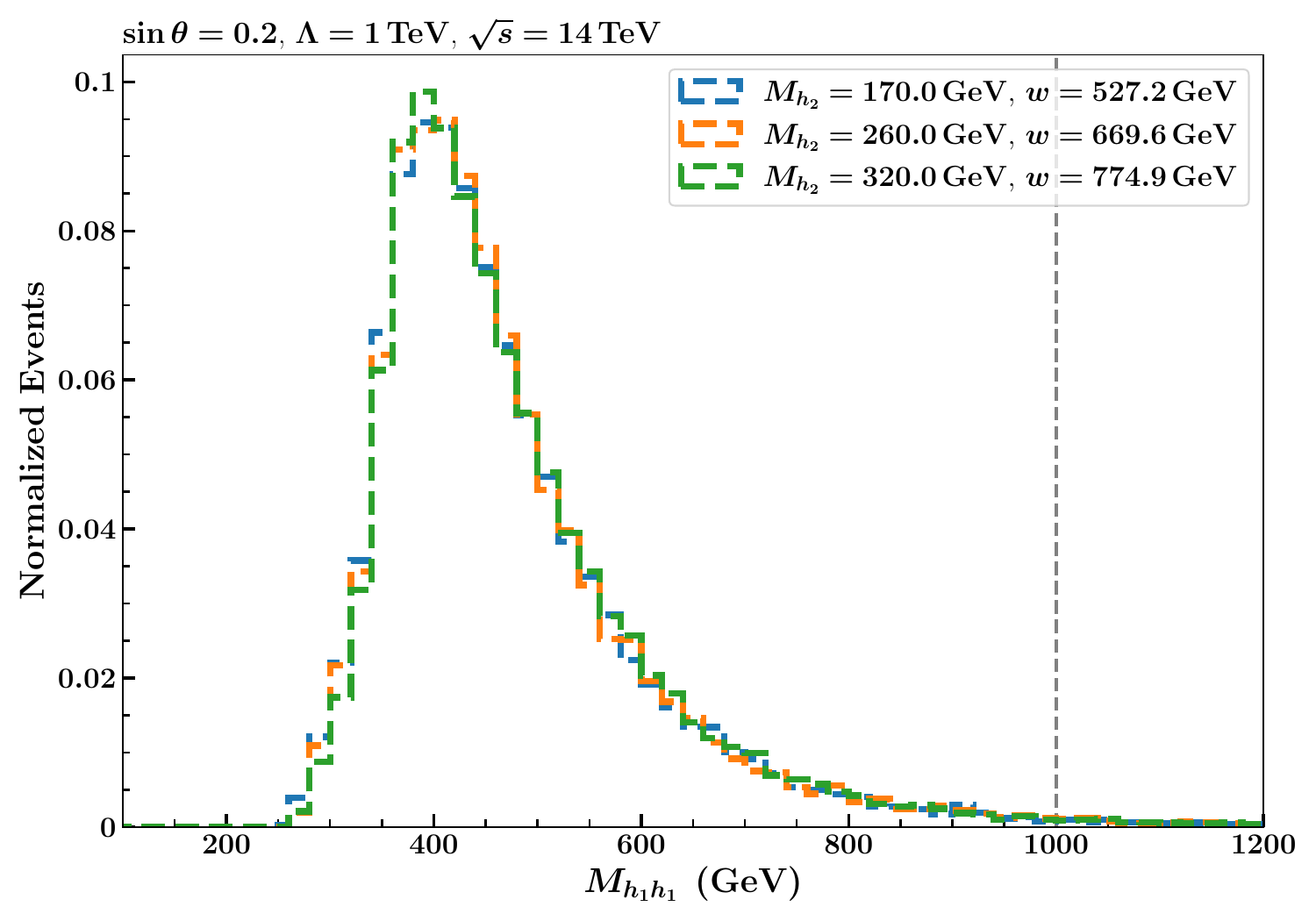}
		\caption{}
		\label{mh1h1_low_w}
	\end{subfigure}
    \caption{(a) Normalized di-Higgs invariant mass  ($M_{h_1h_1}$) distributions obtained for $pp \to h_1 h_1$ process at $\sqrt{s}=14~\text{TeV}$. (b) Represents the same but with different set  of singlet scalar VEV that leads to vanishing $\lambda_{h_2 h_1 h_1}$ coupling. The gray dashed vertical line denotes the cutoff scale $\Lambda = 1~\text{TeV}$.}
	\label{mh1h1}
\end{figure}

\begin{figure}[h!]
   \begin{subfigure}[b]{0.48\textwidth}
		\centering
	\includegraphics[width = \textwidth]{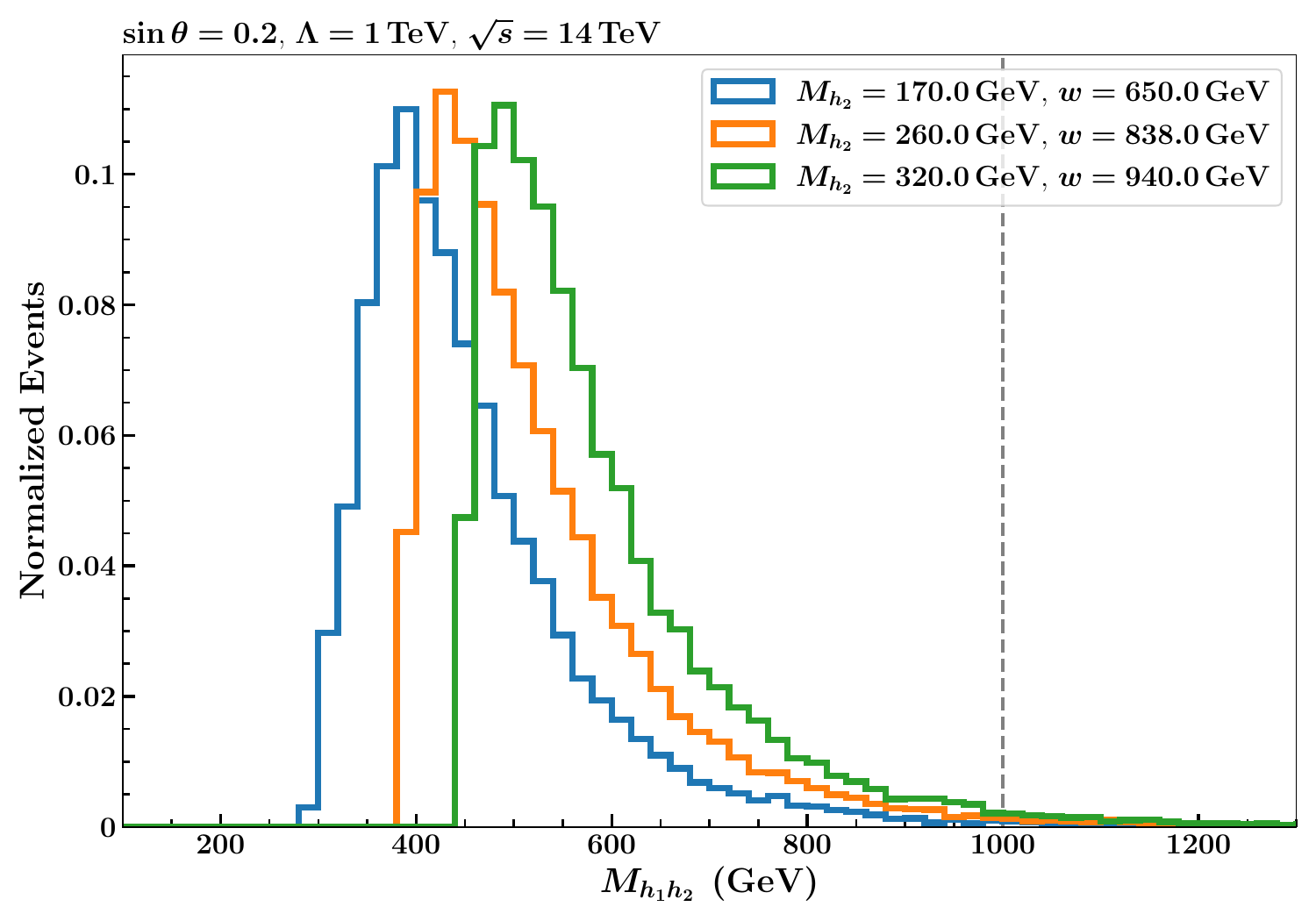}
	\caption{}
	\label{mh1h2_highw}
	\end{subfigure}
	\hfill
\begin{subfigure}[b]{0.48\textwidth}
		\centering
		\includegraphics[width=\textwidth]{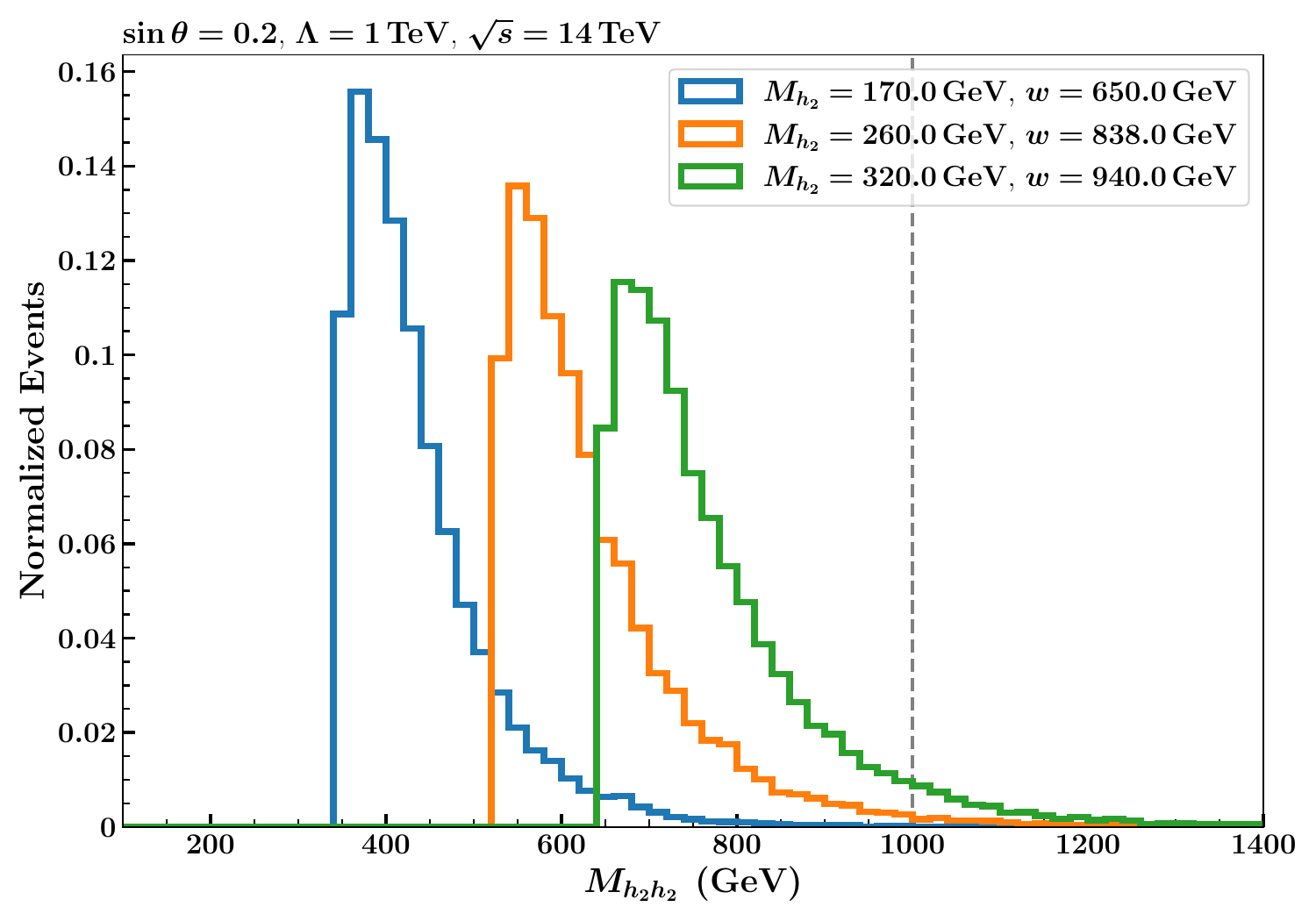}
		\caption{}
		\label{mh1h2_low_w}
	\end{subfigure}
    \caption{Normalized di-scalar invariant mass ( $M_{h_1h_2}$ and $M_{h_2 h_2}$) distributions obtained for  $pp \to h_1 h_2, \ h_2 h_2$  process at $\sqrt{s}=14~\text{TeV}$. The gray dashed vertical line denotes the cutoff scale $\Lambda = 1~\text{TeV}$.}
    \label{mh1h2}
\end{figure}

Figures~\ref{mh1h1} and \ref{mh1h2}  show normalized invariant mass distributions of the $h_i h_j$ pair to demonstrate the validity of the EFT framework. It is evident that most of the events correspond to $M_{h_i h_j} < \Lambda$ with a very small fraction of events ($ 0.05\% - 5.5\%$) flowing  beyond the cut-off scale $\Lambda$, that too depending on the mass of the additional singlet scalar and the type of the final states. The corresponding fraction of events is larger when the final state contains at least one heavy scalar. In particular, Figure~\ref{mh1h1} also highlights the indirect effect of the dimension-six operator on the di-Higgs invariant mass distribution. Assuming the same set of $M_{h_2}-\sin\theta$ values, there exist two classes of distributions for different choices of $w$: one exhibiting a resonance peak (\ref{mh1h1_highw}) and the other without it (\ref{mh1h1_low_w}) for $M_{h_2}>2 M_{h_1}$. The absence of resonant peaks in some distributions can be attributed to the particular values of the singlet scalar VEV ($w_2$) for which $\lambda_{h_2h_1h_1}$ coupling vanishes (Figure~\ref{lambda211_w} in Appendix). This phenomenon is a characteristic and distinguishing signature of the dimension-six operator considered in this work.

\begin{figure}[H]
	\centering
	\begin{subfigure}[b]{0.48\textwidth}
		\centering
		\includegraphics[width=\textwidth]{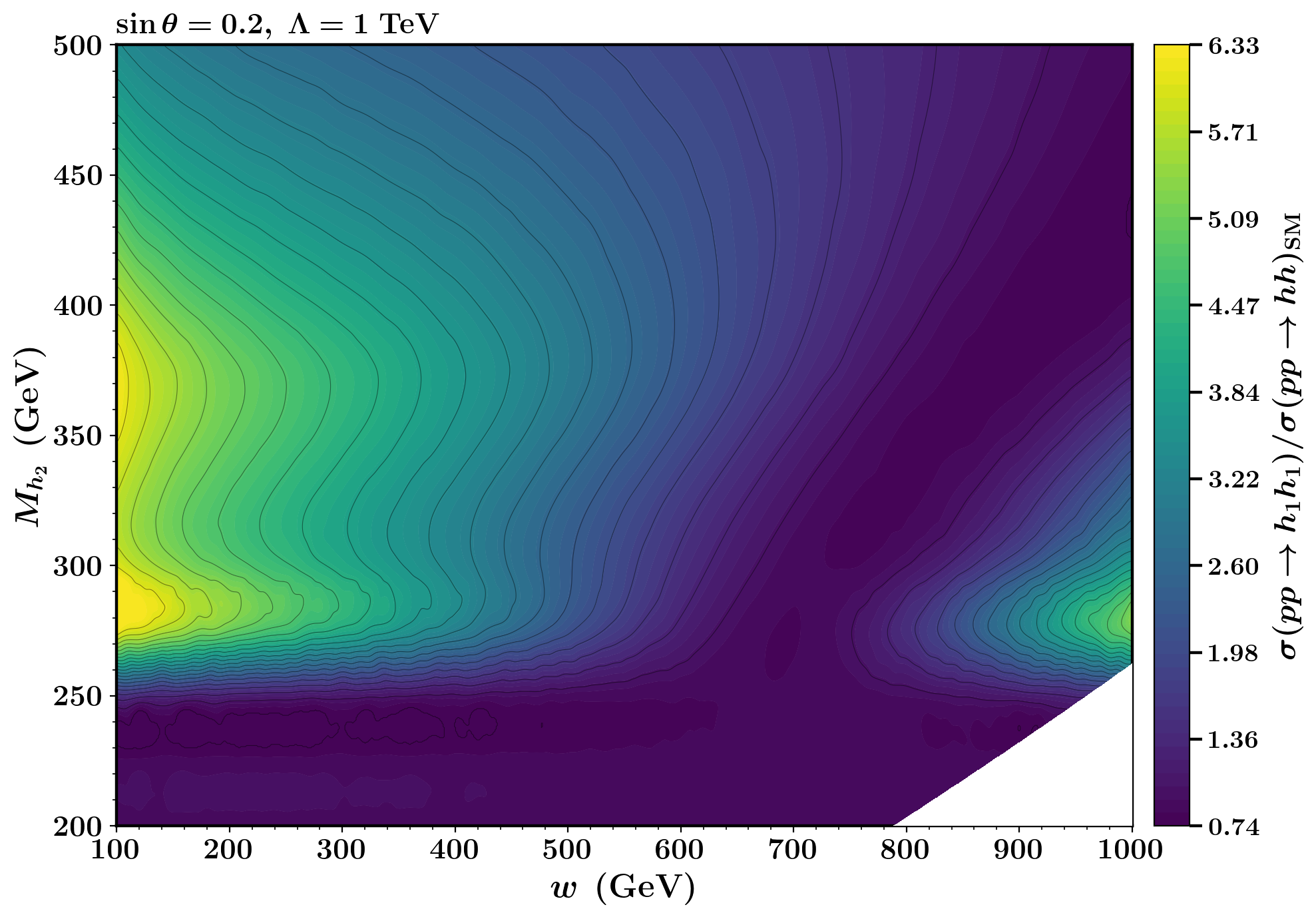}
		\caption{}
        \label{h1h1scan1}
	\end{subfigure}
	\hfill
	\begin{subfigure}[b]{0.48\textwidth}
		\centering
		\includegraphics[width=\textwidth]{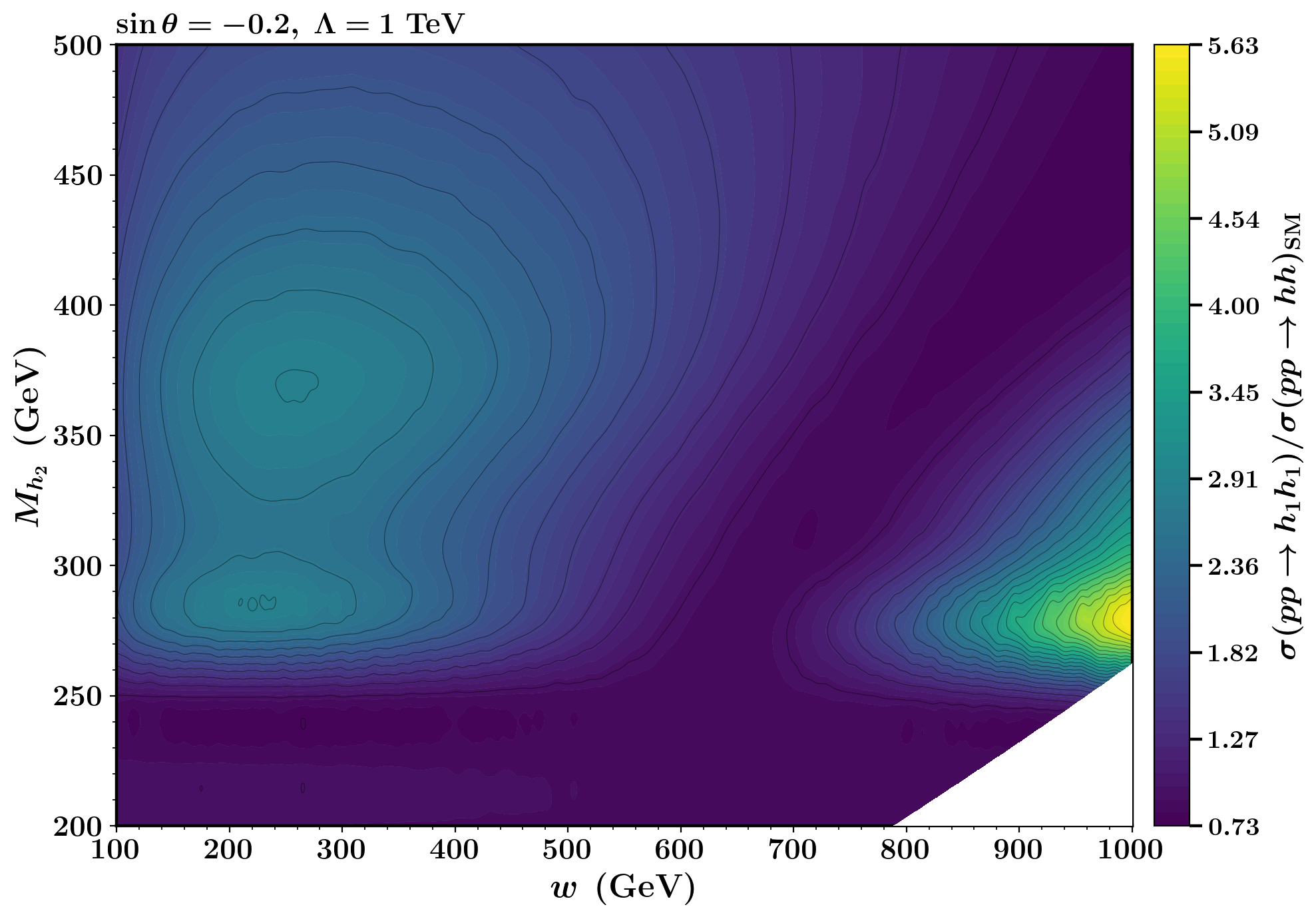}
		\caption{}
        \label{h1h1scan2}
	\end{subfigure}
    \caption{Ratio of the di-Higgs cross-section, $\frac{\sigma(pp\to h_1 h_1)}{\sigma(pp \to hh)_{SM}}$ in the $w-M_{h_2}$ plane for two different values of: (a) $\sin \theta= 0.2$ and (b) $\sin\theta =-0.2$, assuming $\Lambda=1$ TeV. The value of the ratio is represented in the color gradient. The white shaded region in the bottom right corner is excluded by the theoretical constraints discussed in Section~\ref{thcons}.}
    \label{h1h1scan}
    \end{figure}

We have also tabulated the cross-sections for $pp \to h_i h_j$ production at the LHC for some representative benchmark points ($M_{h_2}, \sin\theta, w$) in Table~\ref{tab:xsec_13TeV} and \ref{tab:xsec_14TeV} at $13$ and $14$ TeV LHC center of mass energies, respectively, with the requirement $M_{h_i h_j}< \Lambda$. The singlet scalar mass $M_{h_2}$ is taken in the range $170$ GeV to $340$ GeV, with $\sin\theta = \pm 0.2$ and {\it three} different values of $w$ are considered for each $M_{h_2}$ and $\sin\theta$. As mentioned earlier, the first and third choice of $w$ yield comparable values of the $\lambda_{h_2 h_1 h_1}$ coupling, but with opposite {\it sign} (Figure~\ref{lambda211_w}). Therefore, the cross-sections for these two different choices of $w$ are determined by different interference patterns among various contributing Feynman diagrams. An additional interplay among these diagrams arises due to the variation of the $\lambda_{h_1 h_1 h_1}$ coupling as a function of $w$, as can be seen in Figure~\ref{klambda2}. This feature is clearly visible in the quoted di-Higgs cross-sections in Table~\ref{tab:xsec_13TeV} and \ref{tab:xsec_14TeV}.
One can see an enhancement in the di-Higgs cross-sections by a factor $2-4$ depending on the singlet scalar mass and the VEV of the scalar singlet. In contrast, the predicted di-Higgs cross-sections remain close to the corresponding SM value for the the choices of $w$ that leads $\lambda_{h_2 h_1 h_1}=0$. This behaviour is also captured in Figure~\ref{h1h1scan}, where we plot the ratio of the di-Higgs cross-section as predicted in this scenario with respect to the corresponding SM value in $M_{h_2}-w$ plane. One can see that there exists a region of parameter space predicting close to SM-like di-Higgs rate even for $M_{h_2}>2M_{h_1}$.\\

\begin{table}[H]
\small
\centering
\renewcommand{\arraystretch}{1.2}
\begin{tabular}{|c|c|c c c|c|c c c|}
\hline
\multirow{3}{*}{$M_{h_2}$ [GeV]} &
\multicolumn{4}{c|}{$\sin\theta=+0.2$} &
\multicolumn{4}{c|}{$\sin\theta=-0.2$} \\
\cline{2-9}

& \multirow{2}{*}{$w$ [GeV]} & $\sigma_{h_1h_1}$ & $\sigma_{h_1h_2}$ & $\sigma_{h_2h_2}$
& \multirow{2}{*}{$w$ [GeV]} & $\sigma_{h_1h_1}$ & $\sigma_{h_1h_2}$ & $\sigma_{h_2h_2}$ \\

& & {\footnotesize [fb]} & {\footnotesize [fb]} & {\footnotesize [fb]}
& & {\footnotesize [fb]} & {\footnotesize [fb]} & {\footnotesize [fb]} \\
\hline

170 & 650.0 & 28.9 & 4.07 & 2.83 & 660.0 & 29.0 & 4.00 & 2.66 \\
    & 527.2 & 29.3 & 3.37 & 1.18 & 476.5 & 29.4 & 2.95 & 0.535 \\
    & 372.6 & 29.4 & 2.75 & 0.283 & 185.4 & 30.0 & 2.08 & 0.052 \\
\hline

200 & 716.0 & 28.7 & 3.59 & 1.81 & 719.0 & 28.7 & 3.38 & 1.53 \\
    & 571.8 & 28.8 & 2.93 & 0.729 & 521.2 & 29.1 & 2.52 & 0.305 \\
    & 385.7 & 29.1 & 2.41 & 0.171 & 196.2 & 29.6 & 1.70 & 0.061 \\
\hline

240 & 798.0 & 28.5 & 2.71 & 0.906 & 802.0 & 28.7 & 2.49 & 0.741 \\
    & 636.0 & 28.3 & 2.26 & 0.382 & 585.4 & 28.6 & 1.89 & 0.154 \\
    & 425.2 & 28.3 & 1.91 & 0.101 & 211.1 & 28.8 & 1.26 & 0.043 \\
\hline

260 & 838.0 & 76.7 & 2.26 & 0.647 & 841.0 & 95.7 & 2.06 & 0.515 \\
    & 669.6 & 27.9 & 1.92 & 0.280 & 619.1 & 27.8 & 1.60 & 0.114 \\
    & 450.9 & 74.6 & 1.65 & 0.080 & 218.8 & 79.9 & 1.08 & 0.033 \\
\hline

280 & 873.0 & 80.8 & 1.86 & 0.458 & 878.0 & 102 & 1.69 & 0.364 \\
    & 703.1 & 27.5 & 1.61 & 0.206 & 653.5 & 27.9 & 1.35 & 0.085 \\
    & 487.2 & 78.4 & 1.42 & 0.066 & 226.5 & 91.3 & 0.926 & 0.025 \\
\hline

320 & 940.0 & 66.0 & 1.28 & 0.232 & 947.0 & 84.6 & 1.15 & 0.182 \\
    & 774.9 & 26.9 & 1.15 & 0.117 & 724.4 & 27.3 & 0.968 & 0.050 \\
    & 569.2 & 63.6 & 1.05 & 0.046 & 344.9 & 82.4 & 0.771 & 0.004 \\
\hline

340 & 971.0 & 59.8 & 1.07 & 0.166 & 977.0 & 76.7 & 0.968 & 0.128 \\
    & 811.1 & 26.5 & 0.986 & 0.088 & 760.7 & 26.9 & 0.844 & 0.039 \\
    & 615.5 & 57.5 & 0.919 & 0.038 & 423.4 & 74.5 & 0.715 & 0.002 \\
\hline

\end{tabular}
\caption{Cross sections for $pp\to h_1h_1$, $pp\to h_1h_2$ and $pp\to h_2h_2$ at $\sqrt{s}=13$ TeV, requiring $M_{h_i h_j}<\Lambda=1~\mathrm{TeV}$. The quoted cross sections include a $K$-factor of $2.3$~\cite{deFlorian:2014rta}.}
\label{tab:xsec_13TeV}
\end{table}

\begin{table}[h!]
\small
\centering
\renewcommand{\arraystretch}{1.4}
\begin{tabular}{|c|c|c c c|c|c c c|}
\hline
\multirow{3}{*}{$M_{h_2}$ [GeV]} &
\multicolumn{4}{c|}{$\sin\theta=+0.2$} &
\multicolumn{4}{c|}{$\sin\theta=-0.2$} \\
\cline{2-9}

& \multirow{2}{*}{\raisebox{-0.4ex}{$w$ [GeV]}} & $\sigma_{h_1h_1}$ & $\sigma_{h_1h_2}$ & $\sigma_{h_2h_2}$
& \multirow{2}{*}{\raisebox{-0.4ex}{$w$ [GeV]}} & $\sigma_{h_1h_1}$ & $\sigma_{h_1h_2}$ & $\sigma_{h_2h_2}$ \\

& & {\footnotesize [fb]} & {\footnotesize [fb]} & {\footnotesize [fb]}
& & {\footnotesize [fb]} & {\footnotesize [fb]} & {\footnotesize [fb]} \\
\hline

170 & 650.0 & 34.6 & 4.86 & 3.37 & 660.0 & 34.5 & 4.75 & 3.15 \\
    & 527.2 & 34.9 & 4.01 & 1.40 & 476.5 & 35.2 & 3.51 & 0.636 \\
    & 372.6 & 35.1 & 3.28 & 0.336 & 185.4 & 35.8 & 2.48 & 0.062 \\
\hline

200 & 716.0 & 34.2 & 4.28 & 2.16 & 719.0 & 34.4 & 4.03 & 1.83 \\
    & 571.8 & 34.4 & 3.49 & 0.874 & 521.2 & 34.6 & 3.00 & 0.365 \\
    & 385.7 & 34.7 & 2.88 & 0.205 & 196.2 & 35.3 & 2.03 & 0.073 \\
\hline

240 & 798.0 & 33.9 & 3.24 & 1.10 & 802.0 & 34.2 & 2.98 & 0.892 \\
    & 636.0 & 33.9 & 2.70 & 0.458 & 585.4 & 34.1 & 2.27 & 0.186 \\
    & 425.2 & 33.7 & 2.29 & 0.122 & 211.1 & 34.4 & 1.51 & 0.052 \\
\hline

260 & 838.0 & 89.8 & 2.71 & 0.784 & 841.0 & 112 & 2.47 & 0.627 \\
    & 669.6 & 33.4 & 2.30 & 0.339 & 619.1 & 33.1 & 1.92 & 0.139 \\
    & 450.9 & 87.6 & 1.98 & 0.097 & 218.8 & 93.8 & 1.29 & 0.041 \\
\hline

280 & 873.0 & 94.9 & 2.24 & 0.560 & 878.0 & 120 & 2.03 & 0.445 \\
    & 703.1 & 33.0 & 1.94 & 0.252 & 653.5 & 33.3 & 1.62 & 0.105 \\
    & 487.2 & 92.0 & 1.71 & 0.080 & 226.5 & 107 & 1.11 & 0.031 \\
\hline

320 & 940.0 & 77.9 & 1.54 & 0.286 & 947.0 & 99.9 & 1.39 & 0.223 \\
    & 774.9 & 32.1 & 1.39 & 0.143 & 724.4 & 32.6 & 1.17 & 0.062 \\
    & 569.2 & 75.0 & 1.27 & 0.056 & 344.9 & 97.2 & 0.929 & 0.005 \\
\hline

340 & 971.0 & 70.7 & 1.29 & 0.205 & 977.0 & 90.6 & 1.17 & 0.157 \\
    & 811.1 & 31.5 & 1.19 & 0.109 & 760.7 & 32.1 & 1.02 & 0.048 \\
    & 615.5 & 68.0 & 1.11 & 0.047 & 423.4 & 87.9 & 0.864 & 0.002 \\
\hline

\end{tabular}
\caption{Cross sections for $pp\to h_1h_1$, $pp\to h_1h_2$ and $pp\to h_2h_2$ at $\sqrt{s}=14$ TeV, requiring $M_{h_i h_j}<\Lambda=1~\mathrm{TeV}$. The quoted cross sections include a $K$-factor of $2.3$~\cite{deFlorian:2014rta}.}
\label{tab:xsec_14TeV}
\end{table}

\subsection{Triple-Higgs Production}
\begin{figure}[h!]
	\centering
	\begin{subfigure}[b]{0.24\textwidth}
		\centering
		\includegraphics[width = \textwidth]{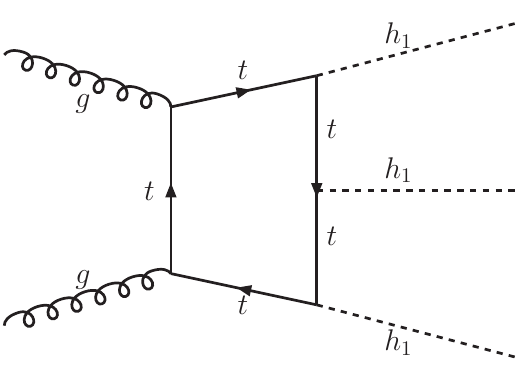}
		\caption{}
		\label{feyn1_1}
	\end{subfigure}
	\hfill
    \begin{subfigure}[b]{0.24\textwidth}
		\centering
		\includegraphics[width = \textwidth]{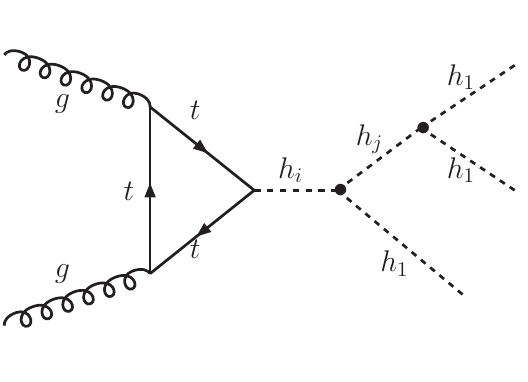}
		\caption{}
		\label{feyn2_1}
	\end{subfigure}
    \hfill
	\begin{subfigure}[b]{0.24\textwidth}
		\centering
		\includegraphics[width= \textwidth]{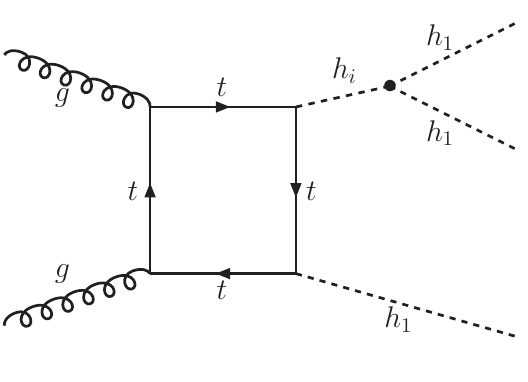}
		\caption{}
		\label{feyn3_1}
	\end{subfigure}
    \hfill
    	\begin{subfigure}[b]{0.24\textwidth}
		\centering
		\includegraphics[width= \textwidth]{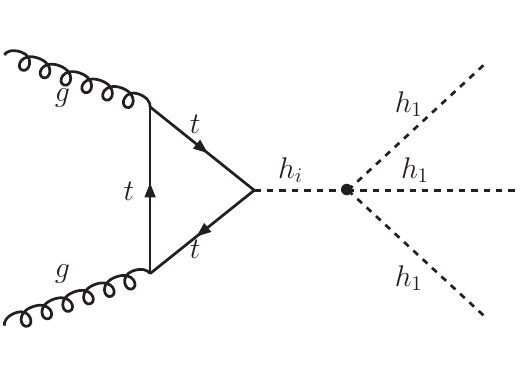}
		\caption{}
		\label{feyn4_1}
	\end{subfigure}
\begin{center}
\begin{subfigure}[b]{0.24\textwidth}
    \centering
    \includegraphics[width=\textwidth]{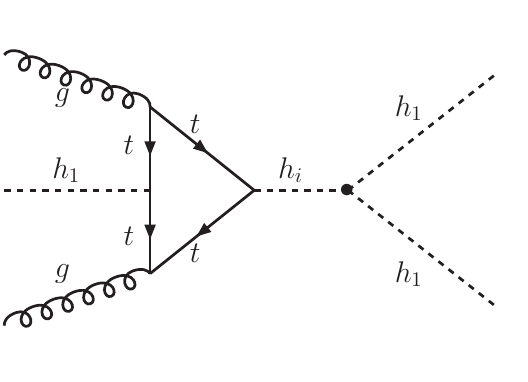}
    \caption{}
\end{subfigure}
\hspace{1cm}
\begin{subfigure}[b]{0.24\textwidth}
    \centering
    \includegraphics[width=\textwidth]{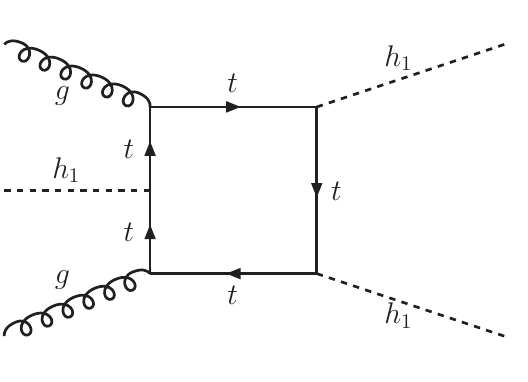}
    \caption{}
\end{subfigure}
\end{center}
	\caption{Representative Feynman diagrams for $pp\to h_1 h_1 h_1$ process. Here, the black dot in the above diagrams represent various modified triple scalar couplings in presence of the effective operator.}
    \label{h1h1h1}
\end{figure}

\begin{figure}[h!]
   \begin{subfigure}[b]{0.48\textwidth}
		\centering
	\includegraphics[width = \textwidth]{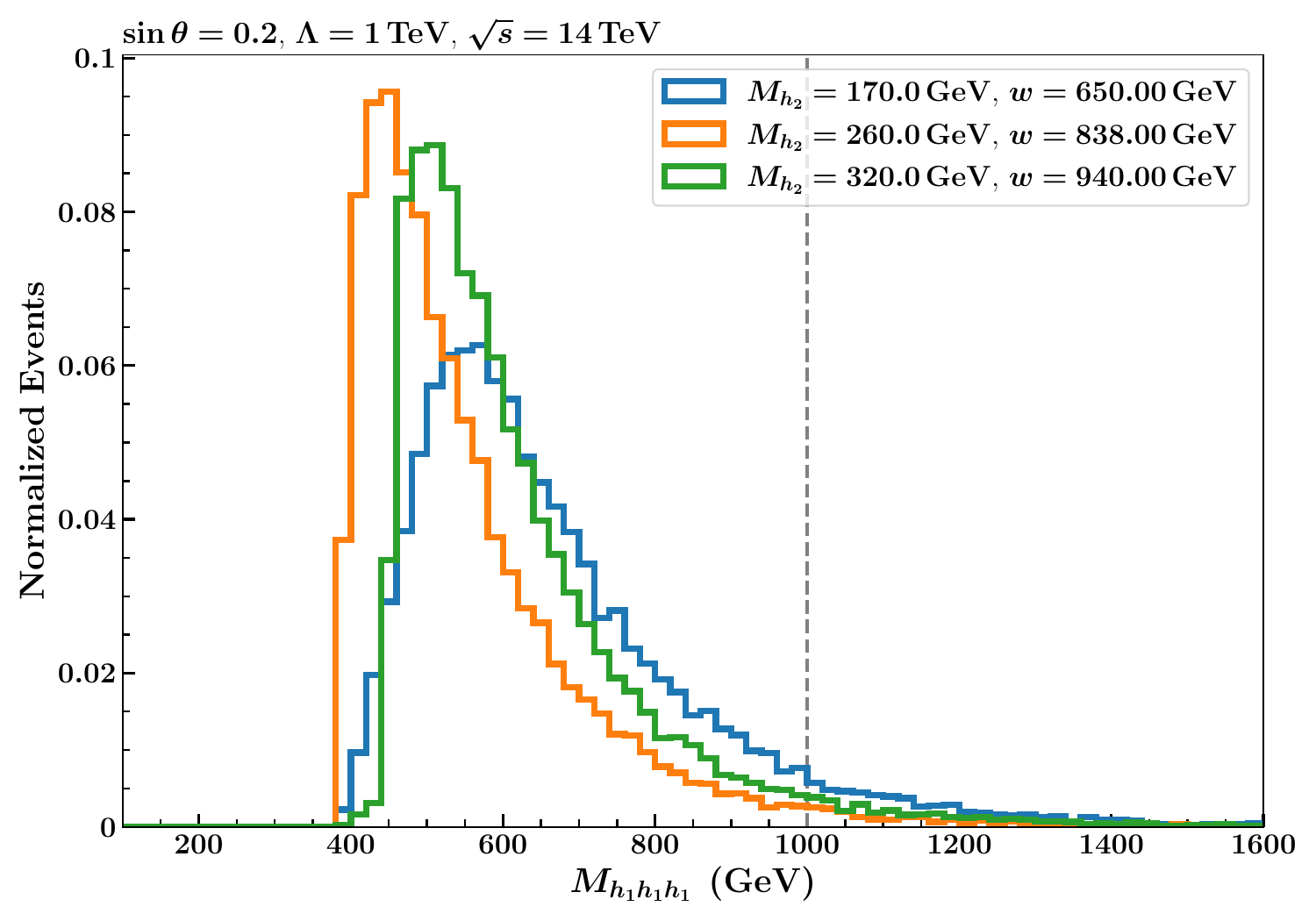}
	\caption{}
	\label{mh1h1h1_highw}
	\end{subfigure}
	\hfill
\begin{subfigure}[b]{0.48\textwidth}
		\centering
		\includegraphics[width=\textwidth]{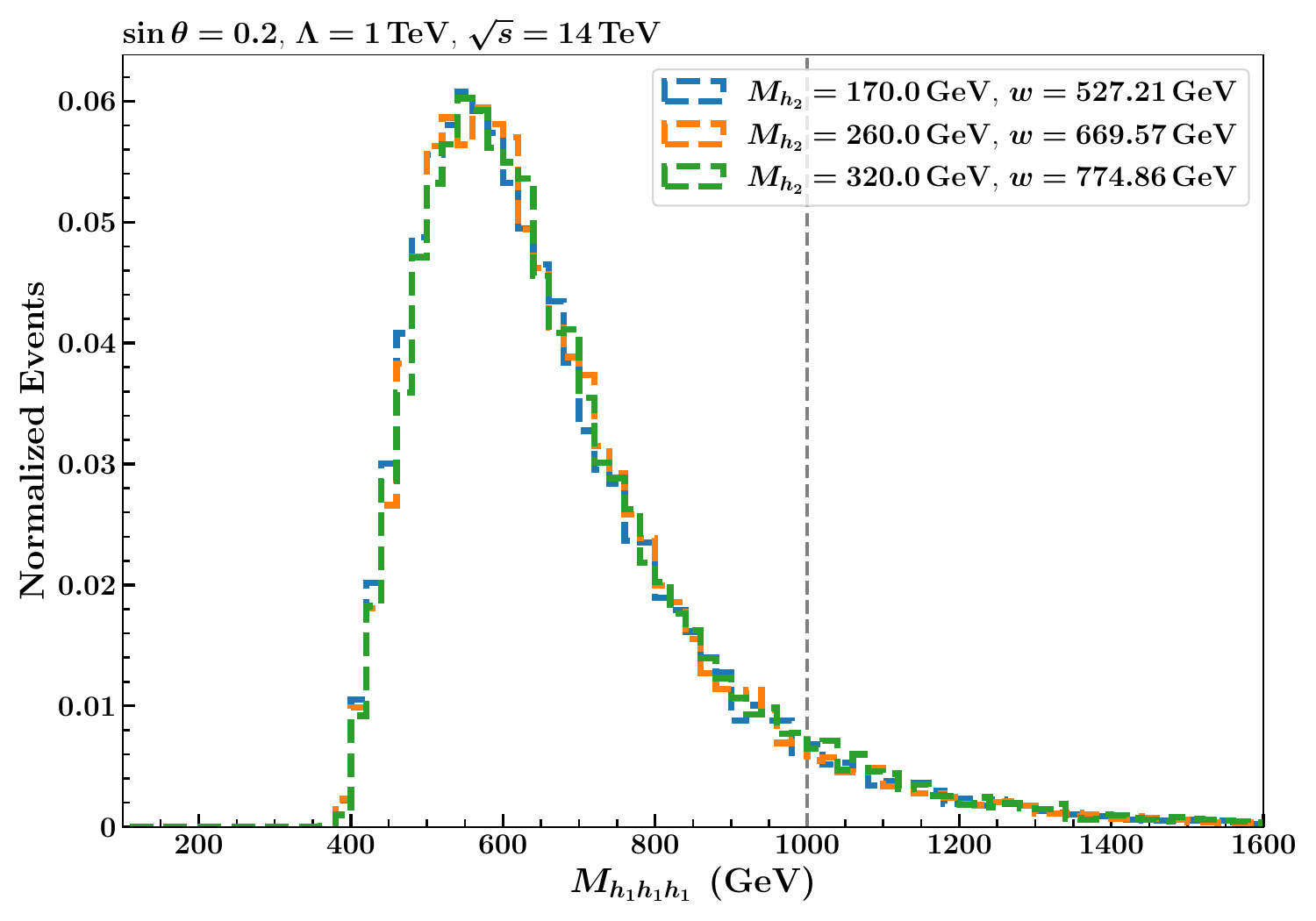}
		\caption{}
		\label{mh1h1h1_low_w}
	\end{subfigure}
    \caption{(a) Normalized triple-Higgs invariant mass distributions ($M_{h_1h_1h_1}$) obtained for $pp \to h_1 h_1 h_1$ process at $\sqrt{s}=14~\text{TeV}$. (b) Represents the same but with different set  of singlet scalar VEV that leads to vanishing $\lambda_{h_2 h_1 h_1}$ coupling. The gray dashed vertical line denotes the cutoff scale $\Lambda = 1~\text{TeV}$.}
    \label{mh1h1h1}
\end{figure}

In this subsection, we discuss the triple-Higgs production which is an important channel to probe new physics beyond the SM, as a part of the multi-scalar searches at the LHC. The next-to-leading-order triple Higgs production cross-section in the SM is of the order of $0.103$ fb at $14$ TeV LHC center of mass energy~\cite{Abouabid:2024gms,ATLAS:2024xcs}. Some representative Feynman diagrams for triple Higgs process in the context of the model described in Section~\ref{model} are shown in Figure~\ref{h1h1h1}. This includes diagrams that are present in the SM plus additional diagrams due to the presence of the singlet scalar field. In Figure~\ref{mh1h1h1}, we plot the triple Higgs invariant mass distributions to validate the EFT framework and to obtain the region of phase space consistent with it. The corresponding cross-sections for all events satisfying $M_{h_1 h_1 h_1}< \Lambda = 1$ TeV are quoted in Table~\ref{tab:h1h1h1_14TeV}. A SM like $K$-factor is also assumed to estimate this cross-sections.

A close look at the Table~\ref{tab:h1h1h1_14TeV} indicates that the triple-Higgs cross-sections also follow a pattern similar to that of the di-Higgs production at the LHC, in particular, with respect to the identical set of values chosen for the singlet scalar VEV. In fact, the behavior of the $pp \to h_1h_1h_1$ cross-sections is also controlled by the same set of triple scalar couplings ($\lambda_{h_i h_1 h_1}$) that are relevant for di-Higgs production. The triple-Higgs cross-sections for $M_{h_2}<2M_{h_1}$ are close to the SM value, as expected, due to the absence of resonant contribution. However, for $M_{h_2}>2M_{h_1}$ there is an enhancement in the triple Higgs cross-section by almost a factor of $5$ ($6$) for $\sin\theta=0.2 \ (-0.2)$ and certain choice of the singlet scalar VEV that ensures a resonance peak in the di-Higgs invariant mass distributions. In the singlet scalar mass range $260-340$ GeV the cross-section gradually decreases with an increase in $M_{h_2}$ due to usual phase space suppression. For $M_{h_2}>3 M_{h_1}$ GeV, there is an additional enhancement in the cross-section due to the resonant triple-Higgs production which can be attributed to the class of Feynman diagram depicted in Figure~\ref{feyn4_1}. For example, at $M_{h_2}= 380$ GeV and $\sin\theta=0.2$ the triple-Higgs cross-section turns out to be 0.261 fb for $w=500$ GeV. This point however does not give rise to SFOEWPT.

\begin{table}[h!]
\small
\centering
\renewcommand{\arraystretch}{1.35}
\begin{tabular}{|c|c|c|c|c|}
\hline
\multirow{2}{*}{$M_{h_2}$ [GeV]} &
\multicolumn{2}{c|}{$\sin\theta=+0.2$} &
\multicolumn{2}{c|}{$\sin\theta=-0.2$} \\
\cline{2-5}
& $w$ [GeV] & $\sigma_{h_1h_1h_1}$ [fb] & $w$ [GeV] & $\sigma_{h_1h_1h_1}$ [fb] \\
\hline

170 & 650.0 & 0.077 & 660.0 & 0.077 \\
    & 527.2 & 0.079 & 476.5 & 0.079 \\
    & 372.6 & 0.080 & 185.4 & 0.081 \\
\hline

200 & 716.0 & 0.076 & 719.0 & 0.077 \\
    & 571.8 & 0.078 & 521.2 & 0.078 \\
    & 385.7 & 0.078 & 196.2 & 0.079 \\
\hline

240 & 798.0 & 0.077 & 802.0 & 0.078 \\
    & 636.0 & 0.075 & 585.4 & 0.076 \\
    & 425.2 & 0.074 & 211.1 & 0.075 \\
\hline

260 & 838.0 & 0.373 & 841.0 & 0.457 \\
    & 669.6 & 0.074 & 619.1 & 0.075 \\
    & 450.9 & 0.309 & 218.8 & 0.239 \\
\hline

280 & 873.0 & 0.362 & 878.0 & 0.445 \\
    & 703.1 & 0.072 & 653.5 & 0.074 \\
    & 487.2 & 0.326 & 226.5 & 0.277 \\
\hline

320 & 940.0 & 0.227 & 947.0 & 0.283 \\
    & 774.9 & 0.069 & 724.4 & 0.071 \\
    & 569.2 & 0.236 & 344.9 & 0.249 \\
\hline

340 & 971.0 & 0.174 & 977.0 & 0.217 \\
    & 811.1 & 0.068 & 760.7 & 0.070 \\
    & 615.5 & 0.196 & 423.4 & 0.217 \\
\hline

\end{tabular}
\caption{Cross sections (in fb) for $pp\to h_1h_1h_1$ at $\sqrt{s}=14$ TeV, requiring $M_{h_1h_1h_1}<\Lambda=1~\mathrm{TeV}$. The quoted cross sections include a $K$-factor of $2.34$ to account for next-to-leading-order effects.}
\label{tab:h1h1h1_14TeV}
\end{table}



\clearpage
\section{Conclusions}
\label{conc}
 
We have considered an effective field theory framework where the SM is augmented with an additional singlet complex scalar field charged under a local $U(1)_D$ gauge group. The tree-level scalar potential contains additional dimension-six operator involving the Higgs and the singlet scalar fields of the form $|H|^2 |\phi|^4$. The complex scalar field is also responsible for the generation of the mass of the $U(1)_D$ gauge boson, referred to as dark photon. We emphasize that the addition of the dimension-six operator allowed by all possible symmetries of the theory gives rise to distinctive features in the context of first order EWPT, resulting GW signal and multi-scalar productions at the LHC. This is in contrast to the conventional singlet scalar extension of the SM in absence of the dimension six term. Importantly, even when a dimension-six operator involving the singlet scalar is present, the phenomenology can differ substantially in terms of the phase transition pattern and the viable parameter space consistent with SFOEWPT depending on whether the singlet scalar acquires a VEV.

We identify the SFOEWPT compatible parameter space in the $M_{h_2}-\sin\theta$ plane while satisfying relevant theoretical and experimental constraints. We further present a detailed prediction of GW power spectrum and asses their sensitivity at the future GW experiments. In addition, we also present cross-sections for multi-scalar production processes at the LHC.
Our main findings can be summarized as follows:
\begin{itemize}
    \item In our setup, the phase transition proceeds predominantly through $(0,u(T)) \to \left(v(T),w(T)\right)$. An important feature of our scenario is that the tree-level potential difference between the local extremum ($0,u$) and the global minima ($v_{_\text{EW}},w$) is controlled by the singlet scalar VEV ($w$) for a fixed cut-off scale $\Lambda$. With increasing $w$ this difference decreases which in turn helps to reduce the critical temperature, therefore, allowing a wider region of viable parameter space in the $M_{h_2}- \sin \theta$ plane consistent with SFOEWPT. In particular, we have shown that a completely new SFOEWPT compatible parameter space (for $M_{h_2}> M_{h_1}$)  opens up in presence of the dimension six operator.
    
    \item In absence of the dimension-six operator one requires large Higgs-singlet portal coupling ($\lambda_{hs}$) or scalar mixing angle ($\sin\theta$) to achieve SFOEWPT~\cite{Curtin:2014jma, Chen:2025ksr, Carena:2019une}. The large value of the former might hamper perturbative unitarity while that of the latter is disfavored by the Higgs-signal strength measurements. The presence of the dimension-six operator helps to weaken the correlation between $\sin\theta$ and $\lambda_{hs}$ in this scenario. Therefore, in this setup  SFOEWPT can be  achieved even for smaller $\sin\theta$ and $|\lambda_{hs}|\sim 1$.
    
    \item Additionally, the \emph{dominant} effect of UV physics on the phase transition dynamics enters as a combination of $\frac{w}{\Lambda}$ and does not necessarily decouple in the large $\Lambda$ limit as long as $w$ is also large (but still $<\Lambda$), keeping the ratio $\frac{w}{\Lambda}$ fixed.
This is in contrast to the scenarios with higher-dimensional operators involving only the Higgs field, or various combinations of the Higgs and additional singlet scalar fields where the latter does not acquire a VEV. It turns out that for $M_{h_2} = 200$ GeV ($250$ GeV) and $|\sin\theta| = 0.2$, the required minimum value of this ratio to achieve SFOEWPT is $0.71$ (0.82).

    \item The parameter space consistent with SFOEWPT can be probed at future gravitational wave facilities as well. In particular, we have shown that the GW power spectrum generated during the phase transition is sensitive to LISA, BBO, DECIGO, $\mu$Ares,  ultimate-DECIGO, DECIGO-corr and ultimate-DECIGO-corr gravitational wave detectors unlike usual singlet scalar extension of the SM with spontaneous $\mathcal{Z}_2$ symmetry breaking. For a fixed mass, mixing angle, and cut-off scale, the peak GW energy density shifts toward lower frequencies and attains a larger amplitude as $w$ increases, thereby improving its sensitivity to future GW experiments.
    
    \item This model also predicts a strong correlation between the SFOEWPT-viable parameter space and enhanced di-Higgs and multiscalar production rates at the LHC. This arises from the fact that achieving SFOEWPT requires a modified scalar potential, which in turn alters the trilinear and quartic scalar couplings. Consequently, measurements of di-Higgs and triple-Higgs cross sections at the LHC can probe the nature of the EWPT. From a purely collider perspective, we also illustrate the sensitivity of the di-Higgs production cross section to the singlet scalar VEV ($w$) for a fixed cut-off scale $\Lambda$ in the presence of the dimension-six operator. In particular, we emphasize that even in the presence of an additional scalar resonance in the mass range $(250\text{–}500)$ GeV, a resonance peak in the di-Higgs invariant mass distribution may be absent for certain values of $w$. This is a very peculiar consequence of the dimension-six operator considered in this work. Moreover, for $M_{h_2}\geq 2 M_{h_1}$ and $|\sin \theta| = 0.2$ the di-Higgs rate at the LHC can be $2-4$ times that predicted in the SM in a region where SFOEWPT is achieved.

\end{itemize}

Although our analysis primarily focuses on the dimension-six operator of the form $|H|^2|\phi|^4$, additional dimension-six operators may also arise from the same UV completion. The inclusion of all possible dimension-six operators allowed by the symmetries of the theory can further enlarge the SFOEWPT-compatible parameter space and/or reduce the required minimum value of the ratio $\frac{w}{\Lambda}$ to achieve SFOEWPT.

\section*{Acknowledgments}
AB acknowledges the University Grants Commission (UGC), Government of India, for providing the NET–JRF fellowship (ID 231620060873). AB also thanks Indra Kumar Banerjee, Koustav Mukherjee, Krishna Tewary and Shivam Verma for their assistance in numerical simulations and helpful discussions; and Soumyadip Sarkar for providing access to high-performance computational facilities. SN acknowledges the funding allotted to
grant number SUR/2022/001404 by the Anusandhan National Research Foundation (formerly the Science and Engineering Research Board (SERB)), Govt. of India.
\clearpage
\appendix
\section{Variation of Lagrangian parameters}
\label{appendix1}

\begin{figure}[h!]
	\centering
	\begin{subfigure}[b]{0.48\textwidth}
		\centering
		\includegraphics[width=\textwidth]{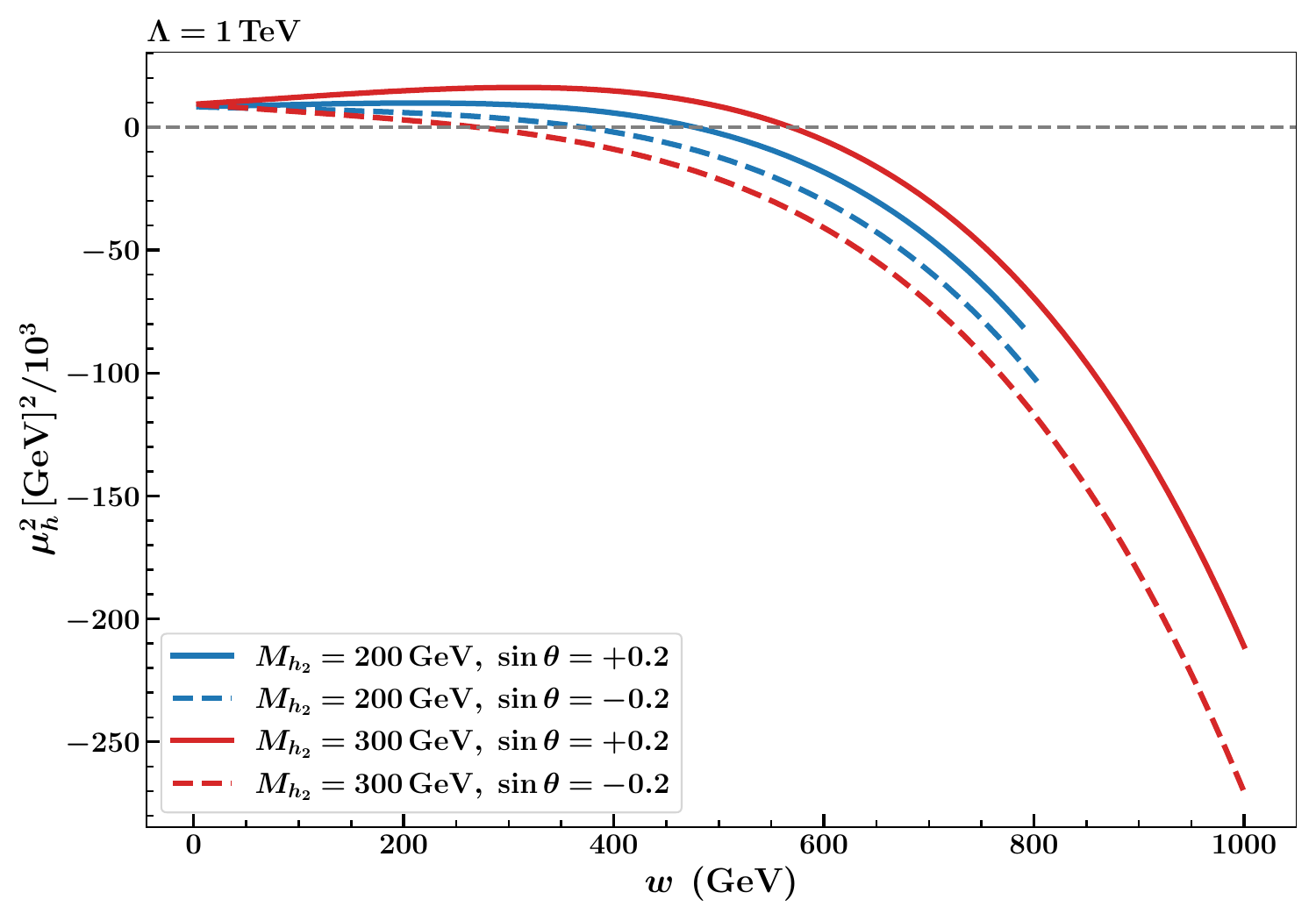}
		\caption{$\mu_h^2$ vs $w$}
	\end{subfigure}
	\hfill
	\begin{subfigure}[b]{0.48\textwidth}
		\centering
		\includegraphics[width=\textwidth]{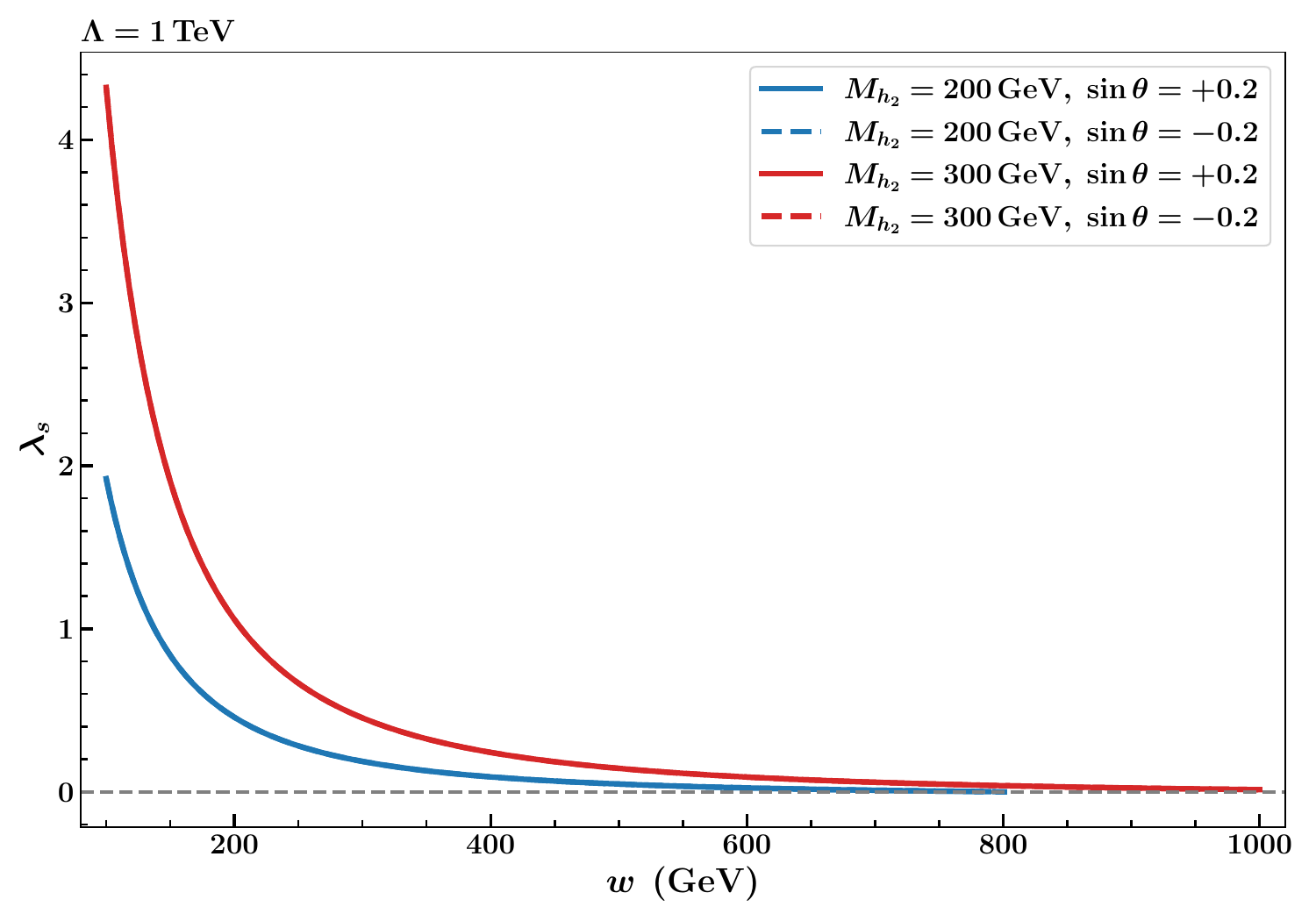}
		\caption{$\lambda_s$ vs $w$}
	\end{subfigure}

    \vspace{0.7cm}

	\begin{subfigure}[b]{0.48\textwidth}
		\centering
		\includegraphics[width=\textwidth]{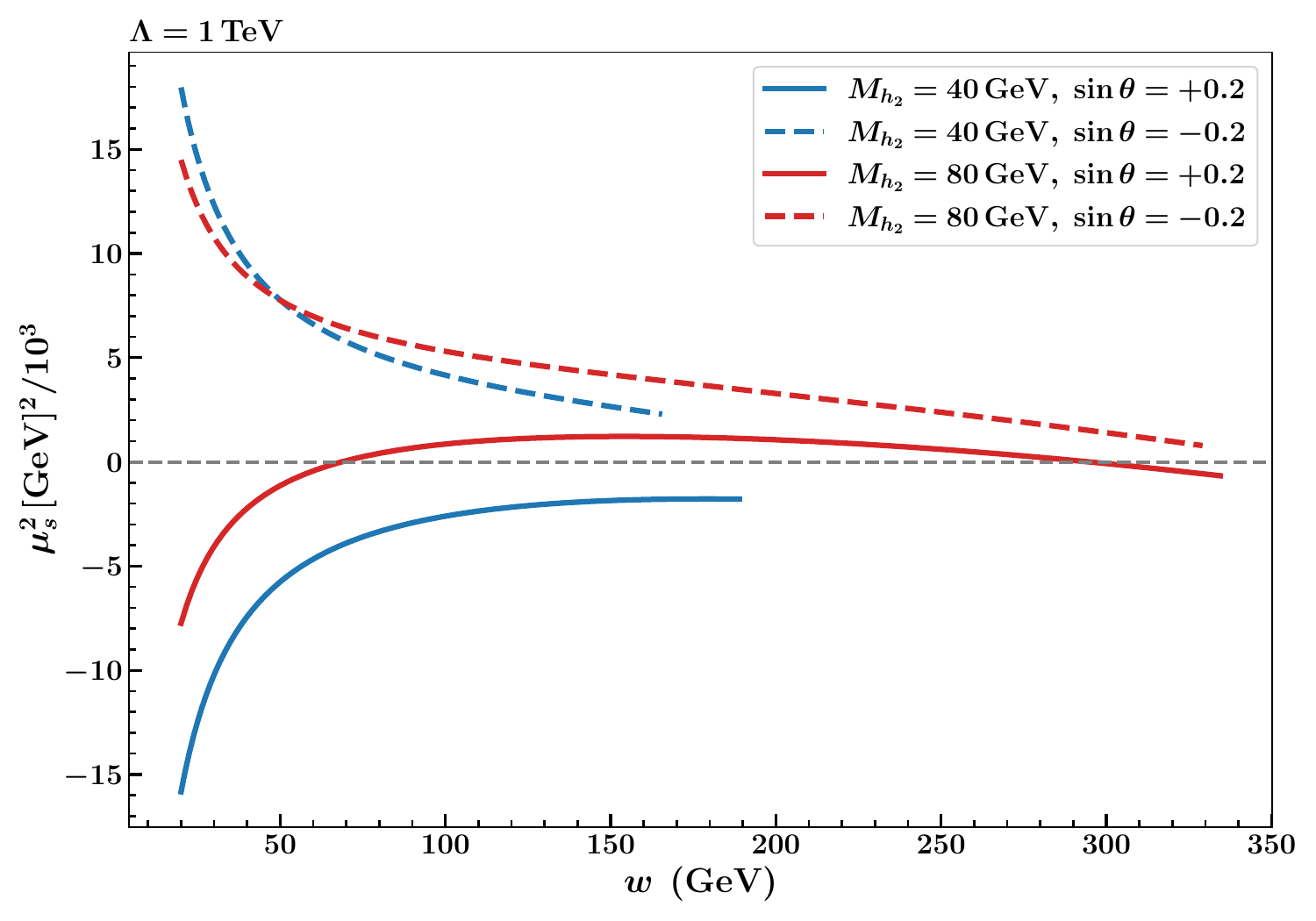}
		\caption{$\mu_s^2$ vs $w$}
        \label{mus2_w_40_80}
	\end{subfigure}
	\hfill
	\begin{subfigure}[b]{0.48\textwidth}
		\centering
		\includegraphics[width=\textwidth]{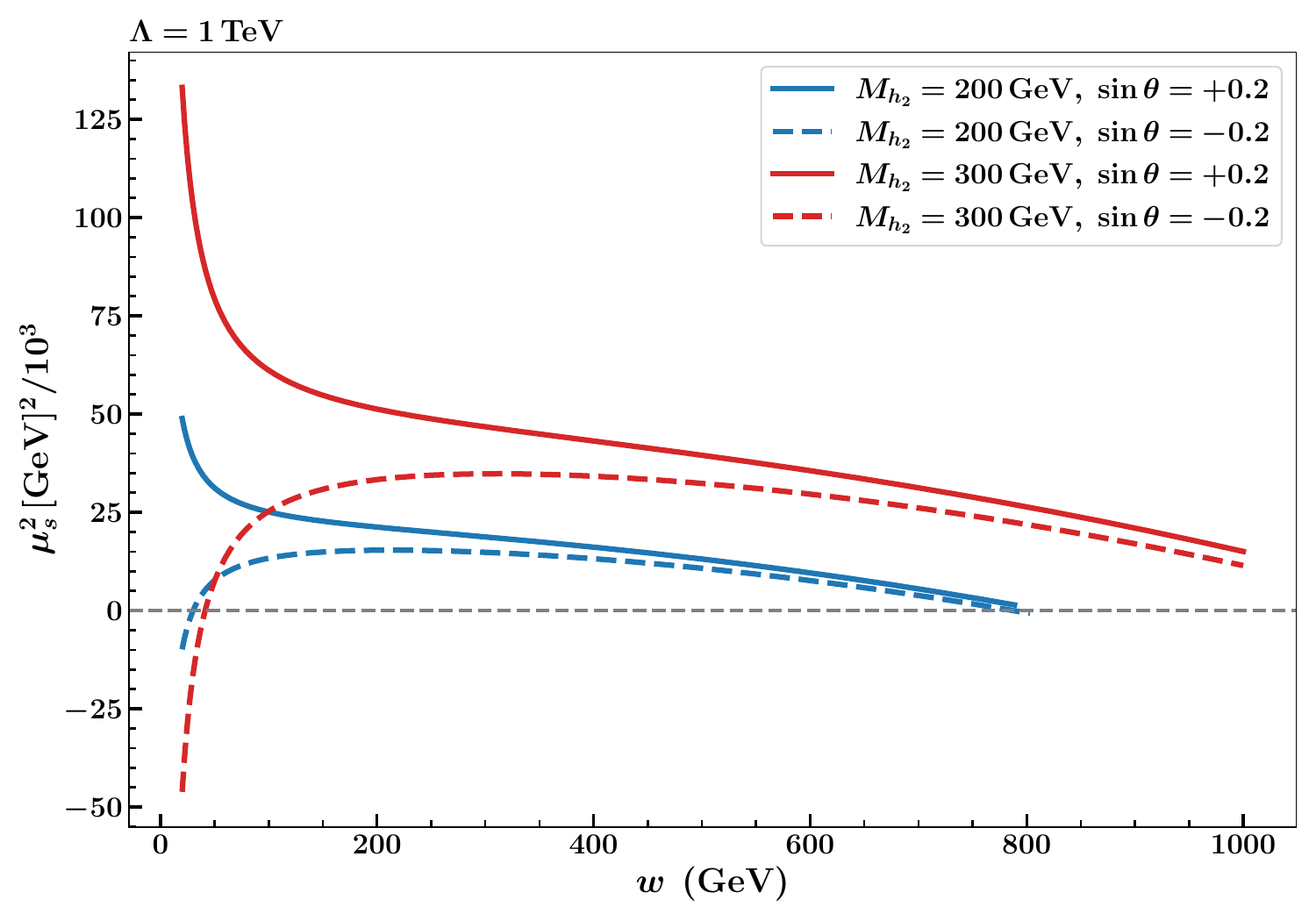}
		\caption{$\mu_s^2$ vs $w$.}
	\end{subfigure}

    \vspace{0.7cm}
    
    \begin{subfigure}[b]{0.48\textwidth}
		\centering
		\includegraphics[width=\textwidth]{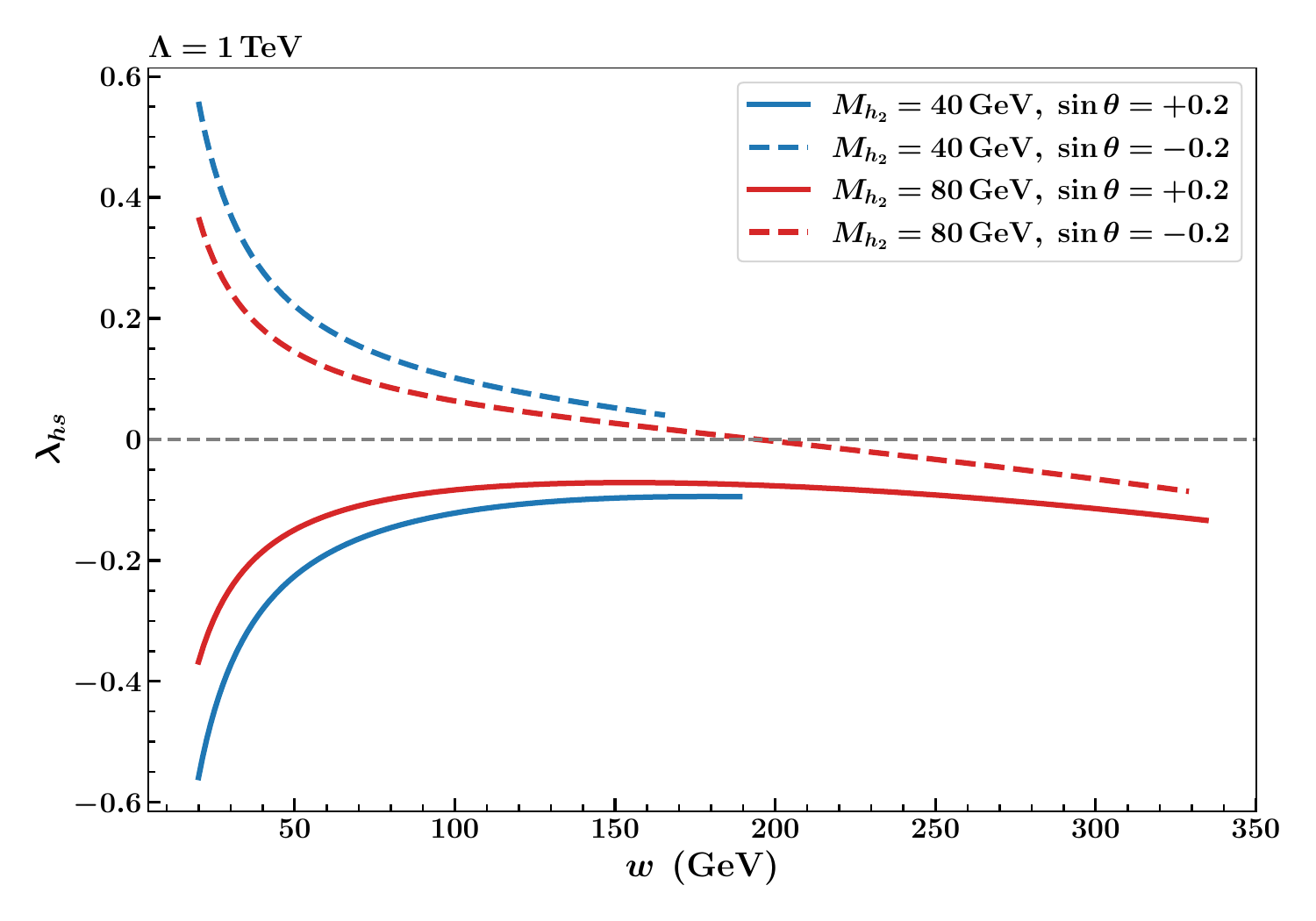}
		\caption{$\lambda_{hs}$ vs $w$}
         \label{lambda_hs_w_40_80}
	\end{subfigure}
	\hfill
	\begin{subfigure}[b]{0.48\textwidth}
		\centering
		\includegraphics[width=\textwidth]{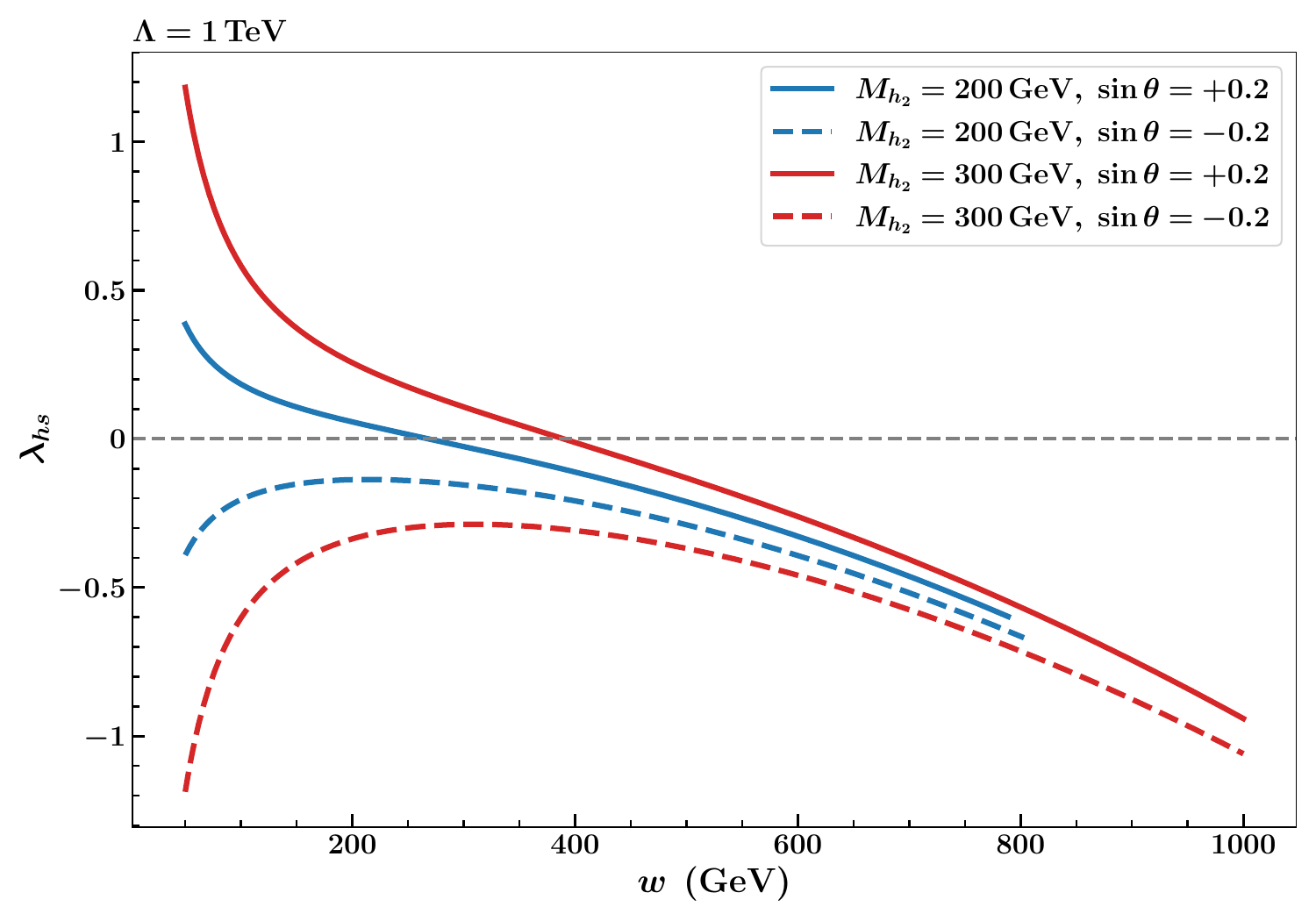}
		\caption{$\lambda_{hs}$ vs $w$}
        \label{lambda_hs_w_200_300}
	\end{subfigure}
	
	\caption{Variation of the relevant Lagrangian parameters as a function of the singlet scalar VEV ($w$) for different choices of $M_{h_2}$ and $\sin\theta$  with fixed $\Lambda$ (1 TeV).}
    \label{lag_pam_w}
\end{figure}

The Lagrangian parameters $\{\mu_h, \mu_s, \lambda_h, \lambda_{s}, \lambda_{hs} \}$ can be expressed in terms of the physical parameters $\{M_{h_1}, M_{h_{2}}, \sin \theta , v_{_{\text{EW}}}, w, \Lambda \}$ as follows :
\begin{align}
	\mu_h^2 &= \frac{1}{2}\left( M_{h_1}^2 \cos^2\theta + M_{h_2}^2 \sin^2\theta \right) + \left(M_{h_2}^2 -M_{h_1}^2 \right)\frac{w}{2 v_{_{\text{EW}}}}\sin\theta \cos\theta - \frac{w^4}{4 \Lambda^2}, \label{muh2}\\
	\lambda_h &= \frac{M_{h_1}^2 \cos^2\theta + M_{h_2}^2 \sin^2\theta}{2 v_{_{\text{EW}}}^2},\label{lambdah} \\
	\mu_s^2 &=\frac{1}{2}\left( M_{h_1}^2 \sin^2\theta + M_{h_2}^2 \cos^2\theta \right)+ \left(M_{h_2}^2 -M_{h_1}^2 \right)\frac{v_{_{\text{EW}}}}{2 w}\sin\theta \cos\theta - \frac{v_{_{\text{EW}}}^2 w^2 }{2 \Lambda^2}, \label{mus2} \\
	\lambda_s &=\frac{1}{2 w^2} \left( M_{h_1}^2 \sin^2\theta + M_{h_2}^2 \cos^2\theta \right) - \frac{v_{_{\text{EW}}}^2}{2 \Lambda^2} ,\label{lambdas} \\
	\lambda_{hs} &= -\frac{w^2}{\Lambda^2} + \frac{\left(-M_{h_1}^2 + M_{h_2}^2\right) \cos\theta \sin\theta}{v_{_{\text{EW}}} w} \label{lambdahs}
\end{align}
The cut-off scale $\Lambda$ which appears in the Lagrangian is one of the free parameters of the theory. In Figure \ref{lag_pam_w}, we plot several Lagrangian parameters against the singlet scalar VEV ($w$) for different choices of its mass $M_{h_2}$ and scalar mixing angle $\sin\theta$.

\section{Triple scalar couplings}
\label{trip}
\subsection*{$\bm{h_1-h_1-h_1}$ coupling}
The $h_1-h_1-h_1$ coupling can be expressed as

\begin{align}
\lambda_{h_1 h_1 h_1}
=
3\frac{M_{h_1}^2}{v_{_{\text{EW}}}}\left[
 \cos^3\theta
-\frac{v_{_{\text{EW}}}}{w}\sin^3\theta
+2\frac{v_{_{\text{EW}}}^2}{M_{h_1}^2}\frac{w^2}{\Lambda^2}\cos\theta \sin^2\theta
\right],
\label{lambda111}
\end{align}
highlighting the additional contribution due to the presence of the dimension-six operator.
\subsection*{$\bm{h_2-h_1-h_1}$ coupling}
\begin{figure}[h!]
	\centering
	\begin{subfigure}[b]{0.48\textwidth}
		\centering
		\includegraphics[width=\textwidth]{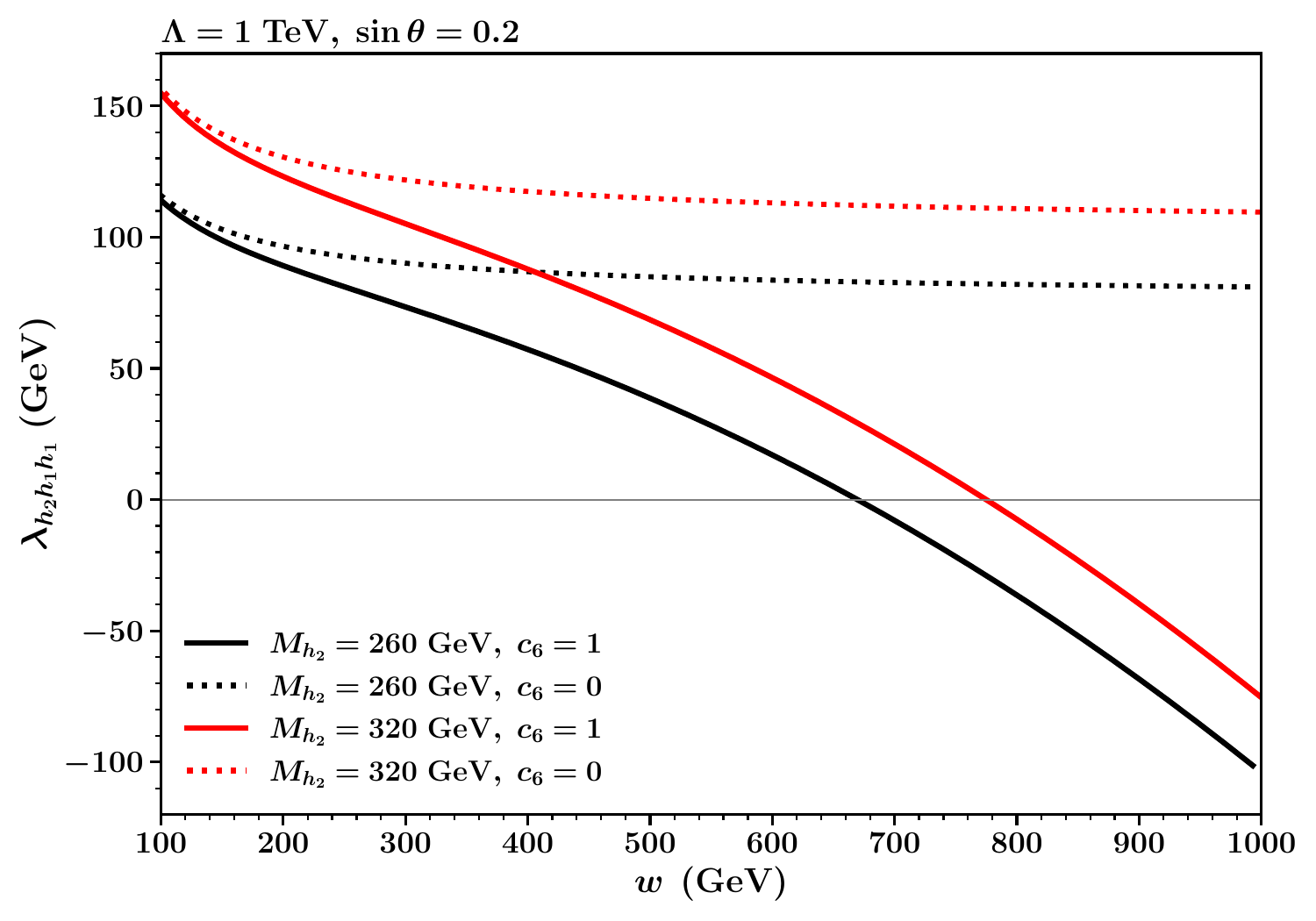}
		\caption{}
	\end{subfigure}
	\hfill
	\begin{subfigure}[b]{0.48\textwidth}
		\centering
		\includegraphics[width=\textwidth]{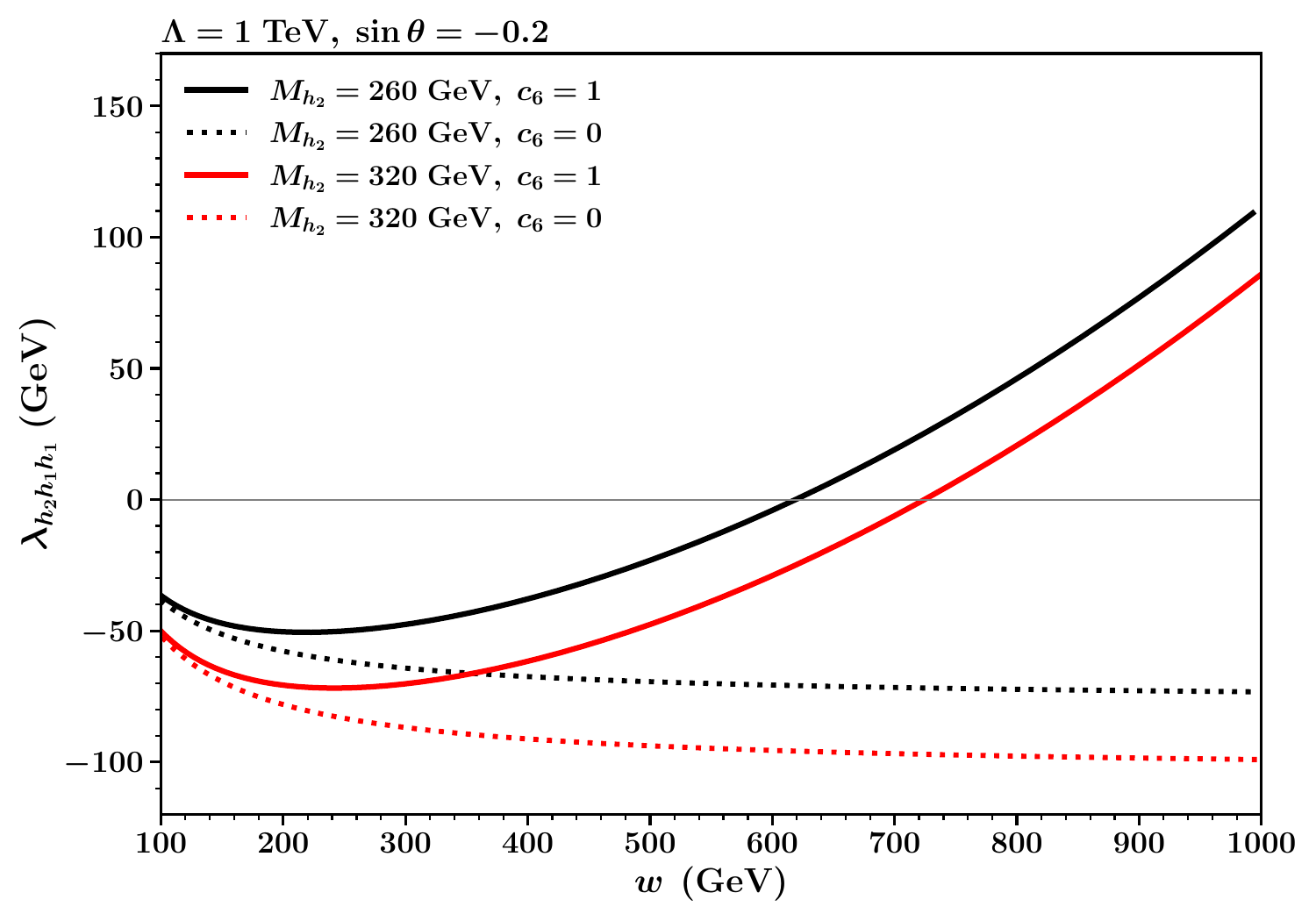}
		\caption{}
	\end{subfigure}
    \caption{Variation of $\lambda_{h_2h_1h_1}$ with singlet scalar VEV ($w$) for (a) $\sin\theta= 0.2$, (b) $\sin\theta= -0.2$ assuming $\Lambda =1$ TeV. Here, the dotted and solid lines represent the variation of this coupling in absence ($c_6=0$) and presence ($c_6=1$) of the dimension-six operator, respectively. }
    \label{lambda211_w}
    \end{figure}

Similarly, the modified $h_2-h_1-h_1$ coupling in presence of the dimension-six operator can be written as 

\begin{align}
\lambda_{h_2 h_1 h_1}
&=
\left(2 M_{h_1}^2 + M_{h_2}^2\right)
\sin\theta \cos\theta
\left(
\frac{\cos\theta}{v_{_{\text{EW}}}}
+
\frac{\sin\theta}{w}
\right)
- 2 v_{_{\text{EW}}}
\frac{w^2}{\Lambda^2}
\,\sin\theta
\left(2\cos^2\theta - \sin^2\theta \right) .
\label{lambda211}
\end{align}
 The variation of this coupling as a function of $w$ is shown in Figure~\ref{lambda211_w}, for different choices of $M_{h_2}$ and $\sin\theta$ with fixed cut-off scale $\Lambda= 1$ TeV .

\clearpage
\bibliographystyle{JHEP}
\bibliography{ref}

\end{document}